# Identification and Development of Therapeutics for COVID-19

This manuscript (permalink) was automatically generated from greenelab/covid19-review@89adbc3 on September 10, 2021. It is also available as a PDF. It represents one section of a larger evolving review on SARS-CoV-2 and COVID-19 available at https://greenelab.github.io/covid19-review/

**This in progress manuscript is not intended for the general public.** This is a review paper that is authored by scientists for an audience of scientists to discuss research that is in progress. If you are interested in guidelines on testing, therapies, or other issues related to your health, you should not use this document. Instead, you should collect information from your local health department, the CDC's guidance, or your own government.

# Authors

- **Halie M. Rando** 0000-0001-7688-1770 rando2 tamefoxtime · Department of Systems Pharmacology and Translational Therapeutics, University of Pennsylvania, Philadelphia, Pennsylvania, United States of America; Department of Biochemistry and Molecular Genetics, University of Colorado School of Medicine, Aurora, Colorado, United States of America; Center for Health AI, University of Colorado School of Medicine, Aurora, Colorado, United States of America · Funded by the Gordon and Betty Moore Foundation (GBMF 4552)

- **Nils Wellhausen** 0000-0001-8955-7582 nilswellhausen · Department of Systems Pharmacology and Translational Therapeutics, University of Pennsylvania, Philadelphia, Pennsylvania, United States of America

- **Soumita Ghosh** 0000-0002-2783-2750 soumitagh · Institute of Translational Medicine and Therapeutics, Perelman School of Medicine, University of Pennsylvania, Philadelphia, Pennsylvania, United States of America

- **Alexandra J. Lee** 0000-0002-0208-3730 ajlee21 · Department of Systems Pharmacology and Translational Therapeutics, University of Pennsylvania, Philadelphia, Pennsylvania, United States of America · Funded by the Gordon and Betty Moore Foundation (GBMF 4552)

- **Anna Ada Dattoli** 0000-0003-1462-831X aadattoli aadattoli · Department of Systems Pharmacology & Translational Therapeutics, Perelman School of Medicine, University of Pennsylvania, Philadelphia, PA 19104, USA

- **Fengling Hu** 0000-0003-1081-5038 hufengling hufengling · Department of Biostatistics, Epidemiology and Informatics, University of Pennsylvania, Philadelphia, Pennsylvania, United States of America

- **James Brian Byrd** 0000-0002-0509-3520 byrdjb thebyrdlab  University of Michigan School of Medicine, Ann Arbor, Michigan, United States of America · Funded by NIH K23HL128909; FastGrants

- **Diane N. Rafizadeh** 0000-0002-2838-067X dianerafi  Perelman School of Medicine, University of Pennsylvania, Philadelphia, Pennsylvania, United States of America; Department of Chemistry, University of Pennsylvania, Philadelphia, Pennsylvania, United States of America · Funded by NIH Medical Scientist Training Program T32 GM07170

- **Ronan Lordan** 0000-0001-9668-3368 RLordan el_ronan  Institute for Translational Medicine and Therapeutics, Perelman School of Medicine, University of Pennsylvania, Philadelphia, PA 19104-5158, USA

- **Yanjun Qi** 0000-0002-5796-7453 qiyanjun  Department of Computer Science, University of Virginia, Charlottesville, VA, United States of America

- **Yuchen Sun** kevinsunofficial  Department of Computer Science, University of Virginia, Charlottesville, VA, United States of America

- **Christian Brueffer** 0000-0002-3826-0989 cbrueffer cbrueffer  Department of Clinical Sciences, Lund University, Lund, Sweden

- **Jeffrey M. Field** 0000-0001-7161-7284 Jeff-Field  Department of Systems Pharmacology and Translational Therapeutics, Perelman School of Medicine, University of Pennsylvania, Philadelphia, PA 19104, USA

- **Marouen Ben Guebila** 0000-0001-5934-966X marouenbg marouenbg  Department of Biostatistics, Harvard School of Public Health, Boston, Massachusetts, United States of America

- **Nafisa M. Jadavji** 0000-0002-3557-7307 nafisajadavji nafisajadavji  Biomedical Science, Midwestern University, Glendale, AZ, United States of America; Department of Neuroscience, Carleton University, Ottawa, Ontario, Canada · Funded by the American Heart Association (20AIREA35050015)

- **Ashwin N. Skelly** 0000-0002-1565-3376 anskelly  Perelman School of Medicine, University of Pennsylvania, Philadelphia, Pennsylvania, United States of America; Institute for Immunology, University of Pennsylvania Perelman School of Medicine, Philadelphia, United States of America · Funded by NIH Medical Scientist Training Program T32 GM07170

- **Bharath Ramsundar** 0000-0001-8450-4262 rbharath rbhar90  The DeepChem Project, https://deepchem.io/

- **Jinhui Wang** 0000-0002-5796-8130 jinhui2  Perelman School of Medicine, University of Pennsylvania, Philadelphia, Pennsylvania, United States of America

- **Rishi Raj Goel** 0000-0003-1715-5191 rishirajgoel rishirajgoel  Institute for Immunology, University of Pennsylvania, Philadelphia, PA, United States of America

- **YoSon Park** 0000-0002-0465-4744 ypar **yoson**  Department of Systems Pharmacology and Translational Therapeutics, University of Pennsylvania, Philadelphia, Pennsylvania, United States of America · Funded by NHGRI R01 HG10067

- **COVID-19 Review Consortium**

- **Simina M. Boca** 0000-0002-1400-3398 SiminaB  Innovation Center for Biomedical Informatics, Georgetown University Medical Center, Washington, District of Columbia, United States of America; Early Biometrics & Statistical Innovation, Data Science & Artificial Intelligence, R & D, AstraZeneca, Gaithersburg, Maryland, United States of America

- **Anthony Gitter** 0000-0002-5324-9833 agitter anthonygitter  Department of Biostatistics and Medical Informatics, University of Wisconsin-Madison, Madison, Wisconsin, United States of America; Morgridge Institute for Research, Madison, Wisconsin, United States of America · Funded by John W. and Jeanne M. Rowe Center for Research in Virology

- **Casey S. Greene** 0000-0001-8713-9213 cgreene GreeneScientist  Department of Systems Pharmacology and Translational Therapeutics, University of Pennsylvania, Philadelphia, Pennsylvania, United States of America; Childhood Cancer Data Lab, Alex's Lemonade Stand Foundation, Philadelphia, Pennsylvania, United States of America; Department of Biochemistry and Molecular Genetics, University of Colorado School of Medicine, Aurora, Colorado, United States of America; Center for Health AI, University of Colorado School of Medicine, Aurora, Colorado, United States of America · Funded by the Gordon and Betty Moore Foundation (GBMF 4552); the National Human Genome Research Institute (R01 HG010067)

**COVID-19 Review Consortium:** Vikas Bansal, John P. Barton, Simina M. Boca, Joel D Boerckel, Christian Brueffer, James Brian Byrd, Stephen Capone, Shikta Das, Anna Ada Dattoli, John J. Dziak, Jeffrey M. Field, Soumita Ghosh, Anthony Gitter, Rishi Raj Goel, Casey S. Greene, Marouen Ben Guebila, Daniel S. Himmelstein, Fengling Hu, Nafisa M. Jadavji, Jeremy P. Kamil, Sergey Knyazev, Likhitha Kolla, Alexandra J. Lee, Ronan Lordan, Tiago Lubiana, Temitayo Lukan, Adam L. MacLean, David Mai, Serghei Mangul, David Manheim, Lucy D'Agostino McGowan, Amruta Naik, YoSon Park, Dimitri Perrin, Yanjun Qi, Diane N. Rafizadeh, Bharath Ramsundar, Halie M. Rando, Sandipan Ray, Michael P. Robson, Vincent Rubinetti, Elizabeth Sell, Lamonica Shinholster, Ashwin N. Skelly, Yuchen Sun, Yusha Sun, Gregory L Szeto, Ryan Velazquez, Jinhui Wang, Nils Wellhausen

Authors with similar contributions are ordered alphabetically.

# 1.1 Abstract

After emerging in China in late 2019, the novel coronavirus SARS-CoV-2 spread worldwide and as of mid-2021 remains a significant threat globally. Only a few coronaviruses are known to infect humans, and only two cause infections similar in severity to SARS-CoV-2: *Severe acute respiratory syndrome-related coronavirus*, a closely related species of SARS-CoV-2 that emerged in 2002, and *Middle East respiratory syndrome-related coronavirus*, which emerged in 2012. Unlike the current pandemic, previous epidemics were controlled rapidly through public health measures, but the body of research investigating severe acute respiratory syndrome and Middle East respiratory syndrome has proven valuable for identifying approaches to treating and preventing novel coronavirus disease 2019 (COVID-19). Building on this research, the medical and scientific communities have responded rapidly to the COVID-19 crisis to identify many candidate therapeutics. The approaches used to identify candidates fall into four main categories: adaptation of clinical approaches to diseases with related pathologies, adaptation based on virological properties, adaptation based on host response, and data-driven identification of candidates based on physical properties or on pharmacological compendia. To date, a small number of therapeutics have already been authorized by regulatory agencies such as the Food and Drug Administration (FDA), while most remain under investigation. The scale of the COVID-19 crisis offers a rare opportunity to collect data on the effects of candidate therapeutics. This information provides insight not only into the management of coronavirus diseases, but also into the relative success of different approaches to identifying candidate therapeutics against an emerging disease.

## 1.2  Importance

The COVID-19 pandemic is a rapidly evolving crisis. With the worldwide scientific community shifting focus onto the SARS-CoV-2 virus and COVID-19, a large number of possible pharmaceutical approaches for treatment and prevention have been proposed. What was known about each of these potential interventions evolved rapidly throughout 2020 and 2021. This fast-paced area of research provides important insight into how the ongoing pandemic can be managed and also demonstrates the power of interdisciplinary collaboration to rapidly understand a virus and match its characteristics with existing or novel pharmaceuticals. As illustrated by the continued threat of viral epidemics during the current millennium, a rapid and strategic response to emerging viral threats can save lives. In this review, we explore how different modes of identifying candidate therapeutics have borne out during COVID-19.

## 2  Introduction

The novel coronavirus *Severe acute respiratory syndrome-related coronavirus 2* (SARS-CoV-2) emerged in late 2019 and quickly precipitated the worldwide spread of novel coronavirus disease 2019 (COVID-19). COVID-19 is associated with symptoms ranging from mild or even asymptomatic to severe, and up to 2% of patients diagnosed with COVID-19 die from COVID-19-related complications such as acute respiratory disease syndrome (ARDS) [1]. As a result, public health efforts have been critical to mitigating the spread of the virus. However, as of mid-2021, COVID-19 remains a significant worldwide concern (Figure 1), with 2021 cases in some regions surging far above the numbers reported during the initial outbreak in early 2020. While a number of vaccines have been developed and approved in different countries starting in late 2020 [2], vaccination efforts have not proceeded at the same pace throughout the world and are not yet close to ending the pandemic.

Due to the continued threat of the virus and the severity of the disease, the identification and development of therapeutic interventions have emerged as significant international priorities. Prior developments during other recent outbreaks of emerging diseases, especially those caused by human coronaviruses (HCoV), have guided biomedical research into the behavior and treatment of this novel coronavirus infection. However, previous emerging HCoV-related disease threats were controlled much more quickly than SARS-CoV-2 through public health efforts (Figure 1). The scale of the COVID-19 pandemic has made the repurposing and development of pharmaceuticals more urgent than in previous coronavirus epidemics.

## 2.1 Lessons from Prior HCoV Outbreaks

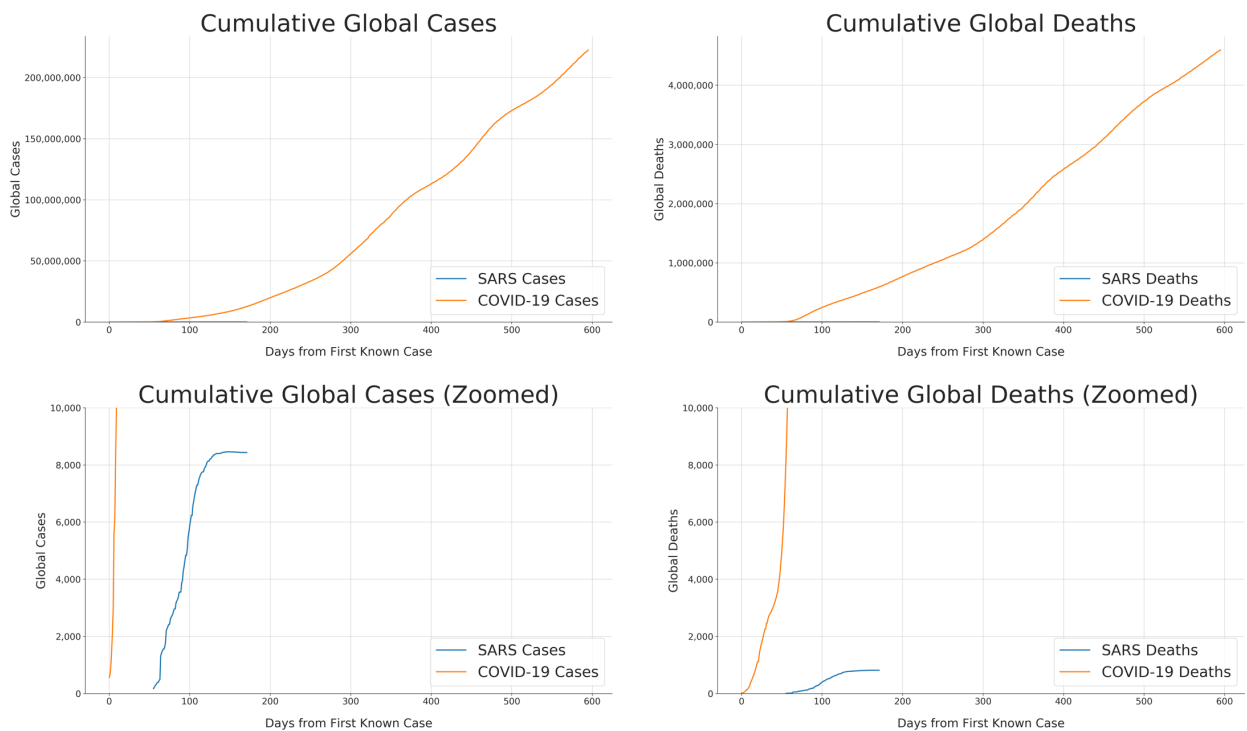

*Figure 1: **Cumulative global incidence of COVID-19 and SARS.** As of September 8, 2021, 222,559,803 COVID-19 cases and 4,596,394 COVID-19 deaths had been reported worldwide since January 22, 2020. A total of 8,432 cases and 813 deaths were reported for SARS from March 17 to July 11, 2003. SARS-CoV-1 was officially contained on July 5, 2003, within 9 months of its appearance [3]. In contrast, SARS-CoV-2 remains a significant global threat nearly two years after its emergence. COVID-19 data are from the COVID-19 Data Repository by the Center for Systems Science and Engineering at Johns Hopkins University [4,5]. SARS data are from the WHO [6] and were obtained from a dataset on GitHub [7]. See https://greenelab.github.io/covid19-review/ for the most recent version of this figure, which is updated daily.*

At first, SARS-CoV-2's rapid shift from an unknown virus to a significant worldwide threat closely paralleled the emergence of *Severe acute respiratory syndrome-related coronavirus* (SARS-CoV-1), which was responsible for the 2002-03 SARS epidemic. The first documented case of COVID-19 was reported in Wuhan, China in November 2019, and the disease quickly spread worldwide in the early months of 2020. In comparison, the first case of SARS was reported in November 2002 in the Guangdong Province of China, and it spread within China and then into several countries across continents during the first half of 2003 [3,8,9]. In fact, genome sequencing quickly revealed the virus causing COVID-19 to be a novel betacoronavirus closely related to SARS-CoV-1 [10].

While similarities between these two viruses are unsurprising given their close phylogenetic relationship, there are also some differences in how the viruses affect humans. SARS-CoV-1 infection is severe, with an estimated case fatality rate (CFR) for SARS of 9.5% [8], while estimates of the CFR associated with COVID-19 are much lower, at up to 2% [1]. SARS-CoV-1 is highly contagious and spread primarily by droplet transmission, with a basic reproduction number ($R_0$) of 4 (i.e., each person infected was estimated to infect four other people) [8]. There is still some controversy whether SARS-CoV-2 is primarily spread by droplets or is primarily airborne [11,12,13,14]. Most estimates of its $R_0$ fall between 2.5 and 3 [1]. Therefore, SARS is thought to be a deadlier and more transmissible disease than COVID-19.

With the 17-year difference between these two outbreaks, there were major differences in the tools available to efforts to organize international responses. At the time that SARS-CoV-1 emerged, no new HCoV had been identified in almost 40 years [9]. The identity of the virus underlying the SARS disease remained unknown until April of 2003, when the SARS-CoV-1 virus was characterized through a worldwide scientific effort spearheaded by the World Health Organization (WHO) [9]. In contrast, the SARS-CoV-2 genomic sequence was released on January 3, 2020 [10], only days after the international community became aware of the novel pneumonia-like illness now known as COVID-19. While SARS-CoV-1 belonged to a distinct lineage from the two other HCoVs known at the time of its discovery [8], SARS-CoV-2 is closely related to SARS-CoV-1 and is a more distant relative of another HCoV characterized in 2012, *Middle East respiratory syndrome-related coronavirus* [15,16]. Significant efforts had been dedicated towards understanding SARS-CoV-1 and MERS-CoV and how they interact with human hosts. Therefore, SARS-CoV-2 emerged under very different circumstances than SARS-CoV-1 in terms of scientific knowledge about HCoVs and the tools available to characterize them.

Despite the apparent advantages for responding to SARS-CoV-2 infections, COVID-19 has caused many orders of magnitude more deaths than SARS did (Figure 1). The SARS outbreak was officially determined to be under control in July 2003, with the success credited to infection management practices such as mask wearing [9]. *Middle East respiratory syndrome-related coronavirus* (MERS-CoV) is still circulating and remains a concern; although the fatality rate is very high at almost 35%, the disease is much less easily transmitted, as its $R_0$ has been estimated to be 1 [8]. The low $R_0$ in combination with public health practices allowed for its spread to be contained [8]. Neither of these trajectories are comparable to that of SARS-CoV-2, which remains a serious threat worldwide over a year and a half after the first cases of COVID-19 emerged (Figure 1).

## 2.2 Potential Approaches to the Treatment of COVID-19

Therapeutic interventions can utilize two approaches: they can either mitigate the effects of an infection that harms an infected person, or they can hinder the spread of infection within a host by disrupting the viral life cycle. The goal of the former strategy is to reduce the severity and risks of an active infection, while for the latter, it is to inhibit the replication of a virus once an individual is infected, potentially freezing disease progression. Additionally, two major approaches can be used to identify interventions that might be relevant to managing an emerging disease or a novel virus: drug repurposing and drug development. Drug repurposing involves identifying an existing compound that may provide benefits in the context of interest [17]. This strategy can focus on either approved or investigational drugs, for which there may be applicable preclinical or safety information [17]. Drug development, on the other hand, provides an opportunity to identify or develop a compound specifically relevant to a particular need, but it is often a lengthy and expensive process characterized by repeated failure [18]. Drug repurposing therefore tends to be emphasized in a situation like the COVID-19 pandemic due to the potential for a more rapid response.

Even from the early months of the pandemic, studies began releasing results from analyses of approved and investigational drugs in the context of COVID-19. The rapid timescale of this response meant that, initially, most evidence came from observational studies, which compare groups of patients who did and did not receive a treatment to determine whether it may have had an effect. This type of study can be conducted rapidly but is subject to confounding. In contrast, randomized controlled trials (RCTs) are the gold-standard method for assessing the effects of an intervention. Here, patients are prospectively and randomly assigned to treatment or control conditions, allowing for much stronger interpretations to be drawn; however, data from these trials take much longer to collect. Both approaches have proven to be important sources of information in the development of a rapid response to the COVID-19 crisis, but as the pandemic draws on and more results become available from RCTs, more definitive answers are becoming available about proposed therapeutics. Interventional clinical trials are currently investigating or have investigated a large number of possible therapeutics and combinations of therapeutics for the treatment of COVID-19 (Figure 2).

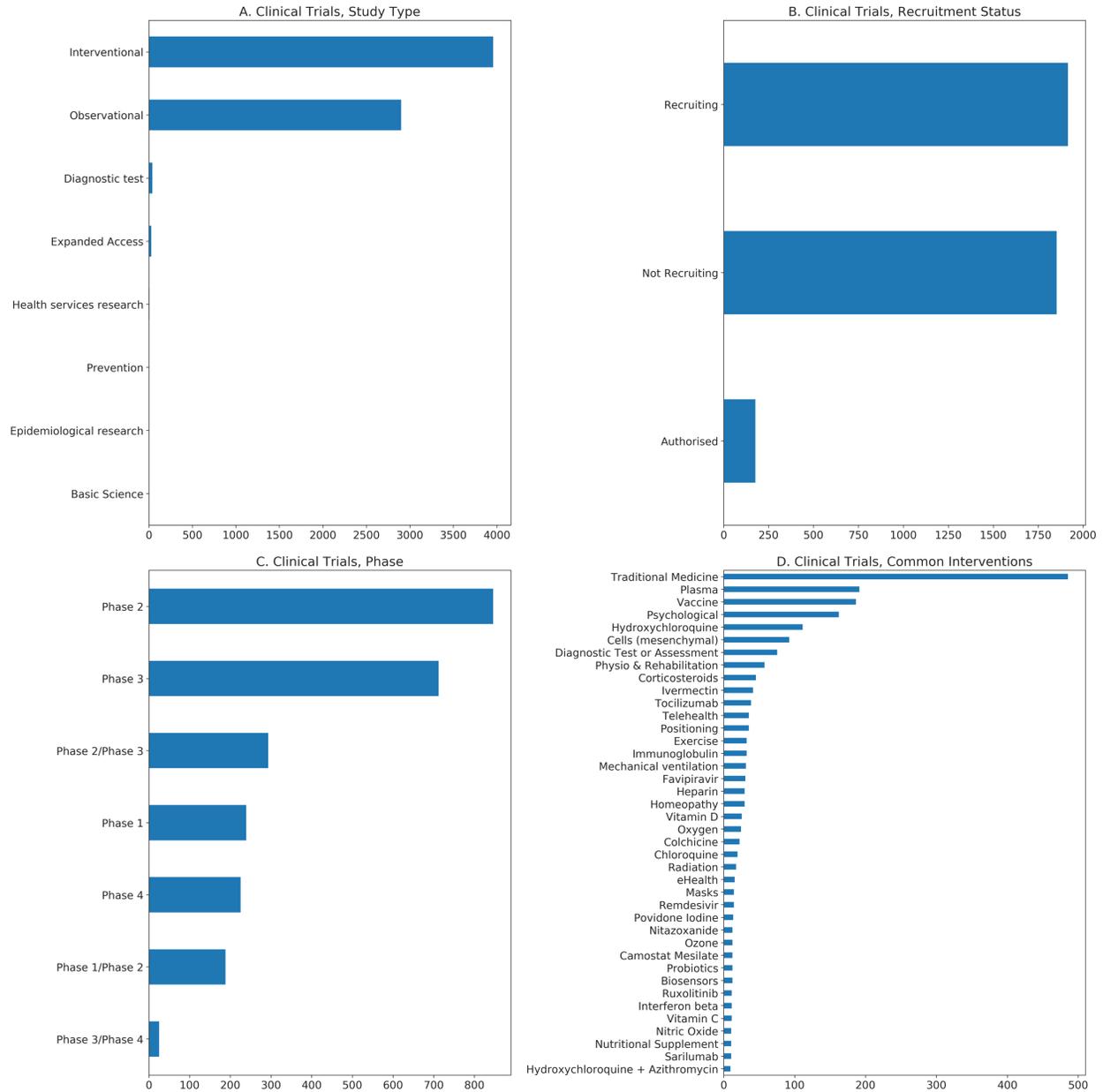

Figure 2: **COVID-19 clinical trials.** Trials data are from the University of Oxford Evidence-Based Medicine Data Lab's COVID-19 TrialsTracker [19]. As of December 31, 2020, there were 6,987 COVID-19 clinical trials of which 3,962 were interventional. The study types include only types used in at least five trials. Only interventional trials are analyzed in the figures depicting status, phase, and intervention. Of the interventional trials, 98 trials had reported results as of December 31, 2020. Recruitment status and trial phase are shown only for interventional trials in which the status or phase is recorded. Common interventions refers to interventions used in at least ten trials. Combinations of interventions, such as hydroxychloroquine with azithromycin, are tallied separately from the individual interventions. See https://greenelab.github.io/covid19-review/ for the most recent version of this figure, which is updated daily.

The purpose of this review is to provide an evolving resource tracking the status of efforts to repurpose and develop drugs for the treatment of COVID-19. We highlight four strategies that provide different paradigms for the identification of potential pharmaceutical treatments. The WHO guidelines [20] and a systematic review [21] are complementary living documents that summarize COVID-19 therapeutics.

# 3 Repurposing Drugs for Symptom Management

A variety of symptom profiles with a range of severity are associated with COVID-19 [1]. In many cases, COVID-19 is not life threatening. A study of COVID-19 patients in a hospital in Berlin, Germany reported that the highest risk of death was associated with infection-related symptoms, such as sepsis, respiratory symptoms such as ARDS, and cardiovascular failure or pulmonary embolism [22]. Similarly, an analysis in Wuhan, China reported that respiratory failure (associated with ARDS) and sepsis/multi-organ failure accounted for 69.5% and 28.0% of deaths, respectively, among 82 deceased patients [23]. COVID-19 is characterized by two phases. The first is the acute response, where an adaptive immune response to the virus is established and in many cases can mitigate viral damage to organs [24]. The second phase characterizes more severe cases of COVID-19. Here, patients experience a cytokine storm, whereby excessive production of cytokines floods into circulation, leading to systemic inflammation, immune dysregulation, and multiorgan dysfunction that can cause multiorgan failure and death if untreated [25]. ARDS-associated respiratory failure can occur during this phase. Cytokine dysregulation was also identified in patients with SARS [26,27].

In the early days of the COVID-19 pandemic, physicians sought to identify potential treatments that could benefit patients, and in some cases shared their experiences and advice with the medical community on social media sites such as Twitter [28]. These on-the-ground treatment strategies could later be analyzed retrospectively in observational studies or investigated in an interventional paradigm through RCTs. Several notable cases involved the use of small-molecule drugs, which are synthesized compounds of low molecular weight, typically less than 1 kilodalton (kDa) [29]. Small-molecule pharmaceutical agents have been a backbone of drug development since the discovery of penicillin in the early twentieth century [30]. It and other antibiotics have long been among the best known applications of small molecules to therapeutics, but biotechnological developments such as the prediction of protein-protein interactions (PPIs) have facilitated advances in precise targeting of specific structures using small molecules [30]. Small molecule drugs today encompass a wide range of therapeutics beyond antibiotics, including antivirals, protein inhibitors, and many broad-spectrum pharmaceuticals.

Many treatments considered for COVID-19 have relied on a broad-spectrum approach. These treatments do not specifically target a virus or particular host receptor, but rather induce broad shifts in host biology that are hypothesized to be potential inhibitors of the virus. This approach relies on the fact that when a virus enters a host, the host becomes the virus's environment. Therefore, the state of the host can also influence the virus's ability to replicate and spread. The administration and assessment of broad-spectrum small-molecule drugs on a rapid time course was feasible because they are often either available in hospitals, or in some cases may also be prescribed to a large number of out-patients. One of the other advantages is that these well-established compounds, if found to be beneficial, are often widely available, in contrast to boutique experimental drugs.

In some cases, prior data was available from experiments examining the response of other HCoVs or HCoV infections to a candidate drug. In addition to non-pharmaceutical interventions such as encouraging non-intubated patients to adopt a prone position [31], knowledge about interactions between HCoVs and the human body, many of which emerged from SARS and MERS research over the past two decades, led to the suggestion that a number of common drugs might benefit COVID-19 patients. However, the short duration and low case numbers of prior outbreaks were less well-suited to the large-scale study of clinical applications than the COVID-19 pandemic is. As a result, COVID-19 has presented the first opportunity to robustly evaluate treatments that were common during prior HCoV outbreaks to determine their clinical efficacy. The first year of the COVID-19 pandemic demonstrated that there are several different trajectories that these clinically suggested, widely available candidates can follow when assessed against a widespread, novel viral threat.

One approach to identifying candidate small molecule drugs was to look at the approaches used to treat SARS and MERS. Treatment of SARS and MERS patients prioritized supportive care and symptom management [8]. Among the clinical treatments for SARS and MERS that were explored, there was generally a lack of evidence indicating whether they were effective. Most of the supportive treatments for SARS were found inconclusive in meta-analysis [32], and a 2004 review reported that not enough evidence was available to make conclusions about most treatments [33]. However, one strategy adopted from prior HCoV outbreaks is currently the best-known treatment for severe cases of COVID-19. Corticosteroids represent broad-spectrum treatments and are a well-known, widely available treatment for pneumonia [34,35,36,37,38,39] that have also been debated as a possible treatment for ARDS [40,41,42,43,44,45]. Corticosteroids were also used and subsequently evaluated as possible supportive care for SARS and MERS. In general, studies and meta-analyses did not identify support for corticosteroids to prevent mortality in these HCoV infections [46,47,48]; however, one found that the effects might be masked by variability in treatment protocols, such as dosage and timing [33]. While the corticosteroids most often used to treat SARS were methylprednisolone and hydrocortisone, availability issues for these drugs at the time led to dexamethasone also being used in North America [49].

Dexamethasone (9α-fluoro-16α-methylprednisolone) is a synthetic corticosteroid that binds to glucocorticoid receptors [50,51]. It functions as an anti-inflammatory agent by binding to glucocorticoid receptors with higher affinity than endogenous cortisol [52]. Dexamethasone and other steroids are widely available and affordable, and they are often used to treat community-acquired pneumonia [53] as well as chronic inflammatory conditions such as asthma, allergies, and rheumatoid arthritis [54,55,56]. Immunosuppressive drugs such as steroids are typically contraindicated in the setting of infection [57], but because COVID-19 results in hyperinflammation that appears to contribute to mortality via lung damage, immunosuppression may be a helpful approach to treatment [58]. A clinical trial that began in 2012 recently reported that dexamethasone may improve outcomes for patients with ARDS [40], but a meta-analysis of a small amount of available data about dexamethasone as a treatment for SARS suggested that it may, in fact, be associated with patient harm [59]. However, the findings in SARS may have been biased by the fact that all of the studies examined were observational and a large number of inconclusive studies were not included [60]. The questions of whether and when to counter hyperinflammation with immunosuppression in the setting of COVID-19 (as in SARS [27]) was an area of intense debate, as the risks of inhibiting antiviral immunity needed to be weighed against the beneficial anti-inflammatory effects [61]. As a result, guidelines early in the pandemic typically recommended avoiding treating COVID-19 patients with corticosteroids such as dexamethasone [59].

Despite this initial concern, dexamethasone was evaluated as a potential treatment for COVID-19 (Appendix 1). Dexamethasone treatment comprised one arm of the multi-site Randomized Evaluation of COVID-19 Therapy (RECOVERY) trial in the United Kingdom [62]. This study found that the 28-day mortality rate was lower in patients receiving dexamethasone than in those receiving standard of care (SOC). However, this finding was driven by differences in mortality among patients who were receiving mechanical ventilation or supplementary oxygen at the start of the study. The report indicated that dexamethasone reduced 28-day mortality relative to SOC in patients who were ventilated (29.3% versus 41.4%) and among those who were receiving oxygen supplementation (23.3% versus 26.2%) at randomization, but not in patients who were breathing independently (17.8% versus 14.0%). These findings also suggested that dexamethasone may have reduced progression to mechanical ventilation, especially among patients who were receiving oxygen support at randomization. Other analyses have supported the importance of disease course in determining the efficacy of dexamethasone: additional results suggest greater potential for patients who have experienced symptoms for at least seven days and patients who were not breathing independently [63]. A meta-analysis that evaluated the results of the RECOVERY trial alongside trials of other corticosteroids, such as hydrocortisone, similarly concluded that corticosteroids may be beneficial to patients with severe COVID-19 who are receiving oxygen supplementation [64]. Thus, it seems likely that dexamethasone is useful for treating inflammation associated with immunopathy or cytokine release syndrome (CRS), which is a condition caused by detrimental overactivation of the immune system [1]. In fact, corticosteroids such as dexamethasone are sometimes used to treat CRS [65]. Guidelines were quickly updated to encourage the use of dexamethasone in severe cases [66], and this affordable and widely available treatment rapidly became a valuable tool against COVID-19 [67], with demand surging within days of the preprint's release [68].

# 4 Approaches Targeting the Virus

Therapeutics that directly target the virus itself hold the potential to prevent people infected with SARS-CoV-2 from developing potentially damaging symptoms (Figure 3). Such drugs typically fall into the broad category of antivirals. Antiviral therapies hinder the spread of a virus within the host, rather than destroying existing copies of the virus, and these drugs can vary in their specificity to a narrow or broad range of viral targets. This process requires inhibiting the replication cycle of a virus by disrupting one of six fundamental steps [69]. In the first of these steps, the virus attaches to and enters the host cell through endocytosis. Then the virus undergoes uncoating, which is classically defined as the release of viral contents into the host cell. Next, the viral genetic material enters the nucleus where it gets replicated during the biosynthesis stage. During the assembly stage, viral proteins are translated, allowing new viral particles to be assembled. In the final step new viruses are released into the extracellular environment. Although antivirals are designed to target a virus, they can also impact other processes in the host and may have unintended effects. Therefore, these therapeutics must be evaluated for both efficacy and safety. As the technology to respond to emerging viral threats has also evolved over the past two decades, a number of candidate treatments have been identified for prior viruses that may be relevant to the treatment of COVID-19.

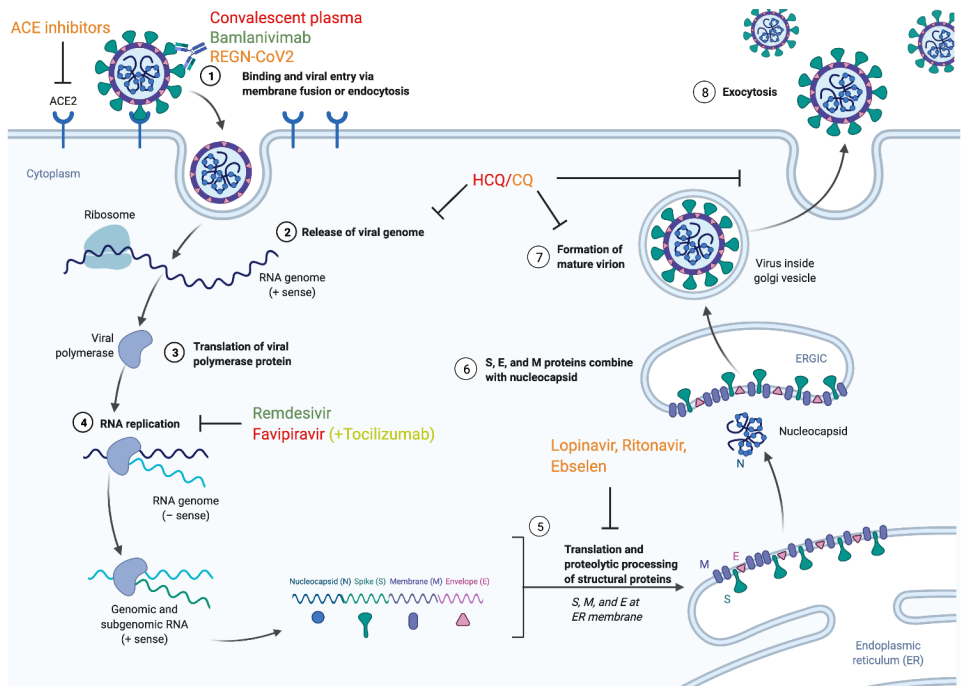

*Figure 3:* **Mechanisms of Action for Potential Therapeutics** *Potential therapeutics currently being studied can target the SARS-CoV-2 virus or modify the host environment through many different mechanisms. Here, the relationships between the virus, host cells, and several therapeutics are visualized. Drug names are color-coded according to the grade assigned to them by the Center for Cytokine Storm Treatment & Laboratory's CORONA Project [70] (Green = A, Lime = B, Orange = C, and Red = D).*

Many antiviral drugs are designed to inhibit the replication of viral genetic material during the biosynthesis step. Unlike DNA viruses, which can use the host enzymes to propagate themselves, RNA viruses like SARS-CoV-2 depend on their own polymerase, the RNA-dependent RNA polymerase (RdRP), for replication [71,72]. RdRP is therefore a potential target for antivirals against RNA viruses. Disruption of RdRP is the proposed mechanism underlying the treatment of SARS and MERS with ribavirin [73]. Ribavirin is an antiviral drug effective against other viral infections that was often used in combination with corticosteroids and sometimes interferon (IFN) medications to treat SARS and MERS [9]. However, analyses of its effects in retrospective and *in vitro* analyses of SARS and the SARS-CoV-1 virus, respectively, have been inconclusive [9]. While IFNs and ribavirin have shown promise in *in vitro* analyses of MERS, their clinical effectiveness remains unknown [9]. The current COVID-19 pandemic has provided an opportunity to assess the clinical effects of these treatments. As one example, ribivarin was also used in the early days of COVID-19, but a retrospective cohort study comparing patients who did and did not receive ribivarin revealed no effect on the mortality rate [74].

Since nucleotides and nucleosides are the natural building blocks for RNA synthesis, an alternative approach has been to explore nucleoside and nucleotide analogs for their potential to inhibit viral replication. Analogs containing modifications to nucleotides or nucleosides can disrupt key processes including replication [75]. A single incorporation does not influence RNA transcription; however, multiple events of incorporation lead to the arrest of RNA synthesis [76]. One candidate antiviral considered for the treatment of COVID-19 is favipiravir (Avigan), also known as T-705, which was discovered by Toyama Chemical Co., Ltd. [77]. It was previously found to be effective at blocking viral amplification in several influenza subtypes as well as other RNA viruses, such as *Flaviviridae* and *Picornaviridae*, through a reduction in plaque formation [78] and viral replication in Madin-Darby canine kidney cells [79]. Favipiravir (6-fluoro-3-hydroxy-2-pyrazinecarboxamide) acts as a purine and purine nucleoside analogue that inhibits viral RNA polymerase in a dose-dependent manner across a range of RNA viruses, including influenza viruses [80,81,82,83,84]. Biochemical experiments showed that favipiravir was recognized as a purine nucleoside analogue and incorporated into the viral RNA template. In 2014, the drug was approved in Japan for the treatment of influenza that was resistant to conventional treatments like neuraminidase inhibitors [85]. Though initial analyses of favipiravir in observational studies of its effects on COVID-19 patients were promising, recent results of two small RCTs suggest that it is unlikely to affect COVID-19 outcomes (Appendix 1).

In contrast, another nucleoside analog, remdesivir, is one of the few treatments against COVID-19 that has received FDA approval. Remdesivir (GS-5734) is an intravenous antiviral that was proposed by Gilead Sciences as a possible treatment for Ebola virus disease. It is metabolized to GS-441524, an adenosine analog that inhibits a broad range of polymerases and then evades exonuclease repair, causing chain termination [86,87,88]. Gilead received an emergency use authorization (EUA) for remdesivir from the FDA early in the pandemic (May 2020) and was later found to reduce mortality and recovery time in a double-blind, placebo-controlled, phase III clinical trial performed at 60 trial sites, 45 of which were in the United States [89,90,91,92]. Subsequently, the WHO Solidarity trial, a large-scale, open-label trial enrolling 11,330 adult in-patients at 405 hospitals in 30 countries around the world, reported no effect of remdesivir on in-hospital mortality, duration of hospitalization, or progression to mechanical ventilation [93]. Therefore, additional clinical trials of remdesivir in different patient pools and in combination with other therapies may be needed to refine its use in the clinic and determine the forces driving these differing results. Remdesivir offers proof of principle that SARS-CoV-2 can be targeted at the level of viral replication, since remdesivir targets the viral RNA polymerase at high potency. Identification of such candidates depends on knowledge about the virological properties of a novel threat. However, the success and relative lack of success, respectively, of remdesivir and favipiravir underscore the fact that drugs with similar mechanisms will not always produce similar results in clinical trials.

# 5 Disrupting Host-Virus Interactions

## 5.1 Interrupting Viral Colonization of Cells

Some of the most widely publicized examples of efforts to repurpose drugs for COVID-19 are broad-spectrum, small-molecule drugs where the mechanism of action made it seem that the drug might disrupt interactions between SARS-CoV-2 and human host cells (Figure 3). However, the exact outcomes of such treatments are difficult to predict *a priori*, and there are several examples where early enthusiasm was not borne out in subsequent trials. One of the most famous examples of an analysis of whether a well-known medication could provide benefits to COVID-19 patients came from the assessment of chloroquine (CQ) and hydroxychloroquine (HCQ), which are used for the treatment and prophylaxis of malaria as well as the treatment of lupus erythematosus and rheumatoid arthritis in adults [94]. These drugs are lysosomotropic agents, meaning they are weak bases that can pass through the plasma membrane. It was thought that they might provide benefits against SARS-CoV-2 by interfering with the digestion of antigens within the lysosome and inhibiting CD4 T-cell stimulation while promoting the stimulation of CD8 T-cells [95]. These compounds also have anti-inflammatory properties [95] and can decrease the production of certain key cytokines involved in the immune response, including interleukin-6 (IL-6) and inhibit the stimulation of Toll-like receptors (TLR) and TLR signaling [95].

*In vitro* analyses reported that CQ inhibited cell entry of SARS-CoV-1 [96] and that both CQ and HCQ inhibited viral replication within cultured cells [97], leading to early hope that it might provide similar therapeutic or protective effects in patients. However, while the first publication on the clinical application of these compounds to the inpatient treatment of COVID-19 was very positive [98], it was quickly discredited [99]. Over the following months, extensive evidence emerged demonstrating that CQ and HCQ offered no benefits for COVID-19 patients and, in fact, carried the risk of dangerous side effects (Appendix 1). The nail in the coffin came when findings from the large-scale RECOVERY trial were released on October 8, 2020. This study enrolled 11,197 hospitalized patients whose physicians believed it would not harm them to participate and used a randomized, open-label design to study the effects of HCQ compared to standard of care (SOC) at 176 hospitals in the United Kingdom [100]. Rates of COVID-19-related mortality did not differ between the control and HCQ arms, but patients receiving HCQ were slightly more likely to die due to cardiac events. Patients who received HCQ also had a longer duration of hospitalization than patients receiving usual care and were more likely to progress to mechanical ventilation or death (as a combined outcome). As a result, enrollment in the HCQ arm of the RECOVERY trial was terminated early [101]. The story of CQ/HCQ therefore illustrates how initial promising *in vitro* analyses can fail to translate to clinical usefulness.

A similar story has arisen with the broad-spectrum, small-molecule anthelmintic ivermectin, which is a synthetic analog of avermectin, a bioactive compound produced by a microorganism known as *Streptomyces avermectinius* and *Streptomyces avermitilis* [102,103]. Avermectin disrupts the ability of parasites to avoid the host immune response by blocking glutamate-gated chloride ion channels in the peripheral nervous system from closing, leading to hyperpolarization of neuronal membranes, disruption of neural transmission, and paralysis

[102,104,105]. Ivermectin has been used since the early 1980s to treat endo- and ecto-parasitic infections by helminths, insects, and arachnids in veterinary contexts [102,106] and since the late 1980s to treat human parasitic infections as well [102,104]. More recent research has indicated that ivermectin might function as a broad-spectrum antiviral by disrupting the trafficking of viral proteins by both RNA and DNA viruses [105,107,108], although most of these studies have demonstrated this effect *in vitro* [108]. The potential for antiviral effects on SARS-CoV-2 were investigated *in vitro*, and ivermectin was found to inhibit viral replication in a cell line derived from Vero cells (Vero-hSLAM) [109]. However, inhibition of viral replication was achieved at concentrations that were much higher than that explored by existing dosage guidelines [110,111], which are likely to be associated with significant side effects due to the increased potential that the compound could cross the mammalian blood-brain barrier [112,113].

Retrospective studies and small RCTs began investigating the effects of standard doses of this low-cost, widely available drug. One retrospective study reported that ivermectin reduced all-cause mortality [114] while another reported no difference in clinical outcomes or viral clearance [115]. Small RCTs enrolling less than 50 patients per arm have also reported a wide array of positive [116,117,118,119,120] and negative results [121,122]. A slightly larger RCT enrolling 115 patients in two arms reported inconclusive results [123]. Hope for the potential of ivermectin peaked with the release of a preprint reporting results of a multicenter, double-blind RCT where a four-day course of ivermectin was associated with clinical improvement and earlier viral clearance in 400 symptomatic patients and 200 close contacts [124]; however, concerns were raised about both the integrity of the data and the paper itself [125,126], and this study was removed by the preprint server Research Square [127]. A similarly sized RCT suggested no effect on the duration of symptoms among 400 patients split evenly across the intervention and control arms [128], and although meta-analyses have reported both null [129,130] and beneficial [131,132,133,134,135,136,137,138] effects of ivermectin on COVID-19 outcomes, the certainty is likely to be low [132]. These findings are potentially biased by a small number of low-quality studies, including the preprint that has been taken down [139], and the authors of one [140] have issued a notice [131] that they will revise their study with the withdrawn study removed. Thus, much like HCQ/CQ, enthusiasm for research that either has not or should not have passed peer review has led to large numbers of patients worldwide receiving treatments that might not have any effect or could even be harmful. Additionally, comments on the now-removed preprint include inquiries into how best to self-administer veterinary ivermectin as a prophylactic [127], and the FDA has posted information explaining why veterinary ivermectin should not be taken by humans concerned about COVID-19 [141]. Ivermectin is now one of several candidate therapeutics being investigated in the large-scale TOGETHER [142] and PRINCIPLE [143] clinical trials. The TOGETHER trial, which previously demonstrated no effect of HCQ and lopinavir-ritonavir [144], released preliminary results in early August 2021 suggesting that ivermectin also has no effect on COVID-19 outcomes [145].

While CQ/HCQ and ivermectin are well-known medications that have long been prescribed in certain contexts, investigation of another well-established type of pharmaceutical was facilitated by the fact that it was already being taken by a large number of COVID-19 patients. Angiotensin-converting enzyme inhibitors (ACEIs) and angiotensin II receptor blockers (ARBs) are among today's most commonly prescribed medications, often being used to control blood pressure [146,147]. In the United States, for example, they are prescribed well over 100,000,000 times annually [148]. Prior to the COVID-19 pandemic, the relationship between ACE2, ACEIs, and SARS had been considered as possible evidence that ACE2 could serve as a therapeutic target [149], and the connection had been explored through *in vitro* and molecular docking analysis [150] but ultimately was not pursued clinically [151]. Data from some animal models suggest that several, but not all, ACEIs and several ARBs increase ACE2 expression in the cells of some organs [152], but clinical studies have not established whether plasma ACE2 expression is increased in humans treated with these medications [153]. In this case, rather than introducing ARBs/ACEIs, a number of analyses have investigated whether discontinuing use affects COVID-19 outcomes. An initial observational study of the association of exposure to ACEIs or ARBs with outcomes in COVID-19 was retracted from the *New England Journal of Medicine* [154] due to concerns related to data availability [155]. As RCTs have become available, they have demonstrated no effect of continuing versus discontinuing ARBs/ACEIs on patient outcomes [156,157] (Appendix 1). Thus, once again, despite a potential mechanistic association with the pathology of SARS-CoV-2 infection, these medications were not found to influence the trajectory of COVID-19 illness.

For medications that are widely known and common, clinical research into their efficacy against a novel threat can be developed very quickly. This feasibility can present a double-edged sword. For example, HCQ and CQ were incorporated into SOC in many countries early in the pandemic and had to be discontinued once their potential to harm COVID-19 patients became apparent [158,159]. Dexamethasone remains the major success story from this category of repurposed drugs and is likely to have saved a large number of lives since summer 2020 [67].

## 5.2 Manipulating the Host Immune Response

Treatments based on understanding a virus and/or how a virus interacts with the human immune system can fall into two categories: they can interact with the innate immune response, which is likely to be a similar response across viruses, or they can be specifically designed to imitate the adaptive immune response to a particular virus. In the latter case, conservation of structure or behavior across viruses enables exploring whether drugs developed for one virus can treat another. During the COVID-19 pandemic, a number of candidate therapeutics have been explored in these categories, with varied success.

Knowledge gained from trying to understand SARS-CoV-1 and MERS-CoV from a fundamental biological perspective and characterize how they interact with the human immune system provides a theoretical basis for identifying candidate therapies. Biologics are a particularly important class of drugs for efforts to address HCoV through this paradigm. They are produced from components of living organisms or viruses, historically primarily from animal tissues [160]. Biologics have become increasingly feasible to produce as recombinant DNA technologies have advanced [160].

There are many differences on the development side between biologics and synthesized pharmaceuticals, such as small molecule drugs. Typically, biologics are orders of magnitude larger than small molecule drugs and are catabolized by the body to their amino acid components [161]. They are often heat sensitive, and their toxicity can vary, as it is not directly associated with the primary effects of the drug; in general, their physiochemical properties are much less understood compared to small molecules [161]. Biologics include significant medical breakthroughs such as insulin for the management of diabetes and vaccines, as well monoclonal antibodies (mAbs) and interferons (IFNs), which can be used to target the host immune response after infection.

mAbs have revolutionized the way we treat human diseases and have become some of the best-selling drugs in the pharmaceutical market in recent years [162]. There are currently 79 FDA approved mAbs on the market, including antibodies for viral infections (e.g. Ibalizumab for *Human immunodeficiency virus* and Palivizumab for *Respiratory syncytial virus*) [162,163]. Virus-specific neutralizing antibodies commonly target viral surface glycoproteins or host structures, thereby inhibiting viral entry through receptor binding interference [164,165]. This interference is predicted to reduce the viral load, mitigate disease, and reduce overall hospitalization. mAbs can be designed for a particular virus, and significant advances have been made in the speed at which new mAbs can be identified and produced. At the time of the SARS and MERS epidemics, interest in mAbs to reduce infection was never realized [166,167], but this allowed for mAbs to quickly be considered among the top candidates against COVID-19.

### 5.2.1 Biologics and the Innate Immune Response

Deaths from COVID-19 often occur when inflammation becomes dysregulated following an immune response to the SARS-CoV-2 virus. Therefore, one potential approach to reducing COVID-19 mortality rates is to manage the inflammatory response in severely ill patients. One candidate therapeutic identified that uses this mechanism is tocilizumab (TCZ). TCZ is a mAb that was developed to manage chronic inflammation caused by the continuous synthesis of the cytokine IL-6 [168]. IL-6 is a pro-inflammatory cytokine belonging to the interleukin family, which is comprised by immune system regulators that are primarily responsible for immune cell differentiation. Often used to treat chronic inflammatory conditions such as rheumatoid arthritis [168], TCZ has become a pharmaceutical of interest for the treatment of COVID-19 because of the role IL-6 plays in this disease. It has also been approved to treat CRS caused by CAR-T treatments [169]. While the secretion of IL-6 can be associated with chronic conditions, IL-6 is a key player in the innate immune response and is secreted by macrophages in response to the detection of pathogen-associated molecular patterns and damage-associated molecular patterns [168]. An analysis of 191 in-patients at two Wuhan hospitals revealed that blood concentrations of IL-6 differed between patients who did and did not recover from COVID-19. Patients who ultimately died had higher IL-6 levels at admission than those who recovered [170]. Additionally, IL-6 levels remained higher throughout the course of hospitalization in the patients who ultimately died [170].

Currently, TCZ is being administered either as a monotherapy or in combination with other treatments in 73 interventional COVID-19 clinical trials (Figure 2). A number of retrospective studies have been conducted in several countries [171,172,173,174,175,176]. In general, these studies have reported a positive effect of TCZ on reducing mortality in COVID-19 patients, although due to their retrospective designs, significant limitations are present in all of them (Appendix 1). It was not until February 11, 2021 that a preprint describing preliminary results of the first RCT of TCZ was released as part of the RECOVERY trial [177]. TCZ was found to reduce 28-day mortality from 33% in patients receiving SOC alone to 29% in those receiving TCZ. Combined analysis of the RECOVERY trial data with data from smaller RCTs suggested a 13% reduction in 28-day mortality [177]. While this initial report did not include the full results expected from the RECOVERY trial, this large-scale, RCT provides strong evidence that TCZ may offer benefits for COVID-19 patients. The RECOVERY trial along with results from several other RCTs [178,179,180,181,182] were cited as support for the EUA issued for TCZ in June 2021 [183]. However, the fact that TCZ suppresses the immune response means that it does carry risks for patients, especially a potential risk of secondary infection (Appendix 1).

TCZ is just one example of a candidate drug targeting the host immune response and specifically excessive inflammation. For example, interferons (IFNs) have also been investigated; these are a family of cytokines critical to activating the innate immune response against viral infections. Synairgen has been investigating a candidate drug, SNG001, which is an IFN-$\beta$-1a formulation to be delivered to the lungs via inhalation [184] that they reported reduced progression to ventilation in a double-blind, placebo-controlled, multi-center study of 101 patients with an average age in the late 50s [185,186]. However, these findings were not supported by the large-scale WHO Solidarity trial, which reported no significant effect of IFN-β-1a on patient survival during hospitalization [93], although differences in the designs of the two studies, and specifically the severity of illness among enrolled patients, may have influenced their divergent outcomes (Appendix 1). Other biologics influencing inflammation are also being explored (Appendix 1). It is also important that studies focused on inflammation as a possible therapeutic target consider the potential differences in baseline inflammation among patients from different backgrounds, which may be caused by differing life experiences (see [187]).

### 5.2.2 Biologics and the Adaptive Immune Response

While TCZ is an example of an mAb focused on managing the innate immune response, other treatments are more specific, targeting the adaptive immune response after an infection. In some cases, treatments can utilize biologics obtained directly from recovered individuals. From the very early days of the COVID-19 pandemic, polyclonal antibodies from convalescent plasma were investigated as a potential treatment for COVID-19 [188,189]. Convalescent plasma was used in prior epidemics including SARS, Ebola Virus Disease, and even the 1918 Spanish Influenza [188,190]. Use of convalescent plasma transfusion (CPT) over more than a century has aimed to reduce symptoms and improve mortality in infected people [190], possibly by accelerating viral clearance [188]. However, it seems unlikely that this classic treatment confers any benefit for COVID-19 patients. Several systematic reviews have investigated whether CPT reduced mortality in COVID-19 patients, and although findings from early in the pandemic (up to April 19, 2020) did support use of CPT [190], the tide has shifted as the body of available literature has grown [191]. While titer levels were suggested as a possible determining factor in the success of CPT against COVID-19 [192], the large-scale RECOVERY trial evaluated the effect of administering high-titer plasma specifically and found no effect on mortality or hospital discharge over a 28-day period [193]. These results thus suggest that, despite initial optimism and an EUA from the FDA, CPT is unlikely to be an effective therapeutic for COVID-19.

A different narrative is shaping up around the use of mAbs specifically targeting SARS-CoV-2. During the first SARS epidemic in 2002, neutralizing antibodies (nAbs) were found in SARS-CoV-1-infected patients [194,195]. Several studies following up on these findings identified various S-glycoprotein epitopes as the major targets of nAbs against SARS-CoV-1 [196]. Coronaviruses use trimeric spike (S) glycoproteins on their surface to bind to the host cell, allowing for cell entry [197,198]. Each S glycoprotein protomer is comprised of an S1 domain, also called the receptor binding domain (RBD), and an S2 domain. The S1 domain binds to the host cell while the S2 domain facilitates the fusion between the viral envelope and host cell membranes [196]. The genomic identity between the RBD of SARS-CoV-1 and SARS-CoV-2 is around 74% [199]. Due to this high degree of similarity, preexisting antibodies against SARS-CoV-1 were initially considered candidates for neutralizing activity against SARS-CoV-2. While some antibodies developed against the SARS-CoV-1 spike protein showed cross-neutralization activity with SARS-CoV-2 [200,201], others failed to bind to SARS-CoV-2 spike protein at relevant concentrations [202]. Cross-neutralizing activities were dependent on whether the epitope recognized by the antibodies were conserved between SARS-CoV-1 and SARS-CoV-2 [200].

Technological advances in antibody drug design as well as in structural biology massively accelerated the discovery of novel antibody candidates and the mechanisms by which they interact with the target structure. Within just a year of the structure of the SARS-CoV-2 spike protein being published, an impressive pipeline of monoclonal antibodies targeting SARS-CoV-2 entered clinical trials, with hundreds more candidates in preclinical stages. The first human monoclonal neutralizing antibody specifically against the SARS-CoV-2 S glycoprotein was developed using hybridoma technology [203], where antibody-producing B-cells developed by mice are inserted into myeloma cells to produce a hybrid cell line (the hybridoma) that is grown in culture. The 47D11 antibody clone was able to cross-neutralize SARS-CoV-1 and SARS-CoV-2. This antibody (now ABVV-47D11) has recently entered clinical trials in collaboration with AbbVie. Additionally, an extensive monoclonal neutralizing antibody pipeline has been developed to combat the ongoing pandemic, with over 50 different antibodies in clinical trials [204]. Thus far, the monotherapy sotrovimab and two antibody cocktails (bamlanivimab/estesevimab and casirivimab/imdevimab) have been granted EUAs by the FDA.

One of the studied antibody cocktails consists of bamlanivimab and estesevimab. Bamlanivimab (Ly-CoV555) is a human mAb that was derived from convalescent plasma donated by a recovered COVID-19 patient, evaluated in research by the National Institute of Allergy and Infectious Diseases (NIAID), and subsequently developed by AbCellera and Eli Lilly. The neutralizing activity of bamlanivimab was initially demonstrated *in vivo* using a nonhuman primate model [205]. Based on these positive preclinical data, Eli Lilly initiated the first human clinical trial for a monoclonal antibody against SARS-CoV-2. The phase 1 trial, which was conducted in hospitalized COVID-19 patients, was completed in August 2020 [206]. Estesevimab (LY-CoV016 or JS-016) is also a monoclonal neutralizing antibody against the spike protein of SARS-CoV-2. It was initially developed by Junshi Biosciences and later licensed and developed through Eli Lilly. A phase 1 clinical trial to assess the safety of etesevimab was completed in October 2020 [207]. Etesevimab was shown to bind a different epitope on the spike protein than bamlanivimab, suggesting that the two antibodies used as a combination therapy would further enhance their clinical use compared to a monotherapy [208]. To assess the efficacy and safety of bamlanivimab alone or in combination with etesevimab for the treatment of COVID-19, a phase 2/3 trial (BLAZE-1) [209] was initiated. The interim analysis of the phase 2 portion suggested that bamlanivimab alone was able to accelerate the reduction in viral load [210]. However, more recent data suggests that only the bamlanivimab/etesevimab combination therapy is able to reduce viral load in COVID-19 patients [208]. Based on this data, the combination therapy received an EUA for COVID-19 from the FDA in February 2021 [211].

A second therapy is comprised of casirivimab and imdevimab (REGN-COV2). Casirivimab (REGN10933) and imdevimab (REGN10987) are two monoclonal antibodies against the SARS-CoV-2 spike protein. They were both developed by Regeneron in a parallel high-throughput screening (HTS) to identify neutralizing antibodies from either humanized mice or patient-derived convalescent plasma [212]. In these efforts, multiple antibodies were characterized for their ability to bind and neutralize the SARS-CoV-2 spike protein. The investigators hypothesized that an antibody cocktail, rather than each individual antibody, could increase the therapeutic efficacy while minimizing the risk for virus escape. Therefore, the authors tested pairs of individual antibodies for their ability to simultaneously bind the RBD of the spike protein. Based on this data, casirivimab and imdevimab were identified as the lead antibody pair, resulting in the initiation of two clinical trials [213,214]. Data from this phase 1-3 trial published in the *New England Journal of Medicine* shows that the REGN-COV2 antibody cocktail reduced viral load, particularly in patients with high viral load or whose endogenous immune response had not yet been initiated [215]. However, in patients who already initiated an immune response, exogenous addition of REGN-COV2 did not improve the endogenous immune response. Both doses were well tolerated with no serious events related to the antibody cocktail. Based on this data, the FDA granted an EUA for REGN-COV2 in patients with mild to moderate COVID-19 who are at risk of developing severe disease [216]. Ongoing efforts are trying to evaluate the efficacy of REGN-COV2 to improve clinical outcomes in hospitalized patients [213].

Sotrovimab is the most recent mAb to receive an EUA. It was identified in the memory B cells of a 2003 survivor of SARS [217] and was found to be cross-reactive with SARS-CoV-2 [201]. This cross-reactivity is likely attributable to conservation within the epitope, with 17 out of 22 residues conserved between the two viruses, four conservatively substituted, and one semi-conservatively substituted [201]. In fact, these residues are highly conserved among sarbecoviruses, a clade that includes SARS-CoV-1 and SARS-CoV-2 [201]. This versatility has led to it being characterized as a "super-antibody" [218], a potent, broadly neutralizing antibody [219]. Interim analysis of data from a clinical trial [220] reported high safety and efficacy of this mAb in 583 COVID-19 patients [221]. Compared to placebo, sotrovimab was found to be 85% more effective in reducing progression to the primary endpoint, which was the proportion of patients who, within 29 days, were either hospitalized for more than 24 hours or died. Additionally, rates of adverse events were comparable, and in some cases lower, among patients receiving sotrovimab compared to patients receiving a placebo. Sotrovimab therefore represents a mAb therapeutic that is effective against SARS-CoV-2 and may also be effective against other sarbecoviruses.

Several potential limitations remain in the application of mAbs to the treatment of COVID-19. One of the biggest challenges is identifying antibodies that not only bind to their target, but also prove to be beneficial for disease management. Currently, use of mAbs is limited to people with mild to moderate disease that are not hospitalized, and it has yet to be determined whether they can be used as a successful treatment option for severe COVID-19 patients. While preventing people from developing severe illness provides significant benefits, patients with severe illness are at the greatest risk of death, and therefore therapeutics that provide benefits against severe illness are particularly desirable. It remains to be seen whether mAbs confer any benefits for patients in this category.

Another concern about therapeutics designed to amplify the response to a specific viral target is that they may need to be modified as the virus evolves. With the ongoing global spread of new SARS-CoV-2 variants, there is a growing concern that mutations in SARS-CoV-2 spike protein could escape antibody neutralization, thereby reducing the efficacy of monoclonal antibody therapeutics and vaccines. A comprehensive mutagenesis screen recently identified several amino acid substitutions in the SARS-CoV-2 spike protein that can prevent antibody neutralization [222]. While some mutations result in resistance to only one antibody, others confer broad resistance to multiple mAbs as well as polyclonal human sera, suggesting that some amino acids are "hotspots" for antibody resistance. However, it was not investigated whether the resistance mutations identified result in a fitness advantage. Accordingly, an impact on neutralizing efficiency has been reported for the B.1.1.7 (Alpha) variant first identified in the UK and the B.1.351 (Beta) variant first identified in in South Africa [223,224,225]. As of June 25, 2021, the CDC recommended a pause in the use of bamlanivimab and etesevimab due to decreased efficacy against the P.1 (Gamma) and B.1.351 (Beta) variants of SARS-CoV-2 [226]. While the reported impact on antibody neutralization needs to be confirmed *in vivo*, it suggests that some adjustments to therapeutic antibody treatments may be necessary to maintain the efficacy that was reported in previous clinical trials.

Several strategies have been employed to try to mitigate the risk of diminished antibody neutralization. Antibody cocktails such as those already holding an EUA may help overcome the risk for attenuation of the neutralizing activity of a single monoclonal antibody. These cocktails consist of antibodies that recognize different epitopes on the spike protein, decreasing the likelihood that a single amino acid change can cause resistance to all antibodies in the cocktail. However, neutralizing resistance can emerge even against an antibody cocktail if the individual antibodies target subdominant epitopes [224]. Another strategy is to develop broadly neutralizing antibodies that target structures that are highly conserved, as these are less likely to mutate [227,228] or to target epitopes that are insensitive to mutations [229]. Sotrovimab, one such "super-antibody", is thought to be somewhat robust to neutralization escape [230] and has been found to be effective against all variants assessed as of August 12, 2021 [231]. Another antibody (ADG-2) targets a highly conserved epitope that overlaps the hACE2 binding site of all clade 1 sarbecoviruses [232]. Prophylactic administration of ADG-2 in an immunocompetent mouse model of COVID-19 resulted in protection against viral replication in the lungs and respiratory burden. Since the epitope targeted by ADG-2 represents an Achilles' heel for clade 1 sarbecoviruses, this antibody, like sotrovimab, might be a promising candidate against all circulating variants as well as emerging SARS-related coronaviruses. To date, it has fared well against the Alpha, Beta, Gamma, and Delta variants [231].

The development of mAbs against SARS-CoV-2 has made it clear that this technology is rapidly adaptable and offers great potential for the response to emerging viral threats. However, additional investigation may be needed to adapt mAb treatments to SARS-CoV-2 as it evolves and potentially to pursue designs that confer benefits for patients at the greatest risk of death. While polyclonal antibodies from convalescent plasma have been evaluated as a treatment for COVID-19, these studies have suggested fewer potential benefits against SARS-CoV-2 than mAbs; convalescent plasma therapy has been thoroughly reviewed elsewhere [188,189]. Thus, advances in biologics for COVID-19 illustrate that an understanding of how the host and virus interact can guide therapeutic approaches. The FDA authorization of two combination mAb therapies, in particular, underscores the potential for this strategy to allow for a rapid response to a novel pathogen. Additionally, while TCZ is not yet as established, this therapy suggests that the strategy of using biologics to counteract the cytokine storm response may provide therapies for the highest-risk patients.

# 6 High-Throughput Screening for Drug Repurposing

The drug development process is slow and costly, and developing compounds specifically targeted to an emerging viral threat is not a practical short-term solution. Screening existing drug compounds for alternative indications is a popular alternative [233,234,235,236]. HTS has been a goal of pharmaceutical development since at least the mid-1980s [237]. Traditionally, phenotypic screens were used to test which compounds would induce a desired change in *in vitro* or *in vivo* models, focusing on empirical, function-oriented exploration naïve to molecular mechanism [238,239,240]. In many cases, these screens utilize large libraries that encompass a diverse set of agents varying in many pharmacologically relevant properties (e.g., [241]). The compounds inducing a desired effect could then be followed up on. Around the turn of the millennium, advances in molecular biology allowed for HTS to shift towards screening for compounds interacting with a specific molecular target under the hypothesis that modulating that target would have a desired effect. These approaches both offer pros and cons, and today a popular view is that they are most effective in combination [238,240,242].

Today, some efforts to screen compounds for potential repurposing opportunities are experimental, but others use computational HTS approaches [233,243]. Computational drug repurposing screens can take advantage of big data in biology [17] and as a result are much more feasible today than during the height of the SARS and MERS outbreaks in the early 2000s and early 2010s, respectively. Advancements in robotics also facilitate the experimental component of HTS [235]. For viral diseases, the goal of drug repurposing is typically to identify existing drugs that have an antiviral effect likely to impede the virus of interest. While both small molecules and biologics can be candidates for repurposing, the significantly lower price of many small molecule drugs means that they are typically more appealing candidates [244].

Depending on the study design, screens vary in how closely they are tied to a hypothesis. As with the candidate therapeutics described above, high-throughput experimental or computational screens can proceed based on a hypothesis. Just as remdesivir was selected as a candidate antiviral because it is a nucleoside analog [245], so too can high-throughput screens select libraries of compounds based on a molecular hypothesis. Likewise, when the library of drugs is selected without basis in a potential mechanism, a screen can be considered hypothesis free [245]. Today, both types of analyses are common both experimentally and computationally. Both strategies have been applied to identifying candidate therapeutics against SARS-CoV-2.

## 6.1 Hypothesis-Driven Screening

Hypothesis-driven screens often select drugs likely to interact with specific viral or host targets or drugs with desired clinical effects, such as immunosuppressants. There are several properties that might identify a compound as a candidate for an emerging viral disease. Drugs that interact with a target that is shared between pathogens (i.e., a viral protease or a polymerase) or between a viral pathogen and another illness (i.e., a cancer drug with antiviral potential) are potential candidates, as are drugs that are thought to interact with additional molecular targets beyond those they were developed for [243]. Such research can be driven by *in vitro* or *in silico* experimentation. Computational analyses depend on identifying compounds that modulate pre-selected proteins in the virus or host. As a result, they build on experimental research characterizing the molecular features of the virus, host, and candidate compounds [236].

One example of the application of this approach to COVID-19 research comes from work on protease inhibitors. Studies have shown that viral proteases play an important role in the life cycle of viruses, including coronaviruses, by modulating the cleavage of viral polyprotein precursors [246]. Several FDA-approved drugs target proteases, such as lopinavir and ritonavir for HIV infection and simeprevir for hepatitis C virus infection. Serine protease inhibitors were previously suggested as possible treatments for SARS and MERS [247]. One early study [197] suggested that camostat mesylate, a protease inhibitor, could block the entry of SARS-CoV-2 into lung cells *in vitro*. Two polyproteins encoded by the SARS-CoV-2 replicase gene, pp1a and pp1ab, are critical for viral replication and transcription [248]. These polyproteins must undergo proteolytic processing, which is usually conducted by main protease ($M^{Pro}$), a 33.8-kDa SARS-CoV-2 protease that is therefore fundamental to viral replication and transcription. Therefore, it was hypothesized that compounds targeting $M^{Pro}$ could be used to prevent or slow the replication of the SARS-CoV-2 virus.

Both computational and experimental approaches facilitated the identification of compounds that might inhibit SARS-CoV-2 $M^{Pro}$. In 2005, computer-aided design facilitated the development of a Michael acceptor inhibitor, now known as N3, to target $M^{Pro}$ of SARS-like coronaviruses [249]. N3 binds in the substrate binding pocket of $M^{Pro}$ in several HCoV [249,250,251,252]. The structure of N3-bound SARS-CoV-2 $M^{Pro}$ has been solved, confirming the computational prediction that N3 would similarly bind in the substrate binding pocket of SARS-CoV-2 [248]. N3 was tested *in vitro* on SARS-CoV-2-infected Vero cells, which belong to a line of cells established from the kidney epithelial cells of an African green monkey, and was found to inhibit SARS-CoV-2 [248]. A library of approximately 10,000 compounds was screened in a fluorescence resonance energy transfer assay constructed using SARS-CoV-2 $M^{Pro}$ expressed in *Escherichia coli* [248].

Six leads were identified in this hypothesis-driven screen. *In vitro* analysis revealed that ebselen had the strongest potency in reducing the viral load in SARS-CoV-2-infected Vero cells [248]. Ebselen is an organoselenium compound with anti-inflammatory and antioxidant properties [253]. Molecular dynamics analysis further demonstrated the potential for ebselen to bind to $M^{Pro}$ and disrupt the protease's enzymatic functions [254]. However, ebselen is likely to be a promiscuous binder, which could diminish its therapeutic potential [248,255], and compounds with higher specificity may be needed to translate this mechanism effectively to clinical trials. In July 2020, phase II clinical trials commenced to assess the effects of SPI-1005, an investigational drug from Sound Pharmaceuticals that contains ebselen [256], on 60 adults presenting with each of moderate [257] and severe [258] COVID-19. Other $M^{Pro}$ inhibitors are also being evaluated in clinical trials [259,260,260]. Pending the results of clinical trials, N3 remains a computationally interesting compound based on both computational and experimental data, but whether these potential effects will translate to the clinic remains unknown.

## 6.2 Hypothesis-Free Screening

Hypothesis-free screens use a discovery-driven approach, where screens are not targeted to specific viral proteins, host proteins, or desired clinical modulation. Hypothesis-free drug screening began twenty years ago with the testing of libraries of drugs experimentally. Today, like many other areas of biology, *in silico* analyses have become increasingly popular and feasible through advances in biological big data [245,261]. Many efforts have collected data about interactions between drugs and SARS-CoV-2 and about the host genomic response to SARS-CoV-2 exposure, allowing for hypothesis-free computational screens that seek to identify new candidate therapeutics. Thus, they utilize a systems biology paradigm to extrapolate the effect of a drug against a virus based on the host interactions with both the virus and the drug [236].

Resources such as the COVID-19 Drug and Gene Set Library, which at the time of its publication contained 1,620 drugs sourced from 173 experimental and computational drug sets and 18,676 human genes sourced from 444 gene sets [262], facilitate such discovery-driven approaches. Analysis of these databases indicated that some drugs had been identified as candidates across multiple independent analyses, including high-profile candidates such as CQ/HCQ and remdesivir [262]. Computational screening efforts can then mine databases and other resources to identify potential PPIs among the host, the virus, and established and/or experimental drugs [263]. Subject matter expertise from human users may be integrated to varying extents depending on the platform (e.g,. [263,264]). These resources have allowed studies to identify potential therapeutics for COVID-19 without an *a priori* reason for selecting them.

One example of a hypothesis-free screen for COVID-19 drugs comes from a PPI-network-based analysis that was published early in the pandemic [265]. Here, researchers cloned the proteins expressed by SARS-CoV-2 *in vitro* and quantified 332 viral-host PPI using affinity purification mass spectrometry [265]. They identified two SARS-CoV-2 proteins (Nsp6 and Orf9c) that interacted with host Sigma-1 and Sigma-2 receptors. Sigma receptors are located in the endoplasmic reticulum of many cell types, and type 1 and 2 Sigma receptors have overlapping but distinct affinities for a variety of ligands [266]. Molecules interacting with the Sigma receptors were then analyzed and found to have an effect on viral infectivity *in vitro* [265]. A follow-up study evaluated the effect of perturbing these 332 proteins in two cells lines, A549 and Caco-2, using knockdown and knockout methods, respectively, and found that the replication of SARS-CoV-2 in cells from both lines was dependent on the expression of *SIGMAR1*, which is the gene that encodes the Sigma-1 receptor [267]. Following these results, drugs interacting with Sigma receptors were suggested as candidates for repurposing for COVID-19 (e.g., [268]). Because many well-known and affordable drugs interact with the Sigma receptors [265,269], they became a major focus of drug repurposing efforts. Some of the drugs suggested by the apparent success of Sigma receptor-targeting drugs were already being investigated at the time. HCQ, for example, forms ligands with both Sigma-1 and Sigma-2 receptors and was already being explored as a candidate therapeutic for COVID-19 [265]. Thus, this computational approach yielded interest in drugs whose antiviral activity was supported by initial *in vitro* analyses.

Follow-up research, however, called into question whether the emphasis on drugs interacting with Sigma receptors might be based on a spurious association [270]. This study built on the prior work by examining whether antiviral activity among compounds correlated with their affinity for the Sigma receptors and found that it did not. The study further demonstrated that cationic amphiphilicity was a shared property among many of the candidate drugs identified through both computational and phenotypic screens and that it was likely to be the source of many compounds' proposed antiviral activity [270]. Cationic amphiphilicity is associated with the induction of phospholipidosis, which is when phospholipids accumulate in the lysosome [271]. Phospholipidosis can disrupt viral replication by inhibiting lipid processing [272] (see discussion of HCQ in Appendix 1). However, phospholipidosis is known to translate poorly from *in vitro* models to *in vivo* models or clinical applications. Thus, this finding suggested that these screens were identifying compounds that shared a physiochemical property rather than a specific target [270]. The authors further demonstrated that antiviral activity against SARS-CoV-2 *in vitro* was correlated with the induction of phospholipidosis for drugs both with and without cationic amphiphilicity [270]. This finding supports the idea that the property of cationic amphicility was being detected as a proxy for the shared effect of phospholipidosis [270]. They demonstrated that phospholipidosis-inducing drugs were not effective at preventing viral propagation *in vivo* in a murine model of COVID-19 [270]. Therefore, removing hits that induce phospholipidosis from computational and *in vitro* experimental repurposing screens (e.g., [273]) may help emphasize those that are more likely to provide clinical benefits. This work illustrates the importance of considering confounding variables in computational screens, a principle that has been incorporated into more traditional approaches to drug development [274].

One drug that acts on Sigma receptors does, however, remain a candidate for the treatment of COVID-19. Several psychotropic drugs target Sigma receptors in the central nervous system and thus attracted interest as potential COVID-19 therapeutics following the findings of two host-virus PPI studies [275]. For several of these drugs, the *in vitro* antiviral activity [267] was not correlated with their affinity for the Sigma-1 receptor [270,275] but was correlated with phospholipidosis [270]. However, fluvoxamine, a selective serotonin reuptake inhibitor that is a particularly potent Sigma-1 receptor agonist [275], has shown promise as a preventative of severe COVID-19 in a preliminary analysis of data from the large-scale TOGETHER trial [145]. As of August 6, 2021, this trial had collected data from over 1,400 patients in the fluvoxamine arm of their study, half of whom received a placebo [145]. Only 74 patients in the fluvoxamine group had progressed to hospitalization for COVID-19 compared to 107 in the placebo group, corresponding to a relative risk of 0.69; additionally, the relative risk of mortality between the two groups was calculated at 0.71. These findings support the results of small clinical trials that have found fluvoxamine to reduce clinical deterioration relative to a placebo [276,277]. However, the ongoing therapeutic potential of fluvoxamine does not contradict the finding that hypothesis-free screening hits can be driven by confounding factors. The authors point out that its relevance would not just be antiviral as it has a potential immunomodulatory mechanism [276]. It has been found to be protective against septic shock in an *in vivo* mouse model [278]. It is possible that fluvoxamine also exerts an antiviral effect [279]. Thus, Sigma-1 receptor activity may contribute to fluvoxamine's potential effects in treating COVID-19, but is not the only mechanism by which this drug can interfere with disease progression.

## 6.3   Potential and Limitations of High-Throughput Analyses

Computational screening allows for a large number of compounds to be evaluated to identify those most likely to display a desired behavior or function. This approach can be guided by a hypothesis or can aim to discover underlying characteristics that produce new hypotheses about the relationship between a host, a virus, and candidate pharmaceuticals. The examples outlined above illustrate that HTS-based evaluations of drug repurposing can potentially provide valuable insights. Computational techniques were used to design compounds targeting $M^{Pro}$ based on an understanding of how this protease aids viral replication, and $M^{Pro}$ inhibitors remain promising candidates [235], although the clinical trial data is not yet available. Similarly, computational analysis correctly identified the Sigma-1 receptor as a protein of interest. Although the process of identifying which drugs might modulate the interaction led to an emphasis on candidates that ultimately have not been supported, fluvoxamine remains an appealing candidate. The difference between the preliminary evidence for fluvoxamine compared to other drugs that interact with Sigma receptors underscores a major critique of hypothesis-free HTS in particular: while these approaches allow for brute force comparison of a large number of compounds against a virus of interest, they lose the element of expertise that is associated with most successes in drug repurposing [245].

There are also practical limitations to these methods. One concern is that computational analyses inherently depend on the quality of the data being evaluated. The urgency of the COVID-19 pandemic led many research groups to pivot towards computational HTS research without familiarity with best practices in this area [235]. As a result, there is an excessive amount of information available from computational studies [280], but not all of it is high-quality. Additionally, the literature used to identify and validate targets can be difficult to reproduce [281], which may pose challenges to target-based experimental screening and to *in silico* screens. Some efforts to repurpose antivirals have focused on host, rather than viral, proteins [236], which might be expected to translate poorly *in vivo* if the targeted proteins serve essential functions in the host. Concerns about the practicality of hypothesis-free screens to gain novel insights are underscored by the fact that very few or possibly no success stories have emerged from hypothesis-free screens over the past twenty years [245]. These findings suggest that data-driven research can be an important component of the drug repurposing ecosystem, but that drug repurposing efforts that proceed without a hypothesis, an emphasis on biological mechanisms, or an understanding of confounding effects may not produce viable candidates.

# 7   Considerations in Balancing Different Approaches

The approaches described here offer a variety of advantages and limitations in responding to a novel viral threat and building on existing bodies of knowledge in different ways. Medicine, pharmacology, basic science (especially virology and immunology), and biological data science can all provide different insights and perspectives for addressing the challenging question of which existing drugs might provide benefits against an emerging viral threat. A symptom management-driven approach allows clinicians to apply experience with related diseases or related symptoms to organize a rapid response aimed at saving the lives of patients already infected with a new disease. Oftentimes, the pharmaceutical agents that are applied are small-molecule, broad-spectrum pharmaceuticals that are widely available and affordable to produce, and they may already be available for other purposes, allowing clinicians to administer them to patients quickly either with an EUA or off-label. In this vein, dexamethasone has emerged as the strongest treatment against severe COVID-19 (Table 1).

Alternatively, many efforts to repurpose drugs for COVID-19 have built on information gained through basic scientific research of HCoV. Understanding how related viruses function has allowed researchers to identify possible pharmacological strategies to disrupt pathogenesis (Figure 3). Some of the compounds identified through these methods include small-molecule antivirals, which can be boutique and experimental medications like remdesivir (Table 1). Other candidate drugs that intercept host-pathogen interactions include biologics, which imitate the function of endogenous host compounds. Most notably, several mAbs that have been developed (casirivimab, imdevimab, bamlanivimab and etesevimab) or repurposed (sotrovimab, tocilizumab) have now been granted EUAs (Table 1). Although not discussed here, several vaccine development programs have also met huge success using a range of strategies [2].

*Table 1: Summary table of candidate therapeutics examined in this manuscript. "Grade" is the rating given to each treatment by the Systematic Tracker of Off-label/Repurposed Medicines Grades (STORM) maintained by the Center for Cytokine Storm Treatment & Laboratory (CSTL) at the University of Pennsylvania [70]. A grade of A indicates that a treatment is considered effective, B that all or most RCTs have shown positive results, C that RCT data are not yet available, and D that multiple RCTs have produced negative results. Treatments not in the STORM database are indicated as N/A. FDA status is also provided where available. The evidence available is based on the progression of the therapeutic through the pharmaceutical development pipeline, with RCTs as the most informative source of evidence. The effectiveness is summarized based on the current available evidence; large trials such as RECOVERY and Solidarity are weighted heavily in this summary. This table was last updated on August 20, 2021.*

| Treatment | Grade | Category | FDA Status | Evidence Available | Suggested Effectiveness |
|---|---|---|---|---|---|
| Dexamethasone | A | Small molecule, broad spectrum | Used off-label | RCT | Supported: RCT shows improved outcomes over SOC, especially in severe cases such as CRS |
| Remdesivir | A | Small molecule, antiviral, adenosine analog | Approved for COVID-19 (and EUA for combination | RCT | Mixed: Conflicting evidence from large WHO-led Solidarity trial vs US-focused RCT and other studies |

| Treatment | Grade | Category | FDA Status | Evidence Available | Suggested Effectiveness |
|---|---|---|---|---|---|
| | | | with baricitinib) | | |
| Tocilizumab | A | Biologic, monoclonal antibody | EUA | RCT | Mixed: It appears that TCZ may work well in combination with dexamethasone in severe cases, but not as monotherapy |
| Sotrovimab | N/A | Biological, monoclonal antibody | EUA | RCT | Supported: Phase 2/3 clinical trial showed reduced hospitalization/death |
| Bamlanivimab and etesevimab | B & N/A | Biologic, monoclonal antibodies | EUA | RCT | Supported: Phase 2 clinical trial showed reduction in viral load, but FDA pause recommended because may be less effective against Delta variant |
| Casirivimab and imdevimab | N/A | Biologic, monoclonal antibodies | EUA | RCT | Supported: Reduced viral load at interim analysis |
| Fluvoxamine | B | Small-molecule, Sigma-1 receptor agonist | N/A | RCT | Supported: Support from two small RCTs and preliminary support from interim analysis of TOGETHER |
| SNG001 | B | Biologic, interferon | None | RCT | Mixed: Support from initial RCT but no effect found in WHO's Solidarity trial |
| $M^{Pro}$ Protease Inhibitors | N/A | Small molecule, protease inhibitor | None | Computational prediction, *in vitro* studies | Unknown |
| ARBs & ACEIs | C | Small molecule, broad spectrum | None | Observational studies and some RCTs | Not supported: Observational study retracted, RCTs suggest no association |
| Favipiravir | D | Small molecule, antiviral, nucleoside analog | None | RCT | Not supported: RCTs do not show significant improvements for individuals taking this treatment, good safety profile |

| Treatment | Grade | Category | FDA Status | Evidence Available | Suggested Effectiveness |
|---|---|---|---|---|---|
| HCQ/CQ | D | Small molecule, broad spectrum | None | RCT | Not supported, possibly harmful: Non-blinded RCTs showed no improvement over SOC, safety profile may be problematic |
| Convalescent plasma transfusion | D | Biologic, polyclonal antibodies | EUA | RCT | Mixed: Supported in small trials but not in large-scale RECOVERY trial |
| Ivermectin | D | Small molecule, broad spectrum | None | RCT | Mixed: Mixed results from small RCTs, major supporting RCT now withdrawn, preliminary results of large RCT (TOGETHER) suggest no effect on emergency room visits or hospitalization for COVID-19 |

All of the small-molecule drugs evaluated and most of the biologics are repurposed, and thus hinge on a theoretical understanding of how the virus interacts with a human host and how pharmaceuticals can be used to modify those interactions rather than being designed specifically against SARS-CoV-2 or COVID-19. As a result, significant attention has been paid to computational approaches that automate the identification of potentially desirable interactions. However, work in COVID-19 has made it clear that relevant compounds can also be masked by confounds, and spurious associations can drive investment in candidate therapeutics that are unlikely to translate to the clinic. Such spurious hits are especially likely to impact hypothesis-free screens. However, hypothesis-free screens may still be able to contribute to the drug discovery or repurposing ecosystem, assuming the computational arm of HTS follows the same trends seen in its experimental arm. In 2011, a landmark study in drug discovery demonstrated that although more new drugs were discovered using target-based rather than phenotypic approaches, the majority of drugs with a novel molecular mechanism of action (MMOA) were identified in phenotypic screens [282]. This pattern applied only to first-in-class drugs, with most follower drugs produced by target-based screening [239]. These findings suggest that target-based drug discovery is more successful when building on a known MMOA, and that modulating a target is most valuable when the target is part of a valuable MMOA [240]. Building on this, many within the field suggested that mechanism-informed phenotypic investigations may be the most useful approach to drug discovery [238,240,242]. As it stands, data-driven efforts to identify patterns in the results of computational screens allowed researchers to notice the shared property of cationic amphicility among many of the hits from computational screening analyses [270]. While easier said than done, efforts to fill in the black box underlying computational HTS and recognize patterns among the identified compounds aid in moving data-oriented drug repurposing efforts in this direction.

The unpredictable nature of success and failure in drug repurposing for COVID-19 thus highlights one of the tenets of phenotypic screening: there are a lot of "unknown unknowns", and a promising mechanism at the level of an MMOA will not necessarily propagate up to the pathway, cellular, or organismal level [238]. Despite the fact that apparently mechanistically relevant drugs may exist, identifying effective treatments for a new viral disease is extremely challenging. Targets of repurposed drugs are often non-specific, meaning that the MMOA can appear to be relevant to COVID-19 without a therapeutic or prophylactic effect being observed in clinical trials. The difference in the current status of remdesivir and favipiravir as treatments for COVID-19 (Table 1) underscores how difficult to predict whether a specific compound will produce a desired effect, even when the mechanisms are similar. Furthermore, the fact that many candidate COVID-19 therapeutics were ultimately identified because of their shared propensity to induce phospholipidosis underscores how challenging it can be to identify a mechanism *in silico* or *in vitro* that will translate to a successful treatment. While significant progress has been made thus far in the pandemic, the therapeutic landscape is likely to continue to evolve as more results become available from clinical trials and as efforts to develop novel therapeutics for COVID-19 progress.

## 8   Towards the Next HCoV Threat

Only very limited testing of candidate therapies was feasible during the SARS and MERS epidemics, and as a result, few treatments were available at the outset of the COVID-19 pandemic. Even corticosteroids, which were used to treat SARS patients, were a controversial therapeutic prior to the release of the results of the large RECOVERY trial. The scale and duration of the COVID-19 pandemic has made it possible to conduct large, rigorous RCTs such as RECOVERY, Solidarity, TOGETHER, and others. As results from these trials have continued to emerge, it has become clear that small clinical trials often produce spurious results. In the case of HCQ/CQ, the therapeutic had already attracted so much attention based on small, preliminary (and in some cases, methodologically concerning) studies that it took the results of multiple large studies before attention began to be redirected to more promising candidates [283]. In fact, most COVID-19 clinical trials lack the statistical power to reliably test their hypotheses [284,285]. In the face of an urgent crisis like COVID-19, the desire to act quickly is understandable, but it is imperative that studies maintain strict standards of scientific rigor [235,274], especially given the potential dangers of politicization, as illustrated by HCQ/CQ [286]. Potential innovations in clinical trial structure, such as adaptable clinical trials with master protocols [287] or the sharing of data among small clinical trials [285] may help to address future crises and to bolster the results from smaller studies, respectively.

In the long-term, new drugs specific for treatment of COVID-19 may also enter development. Development of novel drugs is likely to be guided by what is known about the pathogenesis and molecular structure of SARS-CoV-2. For example, understanding the various structural components of SARS-CoV-2 may allow for the development of small molecule inhibitors of those components. Crystal structures of the SARS-CoV-2 main protease have been resolved [248,288]. Much work remains to be done to determine further crystal structures of other viral components, understand the relative utility of targeting different viral components, perform additional small molecule inhibitor screens, and determine the safety and efficacy of the potential inhibitors. While still nascent, work in this area is promising. Over the longer term, this approach and others may lead to the development of novel therapeutics specifically for COVID-19 and SARS-CoV-2. Such efforts are likely to prove valuable in managing future emergent HCoV, just as research from the SARS and MERS pandemic has provided a basis for the COVID-19 response.

# 1 Appendix: Identification and Development of Therapeutics for COVID-19

## 1.1 Dexamethasone

In order to understand how dexamethasone reduces inflammation, it is necessary to consider the stress response broadly. In response to stress, corticotropin-releasing hormone stimulates the release of neurotransmitters known as catecholamines, such as epinephrine, and steroid hormones known as glucocorticoids, such as cortisol [289,290]. While catecholamines are often associated with the fight-or-flight response, the specific role that glucocorticoids play is less clear, although they are thought to be important to restoring homeostasis [291]. Immune challenge is a stressor that is known to interact closely with the stress response. The immune system can therefore interact with the central nervous system; for example, macrophages can both respond to and produce catecholamines [289]. Additionally, the production of both catecholamines and glucocorticoids is associated with inhibition of proinflammatory cytokines such as IL-6, IL-12, and tumor necrosis factor-α (TNF-α) and the stimulation of anti-inflammatory cytokines such as IL-10, meaning that the stress response can regulate inflammatory immune activity [290]. Administration of dexamethasone has been found to correspond to dose-dependent inhibition of IL-12 production, but not to affect IL-10 [292]; the fact that this relationship could be disrupted by administration of a glucocorticoid-receptor antagonist suggests that it is regulated by the receptor itself [292]. Thus, the administration of dexamethasone for COVID-19 is likely to simulate the release of glucocorticoids endogenously during stress, resulting in binding of the synthetic steroid to the glucocorticoid receptor and the associated inhibition of the production of proinflammatory cytokines. In this model, dexamethasone reduces inflammation by stimulating the biological mechanism that reduces inflammation following a threat such as immune challenge.

Initial support for dexamethasone as a treatment for COVID-19 came from the United Kingdom's RECOVERY trial [62], which assigned over 6,000 hospitalized COVID-19 patients to the standard of care (SOC) or treatment (dexamethasone) arms of the trial at a 2:1 ratio. At the time of randomization, some patients were ventilated (16%), others were on non-invasive oxygen (60%), and others were breathing independently (24%). Patients in the treatment arm were administered dexamethasone either orally or intravenously at 6 mg per day for up to 10 days. The primary end-point was the patient's status at 28-days post-randomization (mortality, discharge, or continued hospitalization), and secondary outcomes analyzed included the progression to invasive mechanical ventilation over the same period. The 28-day mortality rate was found to be lower in the treatment group than in the SOC group (21.6% vs 24.6%, *p* < 0.001). However, the effect was driven by improvements in patients receiving mechanical ventilation or supplementary oxygen. One possible confounder is that patients receiving mechanical ventilation tended to be younger than patients who were not receiving respiratory support (by 10 years on average) and to have had symptoms for a longer period. However, adjusting for age did not change the conclusions, although the duration of symptoms was found to be significantly associated with the effect of dexamethasone administration. Thus, this large, randomized, and multi-site, albeit not placebo-controlled, study suggests that administration of dexamethasone to patients who are unable to breathe independently may significantly improve survival outcomes. Additionally, dexamethasone is a widely available and affordable medication, raising the hope that it could be made available to COVID-19 patients globally.

It is not surprising that administration of an immunosuppressant would be most beneficial in severe cases where the immune system was dysregulated towards inflammation. However, it is also unsurprising that care must be taken in administering an immunosuppressant to patients fighting a viral infection. In particular, the concern has been raised that treatment with dexamethasone might increase patient susceptibility to concurrent (e.g., nosocomial) infections [293]. Additionally, the drug could potentially slow viral clearance and inhibit patients' ability to develop antibodies to SARS-CoV-2 [59,293], with the lack of data about viral clearance being put forward as a major limitation of the RECOVERY trial [294]. Furthermore, dexamethasone has been associated with side effects that include psychosis, glucocorticoid-induced diabetes, and avascular necrosis [59], and the RECOVERY trial did not report outcomes with enough detail to be able to determine whether they observed similar complications. The effects of dexamethasone have also been found to differ among populations, especially in high-income versus middle- or low-income countries [295]. However, since the RECOVERY trial's results were released, strategies have been proposed for administering dexamethasone alongside more targeted treatments to minimize the likelihood of negative side effects [293]. Given the available evidence, dexamethasone is currently the most promising treatment for severe COVID-19.

## 1.2 Favipiravir

The effectiveness of favipiravir for treating patients with COVID-19 is currently under investigation. Evidence for the drug inhibiting viral RNA polymerase are based on time-of-drug addition studies that found that viral loads were reduced with the addition of favipiravir in early times post-infection [80,83,84]. An open-label, nonrandomized, before-after controlled study for COVID-19 was recently conducted [296]. The study included 80 COVID-19 patients (35 treated with favipiravir, 45 control) from the isolation ward of the National Clinical Research Center for Infectious Diseases (The Third People's Hospital of Shenzhen), Shenzhen, China. The patients in the control group were treated with other antivirals, such as lopinavir and ritonavir. It should be noted that although the control patients received antivirals, two subsequent large-scale analyses, the WHO Solidarity trial and the Randomized Evaluation of COVID-19 Therapy (RECOVERY) trial, identified no effect of lopinavir or of a lopinavir-ritonavir combination, respectively, on the metrics of COVID-19-related mortality that each assessed [93,297,298]. Treatment was applied on days 2-14; treatment stopped either when viral clearance was confirmed or at day 14. The efficacy of the treatment was measured by, first, the time until viral clearance using Kaplan-Meier survival curves, and, second, the improvement rate of chest computed tomography (CT) scans on day 14 after treatment. The study found that favipiravir increased the speed of recovery, measured as viral clearance from the patient by RT-PCR, with patients receiving favipiravir recovering in four days compared to 11 days for patients receiving antivirals such as lopinavir and ritonavir. Additionally, the lung CT scans of patients treated with favipiravir showed significantly higher improvement rates (91%) on day 14 compared to control patients (62%, *p* = 0.004). However, there were adverse side effects in 4 (11%) favipiravir-treated patients and 25 (56%) control patients. The adverse side effects included diarrhea, vomiting, nausea, rash, and liver and kidney injury. Despite the study reporting clinical improvement in favipiravir-treated patients, several study design issues are problematic and lower confidence in the overall conclusions. For example, the study was neither randomized nor blinded. Moreover, the selection of patients did not take into consideration important factors such as previous clinical conditions or sex, and there was no age categorization. Additionally, it should be noted that this study was temporarily retracted and then restored without an explanation [299].

In late 2020 and early 2021, the first randomized controlled trials of favipiravir for the treatment of COVID-19 released results [300,301,302]. The first [300] used a randomized, controlled, open-label design to compare two drugs, favipiravir and baloxavir marboxil, to SOC alone. Here, SOC included antivirals such as lopinavir/ritonavir and was administered to all patients. The primary endpoint analyzed was viral clearance at day 14. The sample size for this study was very small, with 29 total patients enrolled, and no significant effect of the treatments was found for the primary or any of the secondary outcomes analyzed, which included mortality. The second study [301] was larger, with 96 patients enrolled, and included only individuals with mild to moderate symptoms who were randomized into two groups: one receiving chloroquine (CQ) in addition to SOC, and the other receiving favipiravir in addition to SOC. This study reported a non-significant trend for patients receiving favipiravir to have a shorter hospital stay (13.29 days compared to 15.89 for CQ, *p* = 0.06) and less likelihood of progressing to mechanical ventilation (*p* = 0.118) or to an oxygen saturation < 90% (*p* = 0.129). These results, combined with the fact that favipiravir was being compared to CQ, which is now widely understood to be ineffective for treating COVID-19, thus do not suggest that favipiravir was likely to have had a strong effect on these outcomes. On the other hand, another trial of 60 patients reported a significant effect of favipiravir on viral clearance at four days (a secondary endpoint), but not at 10 days (the primary endpoint) [302]. This study, as well as a prior study of favipiravir [303], also reported that the drug was generally well-tolerated. Thus, in combination, these small studies suggest that the effects of favipiravir as a treatment for COVID-19 cannot be determined based on the available evidence, but additionally, none raise major concerns about the safety profile of the drug.

## 1.3 Remdesivir

At the outset of the COVID-19 pandemic, remdesivir did not have any have any FDA-approved use. A clinical trial in the Democratic Republic of Congo found some evidence of effectiveness against ebola virus disease (EVD), but two antibody preparations were found to be more effective, and remdesivir was not pursued [304]. Remdesivir also inhibits polymerase and replication of the coronaviruses MERS-CoV and SARS-CoV-1 in cell culture assays with submicromolar IC50s [305]. It has also been found to inhibit SARS-CoV-2, showing synergy with CQ *in vitro* [88].

Remdesivir was first used on some COVID-19 patients under compassionate use guidelines [308]. All were in late stages of COVID-19 infection, and initial reports were inconclusive about the drug's efficacy. Gilead Sciences, the maker of remdesivir, led a recent publication that reported outcomes for compassionate use of the drug in 61 patients hospitalized with confirmed COVID-19. Here, 200 mg of remdesivir was administered intravenously on day 1, followed by a further 100 mg/day for 9 days [92]. There were significant issues with the study design, or lack thereof. There was no randomized control group. The inclusion criteria were variable: some patients only required low doses of oxygen, while others required ventilation. The study included many sites, potentially with variable inclusion criteria and treatment protocols. The patients analyzed had mixed demographics. There was a short follow-up period of investigation. Eight patients were excluded from the analysis mainly due to missing post-baseline information; thus, their health was unaccounted for. Therefore, even though the study reported clinical improvement in 68% of the 53 patients ultimately evaluated, due to the significant issues with study design, it could not be determined whether treatment with remdesivir had an effect or whether these patients would have recovered regardless of treatment. Another study comparing 5- and 10-day treatment regimens reported similar results but was also limited because of the lack of a placebo control [309]. These studies did not alter the understanding of the efficacy of remdesivir in treating COVID-19, but the encouraging results provided motivation for placebo-controlled studies.

The double-blind placebo-controlled ACTT-1 trial [89,90] recruited 1,062 patients and randomly assigned them to placebo treatment or treatment with remdesivir. Patients were stratified for randomization based on site and the severity of disease presentation at baseline [89]. The treatment was 200 mg on day 1, followed by 100 mg on days 2 through 10. Data was analyzed from a total of 1,059 patients who completed the 29-day course of the trial, with 517 assigned to remdesivir and 508 to placebo [89]. The two groups were well matched demographically and clinically at baseline. Those who received remdesivir had a median recovery time of 10 days, as compared with 15 days in those who received placebo (rate ratio for recovery, 1.29; 95% confidence interval (CI), 1.12 to 1.49; $p < 0.001$). The Kaplan-Meier estimates of mortality by 14 days were 6.7% with remdesivir and 11.9% with placebo, with a hazard ratio (HR) for death of 0.55 and a 95% CI of 0.36 to 0.83, and at day 29, remdesivir corresponded to 11.4% and the placebo to 15.2% (HR: 0.73; 95% CI, 0.52 to 1.03). Serious adverse events were reported in 131 of the 532 patients who received remdesivir (24.6%) and in 163 of the 516 patients in the placebo group (31.6%). This study also reported an association between remdesivir administration and both clinical improvement and a lack of progression to more invasive respiratory intervention in patients receiving non-invasive and invasive ventilation at randomization [89]. Largely on the results of this trial, the FDA reissued and expanded the EUA for remdesivir for the treatment of hospitalized COVID-19 patients ages twelve and older [310]. Additional clinical trials [88,311,312,313,314] are currently underway to evaluate the use of remdesivir to treat COVID-19 patients at both early and late stages of infection and in combination with other drugs (Figure 2). As of October 22, 2020, remdesivir received FDA approval based on three clinical trials [315].

However, results suggesting no effect of remdesivir on survival were reported by the WHO Solidarity trial [93]. Patients were randomized in equal proportions into four experimental conditions and a control condition, corresponding to four candidate treatments for COVID-19 and SOC, respectively; no placebo was administered. The 2,750 patients in the remdesivir group were administered 200 mg intravenously on the first day and 100 mg on each subsequent day until day 10 and assessed for in-hospital death (primary endpoint), duration of hospitalization, and progression to mechanical ventilation. There were also 2,708 control patients who would have been eligible and able to receive remdesivir were they not assigned to the control group. A total of 604 patients among these two cohorts died during initial hospitalization, with 301 in the remdesivir group and 303 in the control group. The rate ratio of death between these two groups was therefore not significant (0.95, $p$ = 0.50), suggesting that the administration of remdesivir did not affect survival. The two secondary analyses similarly did not find any effect of remdesivir. Additionally, the authors compared data from their study with data from three other studies of remdesivir (including [89]) stratified by supplemental oxygen status. A meta-analysis of the four studies yielded an overall rate ratio for death of 0.91 ($p$ = 0.20). These results thus do not support the previous findings that remdesivir reduced median recovery time and mortality risk in COVID-19 patients.

In response to the results of the Solidarity trial, Gilead, which manufactures remdesivir, released a statement pointing to the fact that the Solidarity trial was not placebo-controlled or double-blind and at the time of release, the statement had not been peer reviewed [316]; these sentiments have been echoed elsewhere [317]. Other critiques of this study have noted that antivirals are not typically targeted at patients with severe illness, and therefore remdesivir could be more beneficial for patients with mild rather than severe cases [298,318]. However, the publication associated with the trial sponsored by Gilead did purport an effect of remdesivir on patients with severe disease, identifying an 11 versus 18 day recovery period (rate ratio for recovery: 1.31, 95% CI 1.12 to 1.52) [89]. Additionally, a smaller analysis of 598 patients, of whom two-thirds were randomized to receive remdesivir for either 5 or 10 days, reported a small effect of treatment with remdesivir for five days relative to standard of care in patients with moderate COVID-19 [319]. These results suggest that remdesivir could improve outcomes for patients with moderate COVID-19, but that additional information would be needed to understand the effects of different durations of treatment. Therefore, the Solidarity trial may point to limitations in the generalizability of other research on remdesivir, especially since the broad international nature of the Solidarity clinical trial, which included countries with a wide range of economic profiles and a variety of healthcare systems, provides a much-needed global perspective in a pandemic [298]. On the other hand, only 62% of patients in the Solidarity trial were randomized on the day of admission or one day afterwards [93], and concerns have been raised that differences in disease progression could influence the effectiveness of remdesivir [298]. Despite the findings of the Solidarity trial, remdesivir remains available for the treatment of COVID-19 in many places. Remdesivir has also been investigated in combination with other drugs, such as baricitinib, which is an inhibitor of Janus kinase 1 and 2 [320]; the FDA has issued an EUA for the combination of remdesivir and baricitinib in adult and pediatric patients [321]. Follow-up studies are needed and, in many cases, are underway to further investigate remdesivir-related outcomes.

Similarly, the extent to which the remdesivir dosing regimen could influence outcomes continues to be under consideration. A randomized, open-label trial compared the effect of remdesivir on 397 patients with severe COVID-19 over 5 versus 10 days [91,309], complementing the study that found that a 5-day course of remdesivir improved outcomes for patients with moderate

COVID-19 but a 10-day course did not [319]. Patients in the two groups were administered 200 mg of remdesivir intravenously on the first day, followed by 100 mg on the subsequent four or nine days, respectively. The two groups differed significantly in their clinical status, with patients assigned to the 10-day group having more severe illness. This study also differed from most because it included not only adults, but also pediatric patients as young as 12 years old. It reported no significant differences across several outcomes for patients receiving a 5-day or 10-day course, when correcting for baseline clinical status. The data did suggest that the 10-day course might reduce mortality in the most severe patients at day 14, but the representation of this group in the study population was too low to justify any conclusions [309]. Thus, additional research is also required to determine whether the dosage and duration of remdesivir administration influences outcomes.

In summary, remdesivir is the first FDA approved anti-viral against SARS-CoV-2 as well as the first FDA approved COVID-19 treatment. Early investigations of this drug established proof of principle that drugs targeting the virus can benefit COVID-19 patients. Moreover, one of the most successful strategies for developing therapeutics for viral diseases is to target the viral replication machinery, which are typically virally encoded polymerases. Small molecule drugs targeting viral polymerases are the backbones of treatments for other viral diseases including human immunodeficiency virus (HIV) and herpes. Notably, the HIV and herpes polymerases are a reverse transcriptase and a DNA polymerase, respectively, whereas SARS-CoV-2 encodes an RdRP, so most of the commonly used polymerase inhibitors are not likely to be active against SARS-CoV-2. In clinical use, polymerase inhibitors show short term benefits for HIV patients, but for long term benefits they must be part of combination regimens. They are typically combined with protease inhibitors, integrase inhibitors, and even other polymerase inhibitors. Remdesivir provides evidence that a related approach may be beneficial for the treatment of COVID-19.

## 1.4 Hydroxychloroquine and Chloroquine

CQ and hydroxychloroquine (HCQ) increase cellular pH by accumulating in their protonated form inside lysosomes [95,322]. This shift in pH inhibits the breakdown of proteins and peptides by the lysosomes during the process of proteolysis [95]. Interest in CQ and HCQ for treating COVID-19 was catalyzed by a mechanism observed in *in vitro* studies of both SARS-CoV-1 and SARS-CoV-2. In one study, CQ inhibited viral entry of SARS-CoV-1 into Vero E6 cells, a cell line that was derived from Vero cells in 1968, through the elevation of endosomal pH and the terminal glycosylation of ACE2 [96]. Increased pH within the cell, as discussed above, inhibits proteolysis, and terminal glycosylation of ACE2 is thought to interfere with virus-receptor binding. An *in vitro* study of SARS-CoV-2 infection of Vero cells found both HCQ and CQ to be effective in inhibiting viral replication, with HCQ being more potent [97]. Additionally, an early case study of three COVID-19 patients reported the presence of antiphospholipid antibodies in all three patients [323]. Antiphospholipid antibodies are central to the diagnosis of the antiphospholipid syndrome, a disorder that HCQ has often been used to treat [324,325,326]. Because the 90% effective concentration ($EC_{90}$) of CQ in Vero E6 cells (6.90 µM) can be achieved in and tolerated by rheumatoid arthritis (RA) patients, it was hypothesized that it might also be possible to achieve the effective concentration in COVID-19 patients [327]. Additionally, clinical trials have reported HCQ to be effective in treating HIV [328] and chronic Hepatitis C

[329]. Together, these studies triggered initial enthusiasm about the therapeutic potential for HCQ and CQ against COVID-19. HCQ/CQ has been proposed both as a treatment for COVID-19 and a prophylaxis against SARS-CoV-2 exposure, and trials often investigated these drugs in combination with azithromycin (AZ) and/or zinc supplementation. However, as more evidence has emerged, it has become clear that HCQ/CQ offer no benefits against SARS-CoV-2 or COVID-19.

## 1.4.1 Trials Assessing Therapeutic Administration of HCQ/CQ

The initial study evaluating HCQ as a treatment for COVID-19 patients was published on March 20, 2020 by Gautret et al. [98]. This non-randomized, non-blinded, non-placebo clinical trial compared HCQ to SOC in 42 hospitalized patients in southern France. It reported that patients who received HCQ showed higher rates of virological clearance by nasopharyngeal swab on days 3-6 when compared to SOC. This study also treated six patients with both HCQ + AZ and found this combination therapy to be more effective than HCQ alone. However, the design and analyses used showed weaknesses that severely limit interpretability of results, including the small sample size and the lack of: randomization, blinding, placebo (no "placebo pill" given to SOC group), Intention-To-Treat analysis, correction for sequential multiple comparisons, and trial pre-registration. Furthermore, the trial arms were entirely confounded by the hospital and there were false negative outcome measurements (see [330]). Two of these weaknesses are due to inappropriate data analysis and can therefore be corrected *post hoc* by recalculating the p-values (lack of Intention-To-Treat analysis and multiple comparisons). However, all other weaknesses are fundamental design flaws and cannot be corrected for. Thus, the conclusions cannot be generalized outside of the study. The International Society of Antimicrobial Chemotherapy, the scientific organization that publishes the journal where the article appeared, subsequently announced that the article did not meet its expected standard for publications [99], although it has not been officially retracted.

Because of the preliminary data presented in this study, HCQ treatment was subsequently explored by other researchers. About one week later, a follow-up case study reported that 11 consecutive patients were treated with HCQ + AZ using the same dosing regimen [331]. One patient died, two were transferred to the intensive care unit (ICU), and one developed a prolonged QT interval, leading to discontinuation of HCQ + AZ administration. As in the Gautret et al. study, the outcome assessed was virological clearance at day 6 post-treatment, as measured from nasopharyngeal swabs. Of the ten living patients on day 6, eight remained positive for SARS-CoV-2 RNA. Like in the original study, interpretability was severely limited by the lack of a comparison group and the small sample size. However, these results stand in contrast to the claims by Gautret et al. that all six patients treated with HCQ + AZ tested negative for SARS-CoV-2 RNA by day 6 post-treatment. This case study illustrated the need for further investigation using robust study design to evaluate the efficacy of HCQ and/or CQ.

On April 10, 2020, a randomized, non-placebo trial of 62 COVID-19 patients at the Renmin Hospital of Wuhan University was released [332]. This study investigated whether HCQ decreased time to fever break or time to cough relief when compared to SOC [332]. This trial found HCQ decreased both average time to fever break and average time to cough relief, defined as mild or no cough. While this study improved on some of the methodological flaws in

Gautret et al. by randomizing patients, it also had several flaws in trial design and data analysis that prevent generalization of the results. These weaknesses include the lack of placebo, lack of correction for multiple primary outcomes, inappropriate choice of outcomes, lack of sufficient detail to understand analysis, drastic disparities between pre-registration [333] and published protocol (including differences in the inclusion and exclusion criteria, the number of experimental groups, the number of patients enrolled, and the outcome analyzed), and small sample size. The choice of outcomes may be inappropriate as both fevers and cough may break periodically without resolution of illness. Additionally, for these outcomes, the authors reported that 23 of 62 patients did not have a fever and 25 of 62 patients did not have a cough at the start of the study, but the authors failed to describe how these patients were included in a study assessing time to fever break and time to cough relief. It is important to note here that the authors claimed "neither the research performers nor the patients were aware of the treatment assignments." This blinding seems impossible in a non-placebo trial because at the very least, providers would know whether they were administering a medication or not, and this knowledge could lead to systematic differences in the administration of care. Correction for multiple primary outcomes can be adjusted *post hoc* by recalculating p-values, but all of the other issues were design and statistical weaknesses that cannot be corrected for. Additionally, disparities between the pre-registered and published protocols raise concerns about experimental design. The design limitations mean that the conclusions cannot be generalized outside of the study.

A second randomized trial, conducted by the Shanghai Public Health Clinical Center, analyzed whether HCQ increased rates of virological clearance at day 7 in respiratory pharyngeal swabs compared to SOC [334]. This trial was published in Chinese along with an abstract in English, and only the English abstract was read and interpreted for this review. The trial found comparable outcomes in virological clearance rate, time to virological clearance, and time to body temperature normalization between the treatment and control groups. The small sample size is one weakness, with only 30 patients enrolled and 15 in each arm. This problem suggests the study is underpowered to detect potentially useful differences and precludes interpretation of results. Additionally, because only the abstract could be read, other design and analysis issues could be present. Thus, though these studies added randomization to their assessment of HCQ, their conclusions should be interpreted very cautiously. These two studies assessed different outcomes and reached differing conclusions about the efficacy of HCQ for treating COVID-19; the designs of both studies, especially with respect to sample size, meant that no general conclusions can be made about the efficacy of the drug.

Several widely reported studies on HCQ also have issues with data integrity and/or provenance. A Letter to the Editor published in *BioScience Trends* on March 16, 2020 claimed that numerous clinical trials have shown that HCQ is superior to control treatment in inhibiting the exacerbation of COVID-19 pneumonia [335]. This letter has been cited by numerous primary literature, review articles, and media alike [336,337]. However, the letter referred to 15 pre-registration identifiers from the Chinese Clinical Trial Registry. When these identifiers are followed back to the registry, most trials claim they are not yet recruiting patients or are currently recruiting patients. For all of these 15 identifiers, no data uploads or links to publications could be located on the pre-registrations. At the very least, the lack of availability of the primary data means the claim that HCQ is efficacious against COVID-19 pneumonia cannot be verified. Similarly, a recent multinational registry analysis [338] analyzed the efficacy of CQ and HCQ with and without a macrolide, which is a class of antibiotics that includes Azithromycin, for the treatment of COVID-19. The study observed 96,032 patients split into a control and four treatment conditions (CQ with and without a macrolide; HCQ with and without a macrolide). They concluded that treatment with CQ or HCQ was associated with increased risk of *de novo* ventricular arrhythmia during hospitalization. However, this study has since been retracted by *The Lancet* due to an inability to validate the data used [339]. These studies demonstrate that increased skepticism in evaluation of the HCQ/CQ and COVID-19 literature may be warranted, possibly because of the significant attention HCQ and CQ have received as possible treatments for COVID-19 and the politicization of these drugs.

Despite the fact that the study suggesting that CQ/HCQ increased risk of ventricular arrhythmia in COVID-19 patients has now been retracted, previous studies have identified risks associated with HCQ/CQ. A patient with systemic lupus erythematosus developed a prolonged QT interval that was likely exacerbated by use of HCQ in combination with renal failure [340]. A prolonged QT interval is associated with ventricular arrhythmia [341]. Furthermore, a separate study [342] investigated the safety associated with the use of HCQ with and without macrolides between 2000 and 2020. The study involved 900,000 cases treated with HCQ and 300,000 cases treated with HCQ + AZ. The results indicated that short-term use of HCQ was not associated with additional risk, but that HCQ + AZ was associated with an enhanced risk of cardiovascular complications (such as a 15% increased risk of chest pain, calibrated HR = 1.15, 95% CI, 1.05 to 1.26) and a two-fold increased 30-day risk of cardiovascular mortality (calibrated HR = 2.19; 95% CI, 1.22 to 3.94). Therefore, whether studies utilize HCQ alone or HCQ in combination with a macrolide may be an important consideration in assessing risk. As results from initial investigations of these drug combinations have emerged, concerns about the efficacy and risks of treating COVID-19 with HCQ and CQ have led to the removal of CQ/HCQ from SOC practices in several countries [343,344]. As of May 25, 2020, WHO had suspended administration of HCQ as part of the worldwide Solidarity Trial [345], and later the final results of this large-scale trial that compared 947 patients administered HCQ to 906 controls revealed no effect on the primary outcome, mortality during hospitalization (rate ratio: 1.19; $p$ = 0.23)

Additional research has emerged largely identifying HCQ/CQ to be ineffective against COVID-19 while simultaneously revealing a number of significant side effects. A randomized, open-label, non-placebo trial of 150 COVID-19 patients was conducted in parallel at 16 government-designated COVID-19 centers in China to assess the safety and efficacy of HCQ [346]. The trial compared treatment with HCQ in conjunction with SOC to SOC alone in 150 infected patients who were assigned randomly to the two groups (75 per group). The primary endpoint of the study was the negative conversion rate of SARS-CoV-2 in 28 days, and the investigators found no difference in this parameter between the groups (estimated difference between SOC plus HCQ and SOC 4.1%; 95% CI, –10.3% to 18.5%). The secondary endpoints were an amelioration of the symptoms of the disease such as axillary temperature ≤36.6°C, SpO2 >94% on room air, and disappearance of symptoms like shortness of breath, cough, and sore throat. The median time to symptom alleviation was similar across different conditions (19 days in HCQ + SOC versus 21 days in SOC, $p$ = 0.97). Additionally, 30% of the patients receiving SOC+HCQ reported adverse outcomes compared to 8.8% of patients receiving only SOC, with the most common adverse outcome in the SOC+HCQ group being diarrhea (10% versus 0% in the SOC group, $p$ = 0.004). However, there are several factors that limit the interpretability of this study.

Most of the enrolled patients had mild-to-moderate symptoms (98%), and the average age was 46. SOC in this study included the use of antivirals (Lopinavir-Ritonavir, Arbidol, Oseltamivir, Virazole, Entecavir, Ganciclovir, and Interferon alfa), which the authors note could influence the results. Thus, they note that an ideal SOC would need to exclude the use of antivirals, but that ceasing antiviral treatment raised ethical concerns at the time that the study was conducted. In this trial, the samples used to test for the presence of the SARS-CoV-2 virus were collected from the upper respiratory tract, and the authors indicated that the use of upper respiratory samples may have introduced false negatives (e.g., [347]). Another limitation of the study that the authors acknowledge was that the HCQ treatment began, on average, at a 16-day delay from the symptom onset. The fact that this study was open-label and lacked a placebo limits interpretation, and additional analysis is required to determine whether HCQ reduces inflammatory response. Therefore, despite some potential areas of investigation identified in *post hoc* analysis, this study cannot be interpreted as providing support for HCQ as a therapeutic against COVID-19. This study provided no support for HCQ against COVID-19, as there was no difference between the two groups in either negative seroconversion at 28 days or symptom alleviation, and in fact, more severe adverse outcomes were reported in the group receiving HCQ.

Additional evidence comes from a retrospective analysis [348] that examined data from 368 COVID-19 patients across all United States Veteran Health Administration medical centers. The study retrospectively investigated the effect of the administration of HCQ (n=97), HCQ + AZ (n=113), and no HCQ (n=158) on 368 patients. The primary outcomes assessed were death and the need for mechanical ventilation. Standard supportive care was rendered to all patients. Due to the low representation of women (N=17) in the available data, the analysis included only men, and the median age was 65 years. The rate of death was 27.8% in the HCQ-only treatment group, 22.1% in the HCQ + AZ treatment group, and 14.1% in the no-HCQ group. These data indicated a statistically significant elevation in the risk of death for the HCQ-only group compared to the no-HCQ group (adjusted HR: 2.61, $p$ = 0.03), but not for the HCQ + AZ group compared to the no-HCQ group (adjusted HR: 1.14; $p$ = 0.72). Further, the risk of ventilation was similar across all three groups (adjusted HR: 1.43, $p$ = 0.48 (HCQ) and 0.43, $p$ = 0.09 (HCQ + AZ) compared to no HCQ). The study thus showed evidence of an association between increased mortality and HCQ in this cohort of COVID-19 patients but no change in rates of mechanical ventilation among the treatment conditions. The study had a few limitations: it was not randomized, and the baseline vital signs, laboratory tests, and prescription drug use were significantly different among the three groups. All of these factors could potentially influence treatment outcome. Furthermore, the authors acknowledge that the effect of the drugs might be different in females and pediatric subjects, since these subjects were not part of the study. The reported result that HCQ + AZ is safer than HCQ contradicts the findings of the previous large-scale analysis of twenty years of records that found HCQ + AZ to be more frequently associated with cardiac arrhythmia than HCQ alone [342]; whether this discrepancy is caused by the pathology of COVID-19, is influenced by age or sex, or is a statistical artifact is not presently known.

Finally, findings from the RECOVERY trial were released on October 8, 2020. This study used a randomized, open-label design to study the effects of HCQ compared to SOC in 11,197 patients at 176 hospitals in the United Kingdom [100]. Patients were randomized into either the control group or one of the treatment arms, with twice as many patients enrolled in the control group as any treatment group. Of the patients eligible to receive HCQ, 1,561 were randomized into the HCQ arm, and 3,155 were randomized into the control arm. The demographics of the HCQ and control groups were similar in terms of average age (65 years), proportion female (approximately 38%), ethnic make-up (73% versus 76% white), and prevalence of pre-existing conditions (56% versus 57% overall). In the HCQ arm of the study, patients received 800 mg at baseline and again after 6 hours, then 400 mg at 12 hours and every subsequent 12 hours. The primary outcome analyzed was all-cause mortality, and patient vital statistics were reported by physicians upon discharge or death, or else at 28 days following HCQ administration if they remained hospitalized. The secondary outcome assessed was the combined risk of progression to invasive mechanical ventilation or death within 28 days. By the advice of an external data monitoring committee, the HCQ arm of the study was reviewed early, leading to it being closed due a lack of support for HCQ as a treatment for COVID-19. COVID-19-related mortality was not affected by HCQ in the RECOVERY trial (rate ratio, 1.09; 95% CI, 0.97 to 1.23; $p$ = 0.15), but cardiac events were increased in the HCQ arm (0.4 percentage points), as was the duration of hospitalization (rate ratio for discharge alive within 28 days: 0.90; 95% CI, 0.83 to 0.98) and likelihood of progression to mechanical ventilation or death (risk ratio 1.14; 95% CI, 1.03 to 1.27). This large-scale study thus builds upon studies in the United States and China to suggest that HCQ is not an effective treatment, and in fact may negatively impact COVID-19 patients due to its side effects. Therefore, though none of the studies have been blinded, examining them together makes it clear that the available evidence points to significant dangers associated with the administration of HCQ to hospitalized COVID-19 patients, without providing any support for its efficacy.

## 1.4.2 HCQ for the Treatment of Mild Cases

One additional possible therapeutic application of HCQ considered was the treatment of mild COVID-19 cases in otherwise healthy individuals. This possibility was assessed in a randomized, open-label, multi-center analysis conducted in Catalonia (Spain) [349]. This analysis enrolled adults 18 and older who had been experiencing mild symptoms of COVID-19 for fewer than five days. Participants were randomized into an HCQ arm (N=136) and a control arm (N=157), and those in the treatment arm were administered 800 mg of HCQ on the first day of treatment followed by 400 mg on each of the subsequent six days. The primary outcome assessed was viral clearance at days 3 and 7 following the onset of treatment, and secondary outcomes were clinical progression and time to complete resolution of symptoms. No significant differences between the two groups were found: the difference in viral load between the HCQ and control groups was 0.01 (95% CI, -0.28 to 0.29) at day 3 and -0.07 (95% CI -0.44 to 0.29) at day 7, the relative risk of hospitalization was 0.75 (95% CI, 0.32 to 1.77), and the difference in time to complete resolution of symptoms was -2 days ($p$ = 0.38). This study thus suggests that HCQ does not improve recovery from COVID-19, even in otherwise healthy adult patients with mild symptoms.

## 1.4.3 Prophylactic Administration of HCQ

An initial study of the possible prophylactic application of HCQ utilized a randomized, double-blind, placebo-controlled design to analyze the administration of HCQ prophylactically [350].

Asymptomatic adults in the United States and Canada who had been exposed to SARS-CoV-2 within the past four days were enrolled in an online study to evaluate whether administration of HCQ over five days influenced the probability of developing COVID-19 symptoms over a 14-day period. Of the participants, 414 received HCQ and 407 received a placebo. No significant difference in the rate of symptomatic illness was observed between the two groups (11.8% HCQ, 14.3% placebo, $p$ = 0.35). The HCQ condition was associated with side effects, with 40.1% of patients reporting side effects compared to 16.8% in the control group ($p < 0.001$). However, likely due to the high enrollment of healthcare workers (66% of participants) and the well-known side effects associated with HCQ, a large number of participants were able to correctly identify whether they were receiving HCQ or a placebo (46.5% and 35.7%, respectively). Furthermore, due to a lack of availability of diagnostic testing, only 20 of the 107 cases were confirmed with a PCR-based test to be positive for SARS-CoV-2. The rest were categorized as "probable" or "possible" cases by a panel of four physicians who were blind to the treatment status. One possible confounder is that a patient presenting one or more symptoms, which included diarrhea, was defined as a "possible" case, but diarrhea is also a common side effect of HCQ. Additionally, four of the twenty PCR-confirmed cases did not develop symptoms until after the observation period had completed, suggesting that the 14-day trial period may not have been long enough or that some participants also encountered secondary exposure events. Finally, in addition to the young age of the participants in this study, which ranged from 32 to 51, there were possible impediments to generalization introduced by the selection process, as 2,237 patients who were eligible but had already developed symptoms by day 4 were enrolled in a separate study. It is therefore likely that asymptomatic cases were over-represented in this sample, which would not have been detected based on the diagnostic criteria used. Therefore, while this study does represent the first effort to conduct a randomized, double-blind, placebo-controlled investigation of HCQ's effect on COVID-19 prevention after SARS-CoV-2 exposure in a large sample, the lack of PCR tests and several other design flaws significantly impede interpretation of the results. However, in line with the results from therapeutic studies, once again no evidence was found suggesting an effect of HCQ against COVID-19.

A second study [351] examined the effect of administering HCQ to healthcare workers as a pre-exposure prophylactic. The primary outcome assessed was the conversion from SARS-CoV-2 negative to SARS-CoV-2 positive status over the 8 week study period. This study was also randomized, double-blind, and placebo-controlled, and it sought to address some of the limitations of the first prophylactic study. The goal was to enroll 200 healthcare workers, preferentially those working with COVID-19 patients, at two hospitals within the University of Pennsylvania hospital system in Philadelphia, PA. Participants were randomized 1:1 to receive either 600 mg of HCQ daily or a placebo, and their SARS-CoV-2 infection status and antibody status were assessed using RT-PCR and serological testing, respectively, at baseline, 4 weeks, and 8 weeks following the beginning of the treatment period. The statistical design of the study accounted for interim analyses at 50 and 100 participants in case efficacy or futility of HCQ for prophylaxis became clear earlier than completion of enrollment. The 139 individuals enrolled comprised a study population that was fairly young (average age 33) and made of largely of people who were white, women, and without pre-existing conditions. At the second interim analysis, more individuals in the treatment group than the control group had contracted COVID-19 (4 versus 3), causing the estimated z-score to fall below the pre-established threshold for futility. As a result, the trial was terminated early, offering additional evidence against the use of HCQ for prophylaxis.

### 1.4.4 Summary of HCQ/CQ Research Findings

Early *in vitro* evidence indicated that HCQ could be an effective therapeutic against SARS-CoV-2 and COVID-19, leading to significant media attention and public interest in its potential as both a therapeutic and prophylactic. Initially it was hypothesized that CQ/HCQ might be effective against SARS-CoV-2 in part because CQ and HCQ have both been found to inhibit the expression of CD154 in T-cells and to reduce TLR signaling that leads to the production of pro-inflammatory cytokines [352]. Clinical trials for COVID-19 have more often used HCQ rather than CQ because it offers the advantages of being cheaper and having fewer side effects than CQ. However, research has not found support for a positive effect of HCQ on COVID-19 patients. Multiple clinical studies have already been carried out to assess HCQ as a therapeutic agent for COVID-19, and many more are in progress. To date, none of these studies have used randomized, double-blind, placebo-controlled designs with a large sample size, which would be the gold standard. Despite the design limitations (which would be more likely to produce false positives than false negatives), initial optimism about HCQ has largely dissipated. The most methodologically rigorous analysis of HCQ as a prophylactic [350] found no significant differences between the treatment and control groups, and the WHO's global Solidarity trial similarly reported no effect of HCQ on mortality [93]. Thus, HCQ/CQ are not likely to be effective therapeutic or prophylactic agents against COVID-19. One case study identified drug-induced phospholipidosis as the cause of death for a COVID-19 patient treated with HCQ [272], suggesting that in some cases, the proposed mechanism of action may ultimately be harmful. Additionally, one study identified an increased risk of mortality in older men receiving HCQ, and administration of HCQ and HCQ + AZ did not decrease the use of mechanical ventilation in these patients [348]. HCQ use for COVID-19 could also lead to shortages for anti-malarial or anti-rheumatic use, where it has documented efficacy. Despite significant early attention, these drugs appear to be ineffective against COVID-19. Several countries have now removed CQ/HCQ from their SOC for COVID-19 due to the lack of evidence of efficacy and the frequency of adverse effects.

## 1.5 ACE Inhibitors and Angiotensin II Receptor Blockers

Several clinical trials testing the effects of ACEIs or ARBs on COVID-19 outcomes are ongoing [353,354,355,356,357,358,359]. Clinical trials are needed because the findings of the various observational studies bearing on this topic cannot be interpreted as indicating a protective effect of the drug [360,361]. Two analyses [353,359] have reported no effect of continuing or discontinuing ARBs and ACEIs on patients admitted to the hospital for COVID-19. The first, known as REPLACE COVID [156], was a randomized, open-label study that enrolled patients who were admitted to the hospital for COVID-19 and were taking an ACEI at the time of admission. They enrolled 152 patients at 20 hospitals across seven countries and randomized them into two arms, continuation (n=75) and discontinuation (n=77). The primary outcome evaluated was a global rank score that integrated several dimensions of illness. The components of this global rank score, such as time to death and length of mechanical ventilation, were evaluated as secondary endpoints. This analysis reported no differences between the two groups in the primary or any of the secondary outcomes.

Similarly, a second study [157] used a randomized, open-label design to examine the effects of continuing versus discontinuing ARBs and ACEIs on patients hospitalized for mild to moderate COVID-19 at 29 hospitals in Brazil. This study enrolled 740 patients but had to exclude one trial site from all analyses due to the discovery of violations of Good Clinical Trial practice and data falsification. After this exclusion, 659 patients remained, with 334 randomized to discontinuation and 325 to continuation. In this study, the primary endpoint analyzed was the number of days that patients were alive and not hospitalized within 30 days of enrollment. The secondary outcomes included death (including in-hospital death separately), number of days hospitalized, and specific clinical outcomes such as heart failure or stroke. Once again, no significant differences were found between the two groups. Initial studies of randomized interventions therefore suggest that ACEIs and ARBs are unlikely to affect COVID-19 outcomes. These results are also consistent with findings from observational studies (summarized in [156]). Additional information about ACE2, observational studies of ACEIs and ARBs in COVID-19, and clinical trials on this topic have been summarized [362]. Therefore, despite the promising potential mechanism, initial results have not provided support for ACEIs and ARBs as therapies for COVID-19.

## 1.6 Tocilizumab

Human IL-6 is a 26-kDa glycoprotein that consists of 184 amino acids and contains two potential N-glycosylation sites and four cysteine residues. It binds to a type I cytokine receptor (IL-6Rα or glycoprotein 80) that exists in both membrane-bound (IL-6Rα) and soluble (sIL-6Rα) forms [363]. It is not the binding of IL-6 to the receptor that initiates pro- and/or anti-inflammatory signaling, but rather the binding of the complex to another subunit, known as IL-6Rβ or glycoprotein 130 (gp130) [363,364]. Unlike membrane-bound IL-6Rα, which is only found on hepatocytes and some types of leukocytes, gp130 is found on most cells [365]. When IL-6 binds to sIL-6Rα, the complex can then bind to a gp130 protein on any cell [365]. The binding of IL-6 to IL-6Rα is termed classical signaling, while its binding to sIL-6Rα is termed trans-signaling [365,366,367]. These two signaling processes are thought to play different roles in health and illness. For example, trans-signaling may play a role in the proliferation of mucosal T-helper TH2 cells associated with asthma, while an earlier step in this proliferation process may be regulated by classical signaling [365]. Similarly, IL-6 is known to play a role in Crohn's Disease via trans-, but not classical, signaling [365]. Both classical and trans-signaling can occur through three independent pathways: the Janus-activated kinase-STAT3 pathway, the Ras/Mitogen-Activated Protein Kinases pathway and the Phosphoinositol-3 Kinase/Akt pathway [363]. These signaling pathways are involved in a variety of different functions, including cell type differentiation, immunoglobulin synthesis, and cellular survival signaling pathways, respectively [363]. The ultimate result of the IL-6 cascade is to direct transcriptional activity of various promoters of pro-inflammatory cytokines, such as IL-1, TFN, and even IL-6 itself, through the activity of NF-κB [363]. IL-6 synthesis is tightly regulated both transcriptionally and post-transcriptionally, and it has been shown that viral proteins can enhance transcription of the IL-6 gene by strengthening the DNA-binding activity between several transcription factors and IL-6 gene-cis-regulatory elements [368]. Therefore, drugs inhibiting the binding of IL-6 to IL-6Rα or sIL-6Rα are of interest for combating the hyperactive inflammatory response characteristic of cytokine release syndrome (CRS) and cytokine storm syndrome (CSS). TCZ is a humanized monoclonal antibody that binds both to the insoluble and soluble receptor of IL-6, providing de facto inhibition of the IL-6 immune cascade. Interest in TCZ as a possible treatment for COVID-19 was piqued by early evidence indicating that COVID-19 deaths may be induced by the hyperactive immune response, often referred to as CRS or CSS [170], as IL-6 plays a key role in this response [369]. The observation of elevated IL-6 in patients who died relative to those who recovered [170] could reflect an over-production of proinflammatory interleukins, suggesting that TCZ could potentially palliate some of the most severe symptoms of COVID-19 associated with increased cytokine production.

This early interest in TCZ as a possible treatment for COVID-19 was bolstered by a very small retrospective study in China that examined 20 patients with severe symptoms in early February 2020 and reported rapid improvement in symptoms following treatment with TCZ [176]. Subsequently, a number of retrospective studies have been conducted in several countries. Many studies use a retrospective, observational design, where they compare outcomes for COVID-19 patients who received TCZ to those who did not over a set period of time. For example, one of the largest retrospective, observational analyses released to date [171], consisting of 1,351 patients admitted to several care centers in Italy, compared the rates at which patients who received TCZ died or progressed to invasive medical ventilation over a 14-day period compared to patients receiving only SOC. Under this definition, SOC could include other drugs such as HCQ, azithromycin, lopinavir-ritonavir or darunavir-cobicistat, or heparin. While this study was not randomized, a subset of patients who were eligible to receive TCZ were unable to obtain it due to shortages; however, these groups were not directly compared in the analysis. After adjusting for variables such as age, sex, and SOFA (sequential organ failure assessment) score, they found that patients treated with TCZ were less likely to progress to invasive medical ventilation and/or death (adjusted HR = 0.61, CI 0.40-0.92, $p$ = 0.020); analysis of death and ventilation separately suggests that this effect may have been driven by differences in the death rate (20% of control versus 7% of TCZ-treated patients). The study reported particular benefits for patients whose $PaO_2/FiO_2$ ratio, also known as the Horowitz Index for Lung Function, fell below a 150 mm Hg threshold. They found no differences between groups administered subcutaneous versus intravenous TCZ.

Another retrospective observational analysis of interest examined the charts of patients at a hospital in Connecticut, USA where 64% of all 239 COVID-19 patients in the study period were administered TCZ based on assignment by a standardized algorithm [172]. They found that TCZ administration was associated with more similar rates of survivorship in patients with severe versus nonsevere COVID-19 at intake, defined based on the amount of supplemental oxygen needed. They therefore proposed that their algorithm was able to identify patients presenting with or likely to develop CRS as good candidates for TCZ. This study also reported higher survivorship in Black and Hispanic patients compared to white patients when adjusted for age. The major limitation with interpretation for these studies is that there may be clinical characteristics that influenced medical practitioners decisions to administer TCZ to some patients and not others. One interesting example therefore comes from an analysis of patients at a single hospital in Brescia, Italy, where TCZ was not available for a period of time [173]. This study compared COVID-19 patients admitted to the hospital before and after March 13, 2020, when the hospital received TCZ. Therefore, patients who would have been eligible for TCZ prior to this arbitrary date did not receive it as treatment, making this retrospective analysis something of a natural experiment. Despite this design, demographic factors did not appear to be consistent between the two groups, and the average age of the control group was older than the TCZ group. The control group also had a higher percentage of males and a higher incidence of comorbidities such as diabetes and heart disease. All the same, the multivariate HR, which adjusted for these clinical and demographic factors, found a significant difference between survival in the two groups (HR=0.035, CI=0.004-0.347, $p$ = 0.004). The study reported improvement of survival outcomes after the addition of TCZ to the SOC regime, with 11 of 23 patients (47.8%) admitted prior to March 13th dying compared to 2 of 62 (3.2%) admitted afterwards (HR=0.035; 95% CI, 0.004 to 0.347; $p$ = 0.004). They also reported a reduced progression to mechanical ventilation in the TCZ group. However, this study also holds a significant limitation: the time delay between the two groups means that knowledge about how to treat the disease likely improved over this timeframe as well. All the same, the results of these observational retrospective studies provide support for TCZ as a pharmaceutical of interest for follow-up in clinical trials.

Other retrospective analyses have utilized a case-control design to match pairs of patients with similar baseline characteristics, only one of whom received TCZ for COVID-19. In one such study, TCZ was significantly associated with a reduced risk of progression to intensive care unit (ICU) admission or death [174]. This study examined only 20 patients treated with TCZ (all but one of the patients treated with TCZ in the hospital during the study period) and compared them to 25 patients receiving SOC. For the combined primary endpoint of death and/or ICU admission, only 25% of patients receiving TCZ progressed to an endpoint compared to 72% in the SOC group ($p$ = 0.002, presumably based on a chi-square test based on the information provided in the text). When the two endpoints were examined separately, progression to invasive medical ventilation remained significant (32% SOC compared to 0% TCZ, $p$ = 0.006) but not for mortality (48% SOC compared to 25% TCZ, $p$ = 0.066). In contrast, a study that compared 96 patients treated with TCZ to 97 patients treated with SOC only in New York City found that differences in mortality did not differ between the two groups, but that this difference did become significant when intubated patients were excluded from the analysis [175]. Taken together, these findings suggest that future clinical trials of TCZ may want to include intubation as an endpoint. However, these studies should be approached with caution, not only because of the small number of patients enrolled and the retrospective design, but also because they performed a large number of statistical tests and did not account for multiple hypothesis testing. In general, caution must be exercised when interpreting subgroup analyses after a primary combined endpoint analysis. These last findings highlight the need to search for a balance between impairing a harmful immune response, such as the one generated during CRS/CSS, and preventing the worsening of the clinical picture of the patients by potential new viral infections. Early meta-analyses and systematic reviews have investigated the available data about TCZ for COVID-19. One meta-analysis [370] evaluated 19 studies published or released as preprints prior to July 1, 2020 and found that the overall trends were supportive of the frequent conclusion that TCZ does improve survivorship, with a significant HR of 0.41 ($p$ < 0.001). This trend improved when they excluded studies that administered a steroid alongside TCZ, with a significant HR of 0.04 ($p$ < 0.001). They also found some evidence for reduced invasive ventilation or ICU admission, but only when excluding all studies except a small number whose estimates were adjusted for the possible bias introduced by the challenges of stringency during the enrollment process. A systematic analysis of sixteen case-control studies of TCZ estimated an odds ratio of mortality of 0.453 (95% CI 0.376–0.547, *p* < 0.001), suggesting possible benefits associated with TCZ treatment [371]. Although these estimates are similar, it is important to note that they are drawing from the same literature and are therefore likely to be affected by the same potential biases in publication. A different systematic review of studies investigating TCZ treatment for COVID-19 analyzed 31 studies that had been published or released as pre-prints and reported that none carried a low risk of bias [372]. Therefore, the present evidence is not likely to be sufficient for conclusions about the efficacy of TCZ.

On February 11, 2021, a preprint describing the first randomized control trial of TCZ was released as part of the RECOVERY trial [177]. Of the 21,550 patients enrolled in the RECOVERY trial at the time, 4,116 adults hospitalized with COVID-19 across the 131 sites in the United Kingdom were assigned to the arm of the trial evaluating the effect of TCZ. Among them, 2,022 were randomized to receive TCZ and 2,094 were randomized to SOC, with 79% of patients in each group available for analysis at the time that the initial report was released. The primary outcome measured was 28-day mortality, and TCZ was found to reduce 28-day mortality from 33% of patients receiving SOC alone to 29% of those receiving TCZ, corresponding to a rate ratio of 0.86 (95% CI 0.77-0.96; *p* = 0.007). TCZ was also significantly associated with the probability of hospital discharge within 28 days for living patients, which was 47% in the SOC group and 54% in the TCZ group (rate ratio 1.22, 95% CI 1.12-1.34, *p* < 0.0001). A potential statistical interaction between TCZ and corticosteroids was observed, with the combination providing greater mortality benefits than TCZ alone, but the authors note that caution is advisable in light of the number of statistical tests conducted. Combining the RECOVERY trial data with data from seven smaller randomized control trials indicates that TCZ is associated with a 13% reduction in 28-day mortality (rate ratio 0.87, 95% CI 0.79-0.96, *p* = 0.005) [177].

There are possible risks associated with the administration of TCZ for COVID-19. TCZ has been used for over a decade to treat RA [373], and a recent study found the drug to be safe for pregnant and breastfeeding women [374]. However, TCZ may increase the risk of developing infections [373], and RA patients with chronic hepatitis B infections had a high risk of hepatitis B virus reactivation when TCZ was administered in combination with other RA drugs [375]. As a result, TCZ is contraindicated in patients with active infections such as tuberculosis [376]. Previous studies have investigated, with varying results, a possible increased risk of infection in RA patients administered TCZ [377,378], although another study reported that the incidence rate of infections was higher in clinical practice RA patients treated with TCZ than in the rates reported by clinical trials [379]. In the investigation of 544 Italian COVID-19 patients, the group treated with TCZ was found to be more likely to develop secondary infections, with 24% compared to 4% in the control group (*p* < 0.0001) [171]. Reactivation of hepatitis B and herpes simplex virus 1 was also reported in a small number of patients in this study, all of whom were receiving TCZ. A July 2020 case report described negative outcomes of two COVID-19 patients after receiving TCZ, including one death; however, both patients were intubated and had entered septic shock prior to receiving TCZ [380], likely indicating a severe level of cytokine production. Additionally, D-dimer and sIL2R levels were reported by one study to increase in patients treated with TCZ, which raised concerns because of the potential association between elevated D-dimer levels and thrombosis and between sIL2R and diseases where T-cell regulation is compromised [172]. An increased risk of bacterial infection was also identified in a systematic review of the literature, based on the unadjusted estimates reported [370]. In the RECOVERY trial, however, only three out of 2,022 participants in the group receiving TCZ developed adverse reactions determined to be associated with the intervention, and no excess deaths were reported [177]. TCZ administration to COVID-19 patients is not without risks and may introduce additional risk of developing secondary infections; however, while caution may be prudent when treating patients who have latent viral infections, the results of the RECOVERY trial indicate that adverse reactions to TCZ are very rare among COVID-19 patients broadly.

In summary, approximately 33% of hospitalized COVID-19 patients develop ARDS [381], which is caused by an excessive early response of the immune system which can be a component of CRS/CSS [172,376]. This overwhelming inflammation is triggered by IL-6. TCZ is an inhibitor of IL-6 and therefore may neutralize the inflammatory pathway that leads to the cytokine storm. The mechanism suggests TCZ could be beneficial for the treatment of COVID-19 patients experiencing excessive immune activity, and the RECOVERY trial reported a reduction in 28-day mortality. Interest in TCZ as a treatment for COVID-19 was also supported by two meta-analyses [370,382], but a third meta-analysis found that all of the available literature at that time carried a risk of bias [372]. Additionally, different studies used different dosages, number of doses, and methods of administration. Ongoing research may be needed to optimize administration of TCZ [383], although similar results were reported by one study for intravenous and subcutaneous administration [171]. Clinical trials that are in progress are likely to provide additional insight into the effectiveness of this drug for the treatment of COVID-19 along with how it should be administered.

## 1.7 Interferons

IFNs are a family of cytokines critical to activating the innate immune response against viral infections. Interferons are classified into three categories based on their receptor specificity: types I, II and III [369]. Specifically, IFNs I (IFN-$\alpha$ and $\beta$) and II (IFN-$\gamma$) induce the expression of antiviral proteins [384]. Among these IFNs, IFN-$\beta$ has already been found to strongly inhibit the replication of other coronaviruses, such as SARS-CoV-1, in cell culture, while IFN-$\alpha$ and $\gamma$ were shown to be less effective in this context [384]. There is evidence that patients with higher susceptibility to ARDS indeed show deficiency in IFN-$\beta$. For instance, infection with other coronaviruses impairs IFN-$\beta$ expression and synthesis, allowing the virus to escape the innate immune response [385]. On March 18 2020, Synairgen plc received approval to start a phase II trial for SNG001, an IFN-$\beta$-1a formulation to be delivered to the lungs via inhalation [184]. SNG001, which contains recombinant interferon beta-1a, was previously shown to be effective in reducing viral load in an *in vivo* model of swine flu and *in vitro* models of other coronavirus infections [386]. In July 2020, a press release from Synairgen stated that SNG001 reduced progression to ventilation in a double-blind, placebo-controlled, multi-center study of 101 patients with an average age in the late 50s [185]. These results were subsequently published in November 2020 [186]. The study reports that the participants were assigned at a ratio of 1:1 to receive either SNG001 or a placebo that lacked the active compound, by inhalation for up to 14 days. The primary outcome they assessed was the change in patients' score on the WHO

Ordinal Scale for Clinical Improvement (OSCI) at trial day 15 or 16. SNG001 was associated with an odds ratio of improvement on the OSCI scale of 2.32 (95% CI 1.07 – 5.04, $p$ = 0.033) in the intention-to-treat analysis and 2.80 (95% CI 1.21 – 6.52, $p$ = 0.017) in the per-protocol analysis, corresponding to significant improvement in the SNG001 group on the OSCI at day 15/16. Some of the secondary endpoints analyzed also showed differences: at day 28, the OR for clinical improvement on the OSCI was 3.15 (95% CI 1.39 – 7.14, $p$ = 0.006), and the odds of recovery at day 15/16 and at day 28 were also significant between the two groups. Thus, this study suggested that IFN-$\beta$1 administered via SNG001 may improve clinical outcomes.

In contrast, the WHO Solidarity trial reported no significant effect of IFN-β-1a on patient survival during hospitalization [93]. Here, the primary outcome analyzed was in-hospital mortality, and the rate ratio for the two groups was 1.16 (95% CI, 0.96 to 1.39; $p$ = 0.11) administering IFN-β-1a to 2050 patients and comparing their response to 2,050 controls. However, there are a few reasons that the different findings of the two trials might not speak to the underlying efficacy of this treatment strategy. One important consideration is the stage of COVID-19 infection analyzed in each study. The Synairgen trial enrolled only patients who were not receiving invasive ventilation, corresponding to a less severe stage of disease than many patients enrolled in the SOLIDARITY trial, as well as a lower overall rate of mortality [387]. Additionally, the methods of administration differed between the two trials, with the SOLIDARITY trial administering IFN-β-1a subcutaneously [387]. The differences in findings between the studies suggests that the method of administration might be relevant to outcomes, with nebulized IFN-β-1a more directly targeting receptors in the lungs. A trial that analyzed the effect of subcutaneously administered IFN-β-1a on patients with ARDS between 2015 and 2017 had also reported no effect on 28-day mortality [388], while a smaller study analyzing the effect of subcutaneous IFN administration did find a significant improvement in 28-day mortality for COVID-19 [389]. At present, several ongoing clinical trials are investigating the potential effects of IFN-β-1a, including in combination with therapeutics such as remdesivir [390] and administered via inhalation [184]. Thus, as additional information becomes available, a more detailed understanding of whether and under which circumstances IFN-β-1a is beneficial to COVID-19 patients should develop.

## 1.8 Potential Avenues of Interest for Therapeutic Development

Given what is currently known about these therapeutics for COVID-19, a number of related therapies beyond those explored above may also prove to be of interest. For example, the demonstrated benefit of dexamethasone and the ongoing potential of tocilizumab for treatment of COVID-19 suggests that other anti-inflammatory agents might also hold value for the treatment of COVID-19. Current evidence supporting the treatment of severe COVID-19 with dexamethasone suggests that the need to curtail the cytokine storm inflammatory response transcends the risks of immunosuppression, and other anti-inflammatory agents may therefore benefit patients in this phase of the disease. While dexamethasone is considered widely available and generally affordable, the high costs of biologics such as tocilizumab therapy may present obstacles to wide-scale distribution of this drug if it proves of value. At the doses used for RA patients, the cost for tocilizumab ranges from $179.20 to $896 per dose for the IV form and $355 for the pre-filled syringe [391]. Several other anti-inflammatory agents used for the treatment of autoimmune diseases may also be able to counter the effects of the cytokine storm induced by the virus, and some of these, such as cyclosporine, are likely to be more cost-effective and readily available than biologics [392]. While tocilizumab targets IL-6, several other inflammatory markers could be potential targets, including TNF-α. Inhibition of TNF-α by a compound such as Etanercept was previously suggested for treatment of SARS-CoV-1 [393] and may be relevant for SARS-CoV-2 as well. Another anti-IL-6 antibody, sarilumab, is also being investigated [394,395]. Baricitinib and other small molecule inhibitors of the Janus-activated kinase pathway also curtail the inflammatory response and have been suggested as potential options for SARS-CoV-2 infections [396]. Baricitinib, in particular, may be able to reduce the ability of SARS-CoV-2 to infect lung cells [397]. Clinical trials studying baricitinib in COVID-19 have already begun in the US and in Italy [398,399]. Identification and targeting of further inflammatory markers that are relevant in SARS-CoV-2 infection may be of value for curtailing the inflammatory response and lung damage.

In addition to immunosuppressive treatments, which are most beneficial late in disease progression, much research is focused on identifying therapeutics for early-stage patients. For example, although studies of HCQ have not supported the early theory-driven interest in this antiviral treatment, alternative compounds with related mechanisms may still have potential. Hydroxyferroquine derivatives of HCQ have been described as a class of bioorganometallic compounds that exert antiviral effects with some selectivity for SARS-CoV-1 *in vitro* [400]. Future work could explore whether such compounds exert antiviral effects against SARS-CoV-2 and whether they would be safer for use in COVID-19.

Another potential approach is the development of antivirals, which could be broad-spectrum, specific to coronaviruses, or targeted to SARS-CoV-2. Development of new antivirals is complicated by the fact that none have yet been approved for human coronaviruses. Intriguing new options are emerging, however. Beta-D-N4-hydroxycytidine is an orally bioavailable ribonucleotide analog showing broad-spectrum activity against RNA viruses, which may inhibit SARS-CoV-2 replication *in vitro* and *in vivo* in mouse models of HCoVs [401]. A range of other antivirals are also in development. Development of antivirals will be further facilitated as research reveals more information about the interaction of SARS-CoV-2 with the host cell and host cell genome, mechanisms of viral replication, mechanisms of viral assembly, and mechanisms of viral release to other cells; this can allow researchers to target specific stages and structures of the viral life cycle. Finally, antibodies against viruses, also known as antiviral monoclonal antibodies, could be an alternative as well and are described in detail in an above section. The goal of antiviral antibodies is to neutralize viruses through either cell-killing activity or blocking of viral replication [402]. They may also engage the host immune response, encouraging the immune system to hone in on the virus. Given the cytokine storm that results from immune system activation in response to the virus, which has been implicated in worsening of the disease, a neutralizing antibody (nAb) may be preferable. Upcoming work may explore the specificity of nAbs for their target, mechanisms by which the nAbs impede the virus, and improvements to antibody structure that may enhance the ability of the antibody to block viral activity.

Some research is also investigating potential therapeutics and prophylactics that would interact with components of the innate immune response. For example, TLRs are pattern recognition receptors that recognize pathogen- and damage-associated molecular patterns and contribute to innate immune recognition and, more generally, promotion of both the innate and adaptive immune responses [403]. In mouse models, poly(I:C) and CpG, which are agonists of Toll-like receptors TLR3 and TLR9, respectively, showed protective effects when administered prior to SARS-CoV-1 infection [404]. Therefore, TLR agonists hold some potential for broad-spectrum prophylaxis.

# 2 Additional Items

## 2.1 Competing Interests

| Author | Competing Interests | Last Reviewed |
|---|---|---|
| Halie M. Rando | None | 2021-01-20 |
| Nils Wellhausen | None | 2020-11-03 |
| Soumita Ghosh | None | 2020-11-09 |
| Alexandra J. Lee | None | 2020-11-09 |
| Anna Ada Dattoli | None | 2020-03-26 |
| Fengling Hu | None | 2020-04-08 |
| James Brian Byrd | Funded by FastGrants to conduct a COVID-19-related clinical trial | 2020-11-12 |
| Diane N. Rafizadeh | None | 2020-11-11 |
| Ronan Lordan | None | 2020-11-03 |
| Yanjun Qi | None | 2020-07-09 |
| Yuchen Sun | None | 2020-11-11 |
| Christian Brueffer | Employee and shareholder of SAGA Diagnostics AB. | 2020-11-11 |
| Jeffrey M. Field | None | 2020-11-12 |
| Marouen Ben Guebila | None | 2021-08-02 |
| Nafisa M. Jadavji | None | 2020-11-11 |
| Ashwin N. Skelly | None | 2020-11-11 |
| Bharath Ramsundar | None | 2020-11-11 |
| Jinhui Wang | None | 2021-01-21 |
| Rishi Raj Goel | None | 2021-01-20 |

| Author | Competing Interests | Last Reviewed |
|---|---|---|
| YoSon Park | Now employed by Pfizer (subsequent to contributions to this project) | 2020-01-22 |
| COVID-19 Review Consortium | None | 2021-01-16 |
| Simina M. Boca | Currently an employee at AstraZeneca, Gaithersburg, MD, USA, may own stock or stock options. Work initially conducted at Georgetown University Medical Center with writing, reviewing, and editing continued while working at AstraZeneca. | 2021-07-01 |
| Anthony Gitter | Filed a patent application with the Wisconsin Alumni Research Foundation related to classifying activated T cells | 2020-11-10 |
| Casey S. Greene | None | 2021-01-20 |

## 2.2 Author Contributions

| Author | Contributions |
|---|---|
| Halie M. Rando | Project Administration, Writing - Original Draft, Writing - Review & Editing |
| Nils Wellhausen | Project Administration, Visualization, Writing - Original Draft, Writing - Review & Editing |
| Soumita Ghosh | Writing - Original Draft |
| Alexandra J. Lee | Writing - Original Draft, Writing - Review & Editing |
| Anna Ada Dattoli | Writing - Original Draft |
| Fengling Hu | Writing - Original Draft, Writing - Review & Editing |
| James Brian Byrd | Writing - Original Draft, Writing - Review & Editing |
| Diane N. Rafizadeh | Project Administration, Writing - Original Draft, Writing - Review & Editing |
| Ronan Lordan | Project Administration, Writing - Original Draft, Writing - Review & Editing |
| Yanjun Qi | Visualization |
| Yuchen Sun | Visualization |
| Christian Brueffer | Project Administration, Writing - Review & Editing |
| Jeffrey M. Field | Writing - Original Draft, Writing - Review & Editing |
| Marouen Ben Guebila | Writing - Original Draft |
| Nafisa M. Jadavji | Supervision, Writing - Original Draft, Writing - Review & Editing |
| Ashwin N. Skelly | Writing - Review & Editing |
| Bharath Ramsundar | Writing - Review & Editing |
| Jinhui Wang | Writing - Original Draft |
| Rishi Raj Goel | Writing - Review & Editing |
| YoSon Park | Writing - Review & Editing |
| COVID-19 Review Consortium | Project Administration |

| Author | Contributions |
|---|---|
| Simina M. Boca | Project Administration, Writing - Review & Editing |
| Anthony Gitter | Project Administration, Software, Visualization, Writing - Review & Editing |
| Casey S. Greene | Project Administration, Writing - Review & Editing |

## 2.3 Acknowledgements

We thank Nick DeVito for assistance with the Evidence-Based Medicine Data Lab COVID-19 TrialsTracker data. We thank Yael Evelyn Marshall who contributed writing (original draft) as well as reviewing and editing of pieces of the text but who did not formally approve the manuscript, as well as Ronnie Russell, who contributed text to and helped develop the structure of the manuscript early in the writing process and Matthias Fax who helped with writing and editing text related to diagnostics. We are also very grateful to James Fraser for suggestions about successes and limitations in the area of computational screening for drug repurposing. We are grateful to the following contributors for reviewing pieces of the text: Nadia Danilova, James Eberwine and Ipsita Krishnan.

# 3 References

1. **Pathogenesis, Symptomatology, and Transmission of SARS-CoV-2 through analysis of Viral Genomics and Structure** Halie M Rando, Adam L MacLean, Alexandra J Lee, Sandipan Ray, Vikas Bansal, Ashwin N Skelly, Elizabeth Sell, John J Dziak, Lamonica Shinholster, Lucy D'Agostino McGowan, … Casey S Greene *arXiv* (2021-02-15) https://arxiv.org/abs/2102.01521

2. **Vaccine Development Strategies for SARS-CoV-2** COVID-19 Review Consortium *Manubot* (2021-02-19) https://greenelab.github.io/covid19-review/v/d9d90fd7e88ef547fb4cbed0ef73baef5fee7fb5/#vaccine-development-strategies-for-sars-cov-2

3. **A Visual Approach for the SARS (Severe Acute Respiratory Syndrome) Outbreak Data Analysis** Jie Hua, Guohua Wang, Maolin Huang, Shuyang Hua, Shuanghe Yang *International Journal of Environmental Research and Public Health* (2020-06-03) https://doi.org/gjqg6z DOI: 10.3390/ijerph17113973 · PMID: 32503333 · PMCID: PMC7312089

4. **COVID-19 Data Repository** Center for Systems Science and Engineering at Johns Hopkins University *GitHub* https://github.com/CSSEGISandData/COVID-19/tree/master/csse_covid_19_data/csse_covid_19_time_series

5. **An interactive web-based dashboard to track COVID-19 in real time** Ensheng Dong, Hongru Du, Lauren Gardner *The Lancet Infectious Diseases* (2020-05) https://doi.org/ggnsjk DOI: 10.1016/s1473-3099(20)30120-1 · PMID: 32087114 · PMCID: PMC7159018

6. **Severe Acute Respiratory Syndrome (SARS)** https://www.who.int/westernpacific/health-topics/severe-acute-respiratory-syndrome

7.      **GitHub - imdevskp/sars-2003-outbreak-data-webscraping-code: repository contains complete WHO data of 2003 outbreak with code used to web scrap, data mung and cleaning** GitHub https://github.com/imdevskp/sars-2003-outbreak-data-webscraping-code

8.      **Three Emerging Coronaviruses in Two Decades** Jeannette Guarner *American Journal of Clinical Pathology* (2020-04) https://doi.org/ggppq3 DOI: 10.1093/ajcp/aqaa029 · PMID: 32053148 · PMCID: PMC7109697

9.      **SARS and MERS: recent insights into emerging coronaviruses** Emmie de Wit, Neeltje van Doremalen, Darryl Falzarano, Vincent J Munster *Nature Reviews Microbiology* (2016-06-27) https://doi.org/f8v5cv DOI: 10.1038/nrmicro.2016.81 · PMID: 27344959 · PMCID: PMC7097822

10.     **A Novel Coronavirus Genome Identified in a Cluster of Pneumonia Cases — Wuhan, China 2019−2020** Wenjie Tan, Xiang Zhao, Xuejun Ma, Wenling Wang, Peihua Niu, Wenbo Xu, George F. Gao, Guizhen Wu, MHC Key Laboratory of Biosafety, National Institute for Viral Disease Control and Prevention, China CDC, Beijing, China, Center for Biosafety Mega-Science, Chinese Academy of Sciences, Beijing, China *China CDC Weekly* (2020) https://doi.org/gg8z47 DOI: 10.46234/ccdcw2020.017

11.     **Airborne Transmission of SARS-CoV-2** Michael Klompas, Meghan A Baker, Chanu Rhee *JAMA* (2020-08-04) https://doi.org/gg4ttq DOI: 10.1001/jama.2020.12458

12.     **Exaggerated risk of transmission of COVID-19 by fomites** Emanuel Goldman *The Lancet Infectious Diseases* (2020-08) https://doi.org/gg6br7 DOI: 10.1016/s1473-3099(20)30561-2 · PMID: 32628907 · PMCID: PMC7333993

13.     **Ten scientific reasons in support of airborne transmission of SARS-CoV-2** Trisha Greenhalgh, Jose L Jimenez, Kimberly A Prather, Zeynep Tufekci, David Fisman, Robert Schooley *The Lancet* (2021-05) https://doi.org/gjqmvq DOI: 10.1016/s0140-6736(21)00869-2 · PMID: 33865497 · PMCID: PMC8049599

14.     **Covid-19 has redefined airborne transmission** Julian W Tang, Linsey C Marr, Yuguo Li, Stephanie J Dancer *BMJ* (2021-04-14) https://doi.org/gj3jh4 DOI: 10.1136/bmj.n913 · PMID: 33853842

15.     **Genomic characterisation and epidemiology of 2019 novel coronavirus: implications for virus origins and receptor binding** Roujian Lu, Xiang Zhao, Juan Li, Peihua Niu, Bo Yang, Honglong Wu, Wenling Wang, Hao Song, Baoying Huang, Na Zhu, … Wenjie Tan *The Lancet* (2020-02) https://doi.org/ggjr43 DOI: 10.1016/s0140-6736(20)30251-8

16.     **Isolation of a Novel Coronavirus from a Man with Pneumonia in Saudi Arabia** Ali M Zaki, Sander van Boheemen, Theo M Bestebroer, Albert DME Osterhaus, Ron AM Fouchier *New England Journal of Medicine* (2012-11-08) https://doi.org/f4czx5 DOI: 10.1056/nejmoa1211721 · PMID: 23075143

17.     **Drug repurposing: progress, challenges and recommendations** Sudeep Pushpakom, Francesco Iorio, Patrick A Eyers, KJane Escott, Shirley Hopper, Andrew Wells,

Andrew Doig, Tim Guilliams, Joanna Latimer, Christine McNamee, … Munir Pirmohamed *Nature Reviews Drug Discovery* (2018-10-12) https://doi.org/gfrbsz DOI: 10.1038/nrd.2018.168 · PMID: 30310233

18. **Drug discovery and development: Role of basic biological research** Richard C Mohs, Nigel H Greig *Alzheimer's & Dementia: Translational Research & Clinical Interventions* (2017-11) https://doi.org/gf92kj DOI: 10.1016/j.trci.2017.10.005 · PMID: 29255791 · PMCID: PMC5725284

19. **Evidence-Based Medicine Data Lab COVID-19 TrialsTracker** Nick DeVito, Peter Inglesby *GitHub* (2020-03-29) https://github.com/ebmdatalab/covid_trials_tracker-covid DOI: 10.5281/zenodo.3732709

20. **A living WHO guideline on drugs for covid-19** Bram Rochwerg, Arnav Agarwal, Reed AC Siemieniuk, Thomas Agoritsas, François Lamontagne, Lisa Askie, Lyubov Lytvyn, Yee-Sin Leo, Helen Macdonald, Linan Zeng, … Per Olav Vandvik *BMJ* (2020-09-04) https://doi.org/ghktgm DOI: 10.1136/bmj.m3379 · PMID: 32887691

21. **Drug treatments for covid-19: living systematic review and network meta-analysis** Reed AC Siemieniuk, Jessica J Bartoszko, Long Ge, Dena Zeraatkar, Ariel Izcovich, Elena Kum, Hector Pardo-Hernandez, Anila Qasim, Juan Pablo Díaz Martinez, Bram Rochwerg, … Romina Brignardello-Petersen *BMJ* (2020-07-30) https://doi.org/ghs8st DOI: 10.1136/bmj.m2980 · PMID: 32732190 · PMCID: PMC7390912

22. **Causes of Death and Comorbidities in Patients with COVID-19** Sefer Elezkurtaj, Selina Greuel, Jana Ihlow, Edward Michaelis, Philip Bischoff, Catarina Alisa Kunze, Bruno Valentin Sinn, Manuela Gerhold, Kathrin Hauptmann, Barbara Ingold-Heppner, … David Horst *Cold Spring Harbor Laboratory* (2020-06-17) https://doi.org/gg926j DOI: 10.1101/2020.06.15.20131540

23. **Clinical characteristics of 82 cases of death from COVID-19** Bicheng Zhang, Xiaoyang Zhou, Yanru Qiu, Yuxiao Song, Fan Feng, Jia Feng, Qibin Song, Qingzhu Jia, Jun Wang *PLOS ONE* (2020-07-09) https://doi.org/gg4sgx DOI: 10.1371/journal.pone.0235458 · PMID: 32645044 · PMCID: PMC7347130

24. **COVID-19 infection: the perspectives on immune responses** Yufang Shi, Ying Wang, Changshun Shao, Jianan Huang, Jianhe Gan, Xiaoping Huang, Enrico Bucci, Mauro Piacentini, Giuseppe Ippolito, Gerry Melino *Cell Death & Differentiation* (2020-03-23) https://doi.org/ggq8td DOI: 10.1038/s41418-020-0530-3 · PMID: 32205856 · PMCID: PMC7091918

25. **Cytokine Storm** David C Fajgenbaum, Carl H June *New England Journal of Medicine* (2020-12-03) https://doi.org/ghnhm7 DOI: 10.1056/nejmra2026131 · PMID: 33264547 · PMCID: PMC7727315

26. **Lung pathology of fatal severe acute respiratory syndrome** John M Nicholls, Leo LM Poon, Kam C Lee, Wai F Ng, Sik T Lai, Chung Y Leung, Chung M Chu, Pak K Hui, Kong L Mak,

Wilna Lim, … JS Malik Peiris *The Lancet* (2003-05) https://doi.org/c8mmbg DOI: 10.1016/s0140-6736(03)13413-7 · PMID: 12781536 · PMCID: PMC7112492

27. **Pro/con clinical debate: Steroids are a key component in the treatment of SARS** Charles D Gomersall, Marcus J Kargel, Stephen E Lapinsky *Critical Care* (2004) https://doi.org/dpjr29 DOI: 10.1186/cc2452 · PMID: 15025770 · PMCID: PMC420028

28. **Content Analysis and Characterization of Medical Tweets During the Early Covid-19 Pandemic** Ross Prager, Michael T Pratte, Rudy R Unni, Sudarshan Bala, Nicholas Ng Fat Hing, Kay Wu, Trevor A McGrath, Adam Thomas, Brent Thoma, Kwadwo Kyeremanteng *Cureus* (2021-02-27) https://doi.org/gjpccg DOI: 10.7759/cureus.13594 · PMID: 33815994 · PMCID: PMC8007019

29. **Small Molecules vs Biologics | Drug Development Differences** Nuventra Pharma Sciences2525 Meridian Parkway, Suite 200 Durham *PK / PD and Clinical Pharmacology Consultants* (2020-05-13) https://www.nuventra.com/resources/blog/small-molecules-versus-biologics/

30. **Drug Discovery: A Historical Perspective** J Drews *Science* (2000-03-17) https://doi.org/d6bvp7 DOI: 10.1126/science.287.5460.1960 · PMID: 10720314

31. **Prone Positioning in Awake, Nonintubated Patients With COVID-19 Hypoxemic Respiratory Failure** Alison E Thompson, Benjamin L Ranard, Ying Wei, Sanja Jelic *JAMA Internal Medicine* (2020-11-01) https://doi.org/gg2pq4 DOI: 10.1001/jamainternmed.2020.3030 · PMID: 32584946 · PMCID: PMC7301298

32. **SARS: Systematic Review of Treatment Effects** Lauren J Stockman, Richard Bellamy, Paul Garner *PLoS Medicine* (2006-09-12) https://doi.org/d7xwh2 DOI: 10.1371/journal.pmed.0030343 · PMID: 16968120 · PMCID: PMC1564166

33. **Current concepts in SARS treatment** Takeshi Fujii, Aikichi Iwamoto, Tetsuya Nakamura, Aikichi Iwamoto *Journal of Infection and Chemotherapy* (2004) https://doi.org/dpmxk2 DOI: 10.1007/s10156-003-0296-9 · PMID: 14991510 · PMCID: PMC7088022

34. **Corticosteroids for pneumonia** Anat Stern, Keren Skalsky, Tomer Avni, Elena Carrara, Leonard Leibovici, Mical Paul *Cochrane Database of Systematic Reviews* (2017-12-13) https://doi.org/gc9cdk DOI: 10.1002/14651858.cd007720.pub3 · PMID: 29236286 · PMCID: PMC6486210

35. **Corticosteroids for pneumonia** Yuanjing Chen, Ka Li, Hongshan Pu, Taixiang Wu *Cochrane Database of Systematic Reviews* (2011-03-16) https://doi.org/cvc92x DOI: 10.1002/14651858.cd007720.pub2 · PMID: 21412908

36. **Corticosteroids in severe pneumonia** O Sibila, C Agusti, A Torres *European Respiratory Journal* (2008-03-19) https://doi.org/bmdrvg DOI: 10.1183/09031936.00154107 · PMID: 18669784

37. **Efficacy of Corticosteroids in the Treatment of Community-Acquired Pneumonia Requiring Hospitalization** Katsunaka Mikami, Masaru Suzuki, Hiroshi Kitagawa, Masaki Kawakami, Nobuaki Hirota, Hiromichi Yamaguchi, Osamu Narumoto, Yoshiko Kichikawa, Makoto Kawai, Hiroyuki Tashimo, … Yoshio Sakamoto *Lung* (2007-08-21) https://doi.org/fk5f5d DOI: 10.1007/s00408-007-9020-3 · PMID: 17710485

38. **Corticosteroids in the Treatment of Community-Acquired Pneumonia in Adults: A Meta-Analysis** Wei Nie, Yi Zhang, Jinwei Cheng, Qingyu Xiu *PLoS ONE* (2012-10-24) https://doi.org/gj3jh5 DOI: 10.1371/journal.pone.0047926 · PMID: 23112872 · PMCID: PMC3480455

39. **Effect of corticosteroids on the clinical course of community-acquired pneumonia: a randomized controlled trial** Silvia Fernández-Serrano, Jordi Dorca, Carolina Garcia-Vidal, Núria Fernández-Sabé, Jordi Carratalà, Ana Fernández-Agüera, Mercè Corominas, Susana Padrones, Francesc Gudiol, Frederic Manresa *Critical Care* (2011) https://doi.org/c8ksgr DOI: 10.1186/cc10103 · PMID: 21406101 · PMCID: PMC3219361

40. **Dexamethasone treatment for the acute respiratory distress syndrome: a multicentre, randomised controlled trial** Jesús Villar, Carlos Ferrando, Domingo Martínez, Alfonso Ambrós, Tomás Muñoz, Juan A Soler, Gerardo Aguilar, Francisco Alba, Elena González-Higueras, Luís A Conesa, … Jesús Villar *The Lancet Respiratory Medicine* (2020-03) https://doi.org/ggpxzc DOI: 10.1016/s2213-2600(19)30417-5

41. **Corticosteroids in acute respiratory distress syndrome: a step forward, but more evidence is needed** Kiran Reddy, Cecilia O'Kane, Daniel McAuley *The Lancet Respiratory Medicine* (2020-03) https://doi.org/gcv2 DOI: 10.1016/s2213-2600(20)30048-5

42. **Nonventilatory Treatments for Acute Lung Injury and ARDS** Carolyn S Calfee, Michael A Matthay *Chest* (2007-03) https://doi.org/bqzn5v DOI: 10.1378/chest.06-1743 · PMID: 17356114 · PMCID: PMC2789489

43. **Corticosteroids in ARDS** GUmberto Meduri, Paul E Marik, Stephen M Pastores, Djillali Annane *Chest* (2007-09) https://doi.org/cjdz2d DOI: 10.1378/chest.07-0714 · PMID: 17873207

44. **Efficacy and Safety of Corticosteroids for Persistent Acute Respiratory Distress Syndrome** New England Journal of Medicine *Massachusetts Medical Society* (2006-04-20) https://doi.org/c3sfcb DOI: 10.1056/nejmoa051693 · PMID: 16625008

45. **Corticosteroids in the prevention and treatment of acute respiratory distress syndrome (ARDS) in adults: meta-analysis** John Victor Peter, Preeta John, Petra L Graham, John L Moran, Ige Abraham George, Andrew Bersten *BMJ* (2008-05-03) https://doi.org/b7qtn2 DOI: 10.1136/bmj.39537.939039.be · PMID: 18434379 · PMCID: PMC2364864

46. **Antiviral agents and corticosteroids in the treatment of severe acute respiratory syndrome (SARS)** WC Yu *Thorax* (2004-08-01) https://doi.org/bks99t DOI: 10.1136/thx.2003.017665 · PMID: 15282381 · PMCID: PMC1747111

47. **Corticosteroid treatment of severe acute respiratory syndrome in Hong Kong** Loretta Yin-Chun Yam, Arthur Chun-Wing Lau, Florence Yuk-Lin Lai, Edwina Shung, Jane Chan, Vivian Wong *Journal of Infection* (2007-01) https://doi.org/dffg65 DOI: 10.1016/j.jinf.2006.01.005 · PMID: 16542729 · PMCID: PMC7112522

48. **Impact of corticosteroid therapy on outcomes of persons with SARS-CoV-2, SARS-CoV, or MERS-CoV infection: a systematic review and meta-analysis** Huan Li, Chongxiang Chen, Fang Hu, Jiaojiao Wang, Qingyu Zhao, Robert Peter Gale, Yang Liang *Leukemia* (2020-05-05) https://doi.org/ggv2rb DOI: 10.1038/s41375-020-0848-3 · PMID: 32372026 · PMCID: PMC7199650

49. **Managing SARS amidst Uncertainty** Richard P Wenzel, Michael B Edmond *New England Journal of Medicine* (2003-05-15) https://doi.org/ddkjnr DOI: 10.1056/nejmp030072 · PMID: 12748313

50. **Synthesis and Pharmacology of Anti-Inflammatory Steroidal Antedrugs** MOmar F Khan, Henry J Lee *Chemical Reviews* (2008-12-10) https://doi.org/cmkrtc DOI: 10.1021/cr068203e · PMID: 19035773 · PMCID: PMC2650492

51. **Drug vignettes: Dexamethasone** The Centre for Evidence-Based Medicine https://www.cebm.net/covid-19/dexamethasone/

52. **Pharmacology of Postoperative Nausea and Vomiting** Eric S Zabirowicz, Tong J Gan *Elsevier BV* (2019) https://doi.org/ghfkjw DOI: 10.1016/b978-0-323-48110-6.00034-x

53. **Potential benefits of precise corticosteroids therapy for severe 2019-nCoV pneumonia** Wei Zhou, Yisi Liu, Dongdong Tian, Cheng Wang, Sa Wang, Jing Cheng, Ming Hu, Minghao Fang, Yue Gao *Signal Transduction and Targeted Therapy* (2020-02-21) https://doi.org/ggqr84 DOI: 10.1038/s41392-020-0127-9 · PMID: 32296012 · PMCID: PMC7035340

54. **16-METHYLATED STEROIDS. I. 16α-METHYLATED ANALOGS OF CORTISONE, A NEW GROUP OF ANTI-INFLAMMATORY STEROIDS** Glen E Arth, David BR Johnston, John Fried, William W Spooncer, Dale R Hoff, Lewis H Sarett *Journal of the American Chemical Society* (2002-05-01) https://doi.org/cj5c82 DOI: 10.1021/ja01545a061

55. **Treatment of Rheumatoid Arthritis with Dexamethasone** Abraham Cohen *JAMA* (1960-10-15) https://doi.org/csfmhc DOI: 10.1001/jama.1960.03030070009002 · PMID: 13694317

56. **Dexamethasone** DailyMed (2007-10-25) https://dailymed.nlm.nih.gov/dailymed/drugInfo.cfm?setid=537b424a-3e07-4c81-978c-1ad99014032a

57. **Prevention of infection caused by immunosuppressive drugs in gastroenterology** Katarzyna Orlicka, Eleanor Barnes, Emma L Culver *Therapeutic Advances in Chronic Disease* (2013-04-22) https://doi.org/ggrqd3 DOI: 10.1177/2040622313485275 · PMID: 23819020 · PMCID: PMC3697844

58.     **COVID-19: consider cytokine storm syndromes and immunosuppression** Puja Mehta, Daniel F McAuley, Michael Brown, Emilie Sanchez, Rachel S Tattersall, Jessica J Manson *The Lancet* (2020-03) https://doi.org/ggnzmc DOI: 10.1016/s0140-6736(20)30628-0

59.     **Clinical evidence does not support corticosteroid treatment for 2019-nCoV lung injury** Clark D Russell, Jonathan E Millar, JKenneth Baillie *The Lancet* (2020-02) https://doi.org/ggks86 DOI: 10.1016/s0140-6736(20)30317-2 · PMID: 32043983 · PMCID: PMC7134694

60.     **On the use of corticosteroids for 2019-nCoV pneumonia** Lianhan Shang, Jianping Zhao, Yi Hu, Ronghui Du, Bin Cao *The Lancet* (2020-02) https://doi.org/ggq356 DOI: 10.1016/s0140-6736(20)30361-5 · PMID: 32122468 · PMCID: PMC7159292

61.     **Immunosuppression for hyperinflammation in COVID-19: a double-edged sword?** Andrew I Ritchie, Aran Singanayagam *The Lancet* (2020-04) https://doi.org/ggq8hs DOI: 10.1016/s0140-6736(20)30691-7 · PMID: 32220278 · PMCID: PMC7138169

62.     **Effect of Dexamethasone in Hospitalized Patients with COVID-19 – Preliminary Report** Peter Horby, Wei Shen Lim, Jonathan Emberson, Marion Mafham, Jennifer Bell, Louise Linsell, Natalie Staplin, Christopher Brightling, Andrew Ustianowski, Einas Elmahi, … RECOVERY Collaborative Group *Cold Spring Harbor Laboratory* (2020-06-22) https://doi.org/dz5x DOI: 10.1101/2020.06.22.20137273

63.     **Dexamethasone in Hospitalized Patients with Covid-19 — Preliminary Report** The RECOVERY Collaborative Group *New England Journal of Medicine* (2020-07-17) https://doi.org/gg5c8p DOI: 10.1056/nejmoa2021436 · PMID: 32678530 · PMCID: PMC7383595

64.     **Corticosteroids for Patients With Coronavirus Disease 2019 (COVID-19) With Different Disease Severity: A Meta-Analysis of Randomized Clinical Trials** Laura Pasin, Paolo Navalesi, Alberto Zangrillo, Artem Kuzovlev, Valery Likhvantsev, Ludhmila Abrahão Hajjar, Stefano Fresilli, Marcus Vinicius Guimaraes Lacerda, Giovanni Landoni *Journal of Cardiothoracic and Vascular Anesthesia* (2021-02) https://doi.org/ghzkp9 DOI: 10.1053/j.jvca.2020.11.057 · PMID: 33298370 · PMCID: PMC7698829

65.     **Current concepts in the diagnosis and management of cytokine release syndrome** Daniel W Lee, Rebecca Gardner, David L Porter, Chrystal U Louis, Nabil Ahmed, Michael Jensen, Stephan A Grupp, Crystal L Mackall *Blood* (2014-07-10) https://doi.org/ggsrwk DOI: 10.1182/blood-2014-05-552729 · PMID: 24876563 · PMCID: PMC4093680

66.     **Corticosteroids in COVID-19 ARDS** Hallie C Prescott, Todd W Rice *JAMA* (2020-10-06) https://doi.org/gg9wsv DOI: 10.1001/jama.2020.16747 · PMID: 32876693

67.     **Dexamethasone: Therapeutic potential, risks, and future projection during COVID-19 pandemic** Sobia Noreen, Irsah Maqbool, Asadullah Madni *European Journal of Pharmacology* (2021-03) https://doi.org/gj4qgn DOI: 10.1016/j.ejphar.2021.173854 · PMID: 33428898 · PMCID: PMC7836247

68. **Covid-19: Demand for dexamethasone surges as RECOVERY trial publishes preprint** Elisabeth Mahase *BMJ* (2020-06-23) https://doi.org/gj4qgp DOI: 10.1136/bmj.m2512 · PMID: 32576548

69. **Introduction to modern virology** NJ Dimmock, AJ Easton, KN Leppard *Blackwell Pub* (2007) ISBN: 9781405136457

70. **CORONA Data Viewer** Castleman Disease Collaborative Network https://cdcn.org/corona-data-viewer/

71. **Coronaviruses** Helena Jane Maier, Erica Bickerton, Paul Britton (editors) *Methods in Molecular Biology* (2015) https://doi.org/ggqfqx DOI: 10.1007/978-1-4939-2438-7 · PMID: 25870870 · ISBN: 9781493924370

72. **The potential chemical structure of anti-SARS-CoV-2 RNA-dependent RNA polymerase** Jrhau Lung, Yu-Shih Lin, Yao-Hsu Yang, Yu-Lun Chou, Li-Hsin Shu, Yu-Ching Cheng, Hung Te Liu, Ching-Yuan Wu *Journal of Medical Virology* (2020-03-18) https://doi.org/ggp6fm DOI: 10.1002/jmv.25761 · PMID: 32167173

73. **Broad-spectrum coronavirus antiviral drug discovery** Allison L Totura, Sina Bavari *Expert Opinion on Drug Discovery* (2019-03-08) https://doi.org/gg74z5 DOI: 10.1080/17460441.2019.1581171 · PMID: 30849247 · PMCID: PMC7103675

74. **Ribavirin therapy for severe COVID-19: a retrospective cohort study** Song Tong, Yuan Su, Yuan Yu, Chuangyan Wu, Jiuling Chen, Sihua Wang, Jinjun Jiang *International Journal of Antimicrobial Agents* (2020-09) https://doi.org/gg5w75 DOI: 10.1016/j.ijantimicag.2020.106114 · PMID: 32712334 · PMCID: PMC7377772

75. **The evolution of nucleoside analogue antivirals: A review for chemists and non-chemists. Part 1: Early structural modifications to the nucleoside scaffold** Katherine L Seley-Radtke, Mary K Yates *Antiviral Research* (2018-06) https://doi.org/gdpn35 DOI: 10.1016/j.antiviral.2018.04.004 · PMID: 29649496 · PMCID: PMC6396324

76. **The Ambiguous Base-Pairing and High Substrate Efficiency of T-705 (Favipiravir) Ribofuranosyl 5′-Triphosphate towards Influenza A Virus Polymerase** Zhinan Jin, Lucas K Smith, Vivek K Rajwanshi, Baek Kim, Jerome Deval *PLoS ONE* (2013-07-10) https://doi.org/f5br92 DOI: 10.1371/journal.pone.0068347 · PMID: 23874596 · PMCID: PMC3707847

77. **Favipiravir** DrugBank (2020-06-12) https://www.drugbank.ca/drugs/DB12466

78. **In Vitro and In Vivo Activities of Anti-Influenza Virus Compound T-705** Y Furuta, K Takahashi, Y Fukuda, M Kuno, T Kamiyama, K Kozaki, N Nomura, H Egawa, S Minami, Y Watanabe, … K Shiraki *Antimicrobial Agents and Chemotherapy* (2002-04) https://doi.org/cndw7n DOI: 10.1128/aac.46.4.977-981.2002 · PMID: 11897578 · PMCID: PMC127093

79. **Efficacy of Orally Administered T-705 on Lethal Avian Influenza A (H5N1) Virus Infections in Mice** Robert W Sidwell, Dale L Barnard, Craig W Day, Donald F Smee, Kevin W Bailey, Min-Hui Wong, John D Morrey, Yousuke Furuta *Antimicrobial Agents and Chemotherapy* (2007-03) https://doi.org/dm9xr2 DOI: 10.1128/aac.01051-06 · PMID: 17194832 · PMCID: PMC1803113

80. **Mechanism of Action of T-705 against Influenza Virus** Yousuke Furuta, Kazumi Takahashi, Masako Kuno-Maekawa, Hidehiro Sangawa, Sayuri Uehara, Kyo Kozaki, Nobuhiko Nomura, Hiroyuki Egawa, Kimiyasu Shiraki *Antimicrobial Agents and Chemotherapy* (2005-03) https://doi.org/dgbwdh DOI: 10.1128/aac.49.3.981-986.2005 · PMID: 15728892 · PMCID: PMC549233

81. **Activity of T-705 in a Hamster Model of Yellow Fever Virus Infection in Comparison with That of a Chemically Related Compound, T-1106** Justin G Julander, Kristiina Shafer, Donald F Smee, John D Morrey, Yousuke Furuta *Antimicrobial Agents and Chemotherapy* (2009-01) https://doi.org/brknds DOI: 10.1128/aac.01074-08 · PMID: 18955536 · PMCID: PMC2612161

82. **In Vitro and In Vivo Activities of T-705 against Arenavirus and Bunyavirus Infections** Brian B Gowen, Min-Hui Wong, Kie-Hoon Jung, Andrew B Sanders, Michelle Mendenhall, Kevin W Bailey, Yousuke Furuta, Robert W Sidwell *Antimicrobial Agents and Chemotherapy* (2007-09) https://doi.org/d98c87 DOI: 10.1128/aac.00356-07 · PMID: 17606691 · PMCID: PMC2043187

83. **Favipiravir (T-705) inhibits in vitro norovirus replication** J Rocha-Pereira, D Jochmans, K Dallmeier, P Leyssen, MSJ Nascimento, J Neyts *Biochemical and Biophysical Research Communications* (2012-08) https://doi.org/f369j7 DOI: 10.1016/j.bbrc.2012.07.034 · PMID: 22809499

84. **T-705 (Favipiravir) Inhibition of Arenavirus Replication in Cell Culture** Michelle Mendenhall, Andrew Russell, Terry Juelich, Emily L Messina, Donald F Smee, Alexander N Freiberg, Michael R Holbrook, Yousuke Furuta, Juan-Carlos de la Torre, Jack H Nunberg, Brian B Gowen *Antimicrobial Agents and Chemotherapy* (2011-02) https://doi.org/cppwsc DOI: 10.1128/aac.01219-10 · PMID: 21115797 · PMCID: PMC3028760

85. **Favipiravir (T-705), a broad spectrum inhibitor of viral RNA polymerase** Yousuke FURUTA, Takashi KOMENO, Takaaki NAKAMURA *Proceedings of the Japan Academy, Series B* (2017) https://doi.org/gbxcxw DOI: 10.2183/pjab.93.027 · PMID: 28769016 · PMCID: PMC5713175

86. **The antiviral compound remdesivir potently inhibits RNA-dependent RNA polymerase from Middle East respiratory syndrome coronavirus** Calvin J Gordon, Egor P Tchesnokov, Joy Y Feng, Danielle P Porter, Matthias Götte *Journal of Biological Chemistry* (2020-04) https://doi.org/ggqm6x DOI: 10.1074/jbc.ac120.013056 · PMID: 32094225

87. **Coronavirus Susceptibility to the Antiviral Remdesivir (GS-5734) Is Mediated by the Viral Polymerase and the Proofreading Exoribonuclease** Maria L Agostini, Erica L

Andres, Amy C Sims, Rachel L Graham, Timothy P Sheahan, Xiaotao Lu, Everett Clinton Smith, James Brett Case, Joy Y Feng, Robert Jordan, … Mark R Denison *mBio* (2018-03-06) https://doi.org/gc45v6 DOI: 10.1128/mbio.00221-18 · PMID: 29511076 · PMCID: PMC5844999

88. **Remdesivir and chloroquine effectively inhibit the recently emerged novel coronavirus (2019-nCoV) in vitro** Manli Wang, Ruiyuan Cao, Leike Zhang, Xinglou Yang, Jia Liu, Mingyue Xu, Zhengli Shi, Zhihong Hu, Wu Zhong, Gengfu Xiao *Cell Research* (2020-02-04) https://doi.org/ggkbsg DOI: 10.1038/s41422-020-0282-0 · PMID: 32020029 · PMCID: PMC7054408

89. **Remdesivir for the Treatment of Covid-19 — Final Report** John H Beigel, Kay M Tomashek, Lori E Dodd, Aneesh K Mehta, Barry S Zingman, Andre C Kalil, Elizabeth Hohmann, Helen Y Chu, Annie Luetkemeyer, Susan Kline, … HClifford Lane *New England Journal of Medicine* (2020-11-05) https://doi.org/dwkd DOI: 10.1056/nejmoa2007764 · PMID: 32445440 · PMCID: PMC7262788

90. **A Multicenter, Adaptive, Randomized Blinded Controlled Trial of the Safety and Efficacy of Investigational Therapeutics for the Treatment of COVID-19 in Hospitalized Adults** National Institute of Allergy and Infectious Diseases (NIAID) *clinicaltrials.gov* (2020-12-05) https://clinicaltrials.gov/ct2/show/NCT04280705

91. **A Phase 3 Randomized Study to Evaluate the Safety and Antiviral Activity of Remdesivir (GS-5734™) in Participants With Severe COVID-19** Gilead Sciences *clinicaltrials.gov* (2020-12-15) https://clinicaltrials.gov/ct2/show/NCT04292899

92. **Compassionate Use of Remdesivir for Patients with Severe Covid-19** Jonathan Grein, Norio Ohmagari, Daniel Shin, George Diaz, Erika Asperges, Antonella Castagna, Torsten Feldt, Gary Green, Margaret L Green, François-Xavier Lescure, … Timothy Flanigan *New England Journal of Medicine* (2020-06-11) https://doi.org/ggrm99 DOI: 10.1056/nejmoa2007016 · PMID: 32275812 · PMCID: PMC7169476

93. **Repurposed Antiviral Drugs for Covid-19 — Interim WHO Solidarity Trial Results** WHO Solidarity Trial Consortium *New England Journal of Medicine* (2020-12-02) https://doi.org/ghnhnw DOI: 10.1056/nejmoa2023184 · PMID: 33264556 · PMCID: PMC7727327

94. **PLAQUENIL - hydroxychloroquine sulfate tablet** DailyMed (2020-08-12) https://dailymed.nlm.nih.gov/dailymed/drugInfo.cfm?setid=34496b43-05a2-45fb-a769-52b12e099341

95. **New concepts in antimalarial use and mode of action in dermatology** Sunil Kalia, Jan P Dutz *Dermatologic Therapy* (2007-07) https://doi.org/fv69cb DOI: 10.1111/j.1529-8019.2007.00131.x · PMID: 17970883 · PMCID: PMC7163426

96. **Chloroquine is a potent inhibitor of SARS coronavirus infection and spread** Martin J Vincent, Eric Bergeron, Suzanne Benjannet, Bobbie R Erickson, Pierre E Rollin, Thomas G

Ksiazek, Nabil G Seidah, Stuart T Nichol *Virology Journal* (2005) https://doi.org/dvbds4 DOI: 10.1186/1743-422x-2-69 · PMID: 16115318 · PMCID: PMC1232869

97. **In Vitro Antiviral Activity and Projection of Optimized Dosing Design of Hydroxychloroquine for the Treatment of Severe Acute Respiratory Syndrome Coronavirus 2 (SARS-CoV-2)** Xueting Yao, Fei Ye, Miao Zhang, Cheng Cui, Baoying Huang, Peihua Niu, Xu Liu, Li Zhao, Erdan Dong, Chunli Song, … Dongyang Liu *Clinical Infectious Diseases* (2020-08-01) https://doi.org/ggpx7z DOI: 10.1093/cid/ciaa237 · PMID: 32150618 · PMCID: PMC7108130

98. **Hydroxychloroquine and azithromycin as a treatment of COVID-19: results of an open-label non-randomized clinical trial** Philippe Gautret, Jean-Christophe Lagier, Philippe Parola, Van Thuan Hoang, Line Meddeb, Morgane Mailhe, Barbara Doudier, Johan Courjon, Valérie Giordanengo, Vera Esteves Vieira, … Didier Raoult *International Journal of Antimicrobial Agents* (2020-07) https://doi.org/dp7d DOI: 10.1016/j.ijantimicag.2020.105949 · PMID: 32205204 · PMCID: PMC7102549

99. **Official Statement from International Society of Antimicrobial Chemotherapy** Andreas Voss (2020-04-03) https://www.isac.world/news-and-publications/official-isac-statement

100. **Effect of Hydroxychloroquine in Hospitalized Patients with Covid-19** The RECOVERY Collaborative Group *New England Journal of Medicine* (2020-11-19) https://doi.org/ghd8c7 DOI: 10.1056/nejmoa2022926 · PMID: 33031652 · PMCID: PMC7556338

101. **No clinical benefit from use of hydroxychloroquine in hospitalised patients with COVID-19 — RECOVERY Trial** https://www.recoverytrial.net/news/statement-from-the-chief-investigators-of-the-randomised-evaluation-of-covid-19-therapy-recovery-trial-on-hydroxychloroquine-5-june-2020-no-clinical-benefit-from-use-of-hydroxychloroquine-in-hospitalised-patients-with-covid-19

102. **The life and times of ivermectin — a success story** Satoshi Ōmura, Andy Crump *Nature Reviews Microbiology* (2004-12) https://doi.org/fftvr8 DOI: 10.1038/nrmicro1048 · PMID: 15550944

103. **Avermectins, New Family of Potent Anthelmintic Agents: Producing Organism and Fermentation** Richard W Burg, Brinton M Miller, Edward E Baker, Jerome Birnbaum, Sara A Currie, Robert Hartman, Yu-Lin Kong, Richard L Monaghan, George Olson, Irving Putter, … Satoshi Ōmura *Antimicrobial Agents and Chemotherapy* (1979-03) https://doi.org/gmd8cj DOI: 10.1128/aac.15.3.361 · PMID: 464561 · PMCID: PMC352666

104. **Ivermectin, 'Wonder drug' from Japan: the human use perspective** Andy CRUMP, Satoshi OMURA *Proceedings of the Japan Academy, Series B* (2011) https://doi.org/cpq4wk DOI: 10.2183/pjab.87.13 · PMID: 21321478 · PMCID: PMC3043740

105. **Ivermectin: enigmatic multifaceted 'wonder' drug continues to surprise and exceed expectations** Andy Crump *The Journal of Antibiotics* (2017-02-15) https://doi.org/gmd8cf DOI: 10.1038/ja.2017.11 · PMID: 28196978

106. **Avermectins, New Family of Potent Anthelmintic Agents: Efficacy of the B1a Component** JR Egerton, DA Ostlind, LS Blair, CH Eary, D Suhayda, S Cifelli, RF Riek, WC Campbell *Antimicrobial Agents and Chemotherapy* (1979-03) https://doi.org/825 DOI: 10.1128/aac.15.3.372 · PMID: 464563 · PMCID: PMC352668

107. **Ivermectin: a systematic review from antiviral effects to COVID-19 complementary regimen** Fatemeh Heidary, Reza Gharebaghi *The Journal of Antibiotics* (2020-06-12) https://doi.org/ghcz8p DOI: 10.1038/s41429-020-0336-z · PMID: 32533071 · PMCID: PMC7290143

108. **Ivermectin, a new candidate therapeutic against SARS-CoV-2/COVID-19** Khan Sharun, Kuldeep Dhama, Shailesh Kumar Patel, Mamta Pathak, Ruchi Tiwari, Bhoj Raj Singh, Ranjit Sah, DKatterine Bonilla-Aldana, Alfonso J Rodriguez-Morales, Hakan Leblebicioglu *Annals of Clinical Microbiology and Antimicrobials* (2020-05-30) https://doi.org/gmhmfg DOI: 10.1186/s12941-020-00368-w · PMID: 32473642 · PMCID: PMC7261036

109. **The FDA-approved drug ivermectin inhibits the replication of SARS-CoV-2 in vitro** Leon Caly, Julian D Druce, Mike G Catton, David A Jans, Kylie M Wagstaff *Antiviral Research* (2020-06) https://doi.org/ggqvsj DOI: 10.1016/j.antiviral.2020.104787 · PMID: 32251768 · PMCID: PMC7129059

110. **Ivermectin and COVID-19: Keeping Rigor in Times of Urgency** Carlos Chaccour, Felix Hammann, Santiago Ramón-García, NRegina Rabinovich *The American Journal of Tropical Medicine and Hygiene* (2020-06-03) https://doi.org/gj6kbh DOI: 10.4269/ajtmh.20-0271 · PMID: 32314704 · PMCID: PMC7253113

111. **The Approved Dose of Ivermectin Alone is not the Ideal Dose for the Treatment of COVID-19** Virginia D Schmith, Jie (Jessie) Zhou, Lauren RL Lohmer *Clinical Pharmacology & Therapeutics* (2020-06-07) https://doi.org/ggvcz2 DOI: 10.1002/cpt.1889 · PMID: 32378737 · PMCID: PMC7267287

112. **Ivermectin as a potential COVID-19 treatment from the pharmacokinetic point of view: antiviral levels are not likely attainable with known dosing regimens** Georgi Momekov, Denitsa Momekova *Biotechnology & Biotechnological Equipment* (2020-06-05) https://doi.org/gj6kbm DOI: 10.1080/13102818.2020.1775118

113. **Relative Neurotoxicity of Ivermectin and Moxidectin in Mdr1ab (−/−) Mice and Effects on Mammalian GABA(A) Channel Activity** Cécile Ménez, Jean-François Sutra, Roger Prichard, Anne Lespine *PLoS Neglected Tropical Diseases* (2012-11-01) https://doi.org/gmhmfh DOI: 10.1371/journal.pntd.0001883 · PMID: 23133688 · PMCID: PMC3486876

114. **Use of Ivermectin Is Associated With Lower Mortality in Hospitalized Patients With Coronavirus Disease 2019** Juliana Cepelowicz Rajter, Michael S Sherman, Naaz Fatteh,

Fabio Vogel, Jamie Sacks, Jean-Jacques Rajter *Chest* (2021-01) https://doi.org/gjr28f DOI: 10.1016/j.chest.2020.10.009 · PMID: 33065103 · PMCID: PMC7550891

115. **Lack of efficacy of standard doses of ivermectin in severe COVID-19 patients** Daniel Camprubí, Alex Almuedo-Riera, Helena Martí-Soler, Alex Soriano, Juan Carlos Hurtado, Carme Subirà, Berta Grau-Pujol, Alejandro Krolewiecki, Jose Muñoz *PLOS ONE* (2020-11-11) https://doi.org/gmhmfj DOI: 10.1371/journal.pone.0242184 · PMID: 33175880 · PMCID: PMC7657540

116. **Ivermectin as an adjunct treatment for hospitalized adult COVID-19 patients: A randomized multi-center clinical trial** Morteza Shakhsi Niaee, Peyman Namdar, Abbas Allami, Leila Zolghadr, Amir Javadi, Amin Karampour, Mehran Varnaseri, Behzad Bijani, Fatemeh Cheraghi, Yazdan Naderi, … Nematollah Gheibi *Asian Pacific Journal of Tropical Medicine* (2021) https://doi.org/gmhmfp DOI: 10.4103/1995-7645.318304

117. **Ivermectin shows clinical benefits in mild to moderate COVID19: a randomized controlled double-blind, dose-response study in Lagos** OE Babalola, CO Bode, AA Ajayi, FM Alakaloko, IE Akase, E Otrofanowei, OB Salu, WL Adeyemo, AO Ademuyiwa, S Omilabu *QJM: An International Journal of Medicine* (2021-02-18) https://doi.org/gmhbg5 DOI: 10.1093/qjmed/hcab035 · PMID: 33599247 · PMCID: PMC7928689

118. **Use of ivermectin in the treatment of Covid-19: A pilot trial** Henrique Pott-Junior, Mônica Maria Bastos Paoliello, Alice de Queiroz Constantino Miguel, Anderson Ferreira da Cunha, Caio Cesar de Melo Freire, Fábio Fernandes Neves, Lucimar Retto da Silva de Avó, Meliza Goi Roscani, Sigrid De Sousa dos Santos, Silvana Gama Florêncio Chachá *Toxicology Reports* (2021) https://doi.org/gmhmdc DOI: 10.1016/j.toxrep.2021.03.003 · PMID: 33723507 · PMCID: PMC7942165

119. **A five-day course of ivermectin for the treatment of COVID-19 may reduce the duration of illness** Sabeena Ahmed, Mohammad Mahbubul Karim, Allen G Ross, Mohammad Sharif Hossain, John D Clemens, Mariya Kibtiya Sumiya, Ching Swe Phru, Mustafizur Rahman, Khalequ Zaman, Jyoti Somani, … Wasif Ali Khan *International Journal of Infectious Diseases* (2021-02) https://doi.org/gjwcdt DOI: 10.1016/j.ijid.2020.11.191 · PMID: 33278625 · PMCID: PMC7709596

120. **Effects of Ivermectin in Patients With COVID-19: A Multicenter, Double-blind, Randomized, Controlled Clinical Trial** Leila Shahbaznejad, Alireza Davoudi, Gohar Eslami, John S Markowitz, Mohammad Reza Navaeifar, Fatemeh Hosseinzadeh, Faeze Sadat Movahedi, Mohammad Sadegh Rezai *Clinical Therapeutics* (2021-06) https://doi.org/gmb256 DOI: 10.1016/j.clinthera.2021.04.007 · PMID: 34052007 · PMCID: PMC8101859

121. **The effect of early treatment with ivermectin on viral load, symptoms and humoral response in patients with non-severe COVID-19: A pilot, double-blind, placebo-controlled, randomized clinical trial** Carlos Chaccour, Aina Casellas, Andrés Blanco-Di Matteo, Iñigo Pineda, Alejandro Fernandez-Montero, Paula Ruiz-Castillo, Mary-Ann Richardson, Mariano Rodríguez-Mateos, Carlota Jordán-Iborra, Joe Brew, … Mirian Fernández-Alonso

*EClinicalMedicine* (2021-02) https://doi.org/gmf4d3 DOI: 10.1016/j.eclinm.2020.100720 · PMID: 33495752 · PMCID: PMC7816625

122. **Ivermectin in mild and moderate COVID-19 (RIVET-COV): a randomized, placebo-controlled trial** Anant Mohan, Pawan Tiwari, Tejas Suri, Saurabh Mittal, Ankit Patel, Avinash Jain, Velpandian T., Ujjwal Kumar Das, Tarun K Bopanna, RM Pandey, … Randeep Guleria *Research Square Platform LLC* (2021-01-30) https://doi.org/gmh3hq DOI: 10.21203/rs.3.rs-191648/v1

123. **Evaluation of Ivermectin as a Potential Treatment for Mild to Moderate COVID-19: A Double-Blind Randomized Placebo Controlled Trial in Eastern India** Ravikirti, Ranjini Roy, Chandrima Pattadar, Rishav Raj, Neeraj Agarwal, Bijit Biswas, Pramod Kumar Manjhi, Deependra Kumar Rai, Shyama, Anjani Kumar, Asim Sarfaraz *Journal of Pharmacy & Pharmaceutical Sciences* (2021-07-15) https://doi.org/gmhmfk DOI: 10.18433/jpps32105 · PMID: 34265236

124. **Efficacy and Safety of Ivermectin for Treatment and prophylaxis of COVID-19 Pandemic** Ahmed Elgazzar, Abdelaziz Eltaweel, Shaimaa Abo Youssef, Basma Hany, Mohy Hafez, Hany Moussa *Research Square Platform LLC* (2020-10-30) https://doi.org/gmhmfr DOI: 10.21203/rs.3.rs-100956/v3

125. **Why Was a Major Study on Ivermectin for COVID-19 Just Retracted?** Grftr News (2021-07-15) https://grftr.news/why-was-a-major-study-on-ivermectin-for-covid-19-just-retracted/

126. **Nick Brown's blog: Some problems in the dataset of a large study of Ivermectin for the treatment of Covid-19** Nick Brown *Nick Brown's blog* (2021-07-15) https://steamtraen.blogspot.com/2021/07/Some-problems-with-the-data-from-a-Covid-study.html

127. **Efficacy and Safety of Ivermectin for Treatment and prophylaxis of COVID-19 Pandemic** Research Square Platform LLC *Research Square Platform LLC* (2020-10-30) https://doi.org/gmhmfm DOI: 10.21203/rs.3.rs-100956/v4

128. **Effect of Ivermectin on Time to Resolution of Symptoms Among Adults With Mild COVID-19** Eduardo López-Medina, Pío López, Isabel C Hurtado, Diana M Dávalos, Oscar Ramirez, Ernesto Martínez, Jesus A Díazgranados, José M Oñate, Hector Chavarriaga, Sócrates Herrera, … Isabella Caicedo *JAMA* (2021-04-13) https://doi.org/gjft3s DOI: 10.1001/jama.2021.3071 · PMID: 33662102 · PMCID: PMC7934083

129. **Ivermectin for the treatment of COVID-19: A systematic review and meta-analysis of randomized controlled trials** Yuani M Roman, Paula Alejandra Burela, Vinay Pasupuleti, Alejandro Piscoya, Jose E Vidal, Adrian V Hernandez *Cold Spring Harbor Laboratory* (2021-05-26) https://doi.org/gmh3hv DOI: 10.1101/2021.05.21.21257595

130. **Outcomes of Ivermectin in the treatment of COVID-19: a systematic review and meta-analysis** Alex Castañeda-Sabogal, Diego Chambergo-Michilot, Carlos J Toro-Huamanchumo, Christian Silva-Rengifo, José Gonzales-Zamora, Joshuan J Barboza *Cold*

*Spring Harbor Laboratory* (2021-01-27) https://doi.org/gmhmdd DOI: 10.1101/2021.01.26.21250420

131. **Expression of Concern: "Meta-analysis of Randomized Trials of Ivermectin to Treat SARS-CoV-2 Infection"** Andrew Hill, Anna Garratt, Jacob Levi, Jonathan Falconer, Leah Ellis, Kaitlyn McCann, Victoria Pilkington, Ambar Qavi, Junzheng Wang, Hannah Wentzel *Open Forum Infectious Diseases* (2021-08-01) https://doi.org/gmhrz6 DOI: 10.1093/ofid/ofab394 · PMID: 34410284 · PMCID: PMC8369353

132. **Ivermectin and mortality in patients with COVID-19: A systematic review, meta-analysis, and meta-regression of randomized controlled trials** Ahmad Fariz Malvi Zamzam Zein, Catur Setiya Sulistiyana, Wilson Matthew Raffaelo, Raymond Pranata *Diabetes & Metabolic Syndrome: Clinical Research & Reviews* (2021-07) https://doi.org/gmhmdb DOI: 10.1016/j.dsx.2021.102186 · PMID: 34237554 · PMCID: PMC8236126

133. **Ivermectin for Prevention and Treatment of COVID-19 Infection: A Systematic Review, Meta-analysis, and Trial Sequential Analysis to Inform Clinical Guidelines** Andrew Bryant, Theresa A Lawrie, Therese Dowswell, Edmund J Fordham, Scott Mitchell, Sarah R Hill, Tony C Tham *American Journal of Therapeutics* (2021-07) https://doi.org/gksqvz DOI: 10.1097/mjt.0000000000001402 · PMID: 34145166 · PMCID: PMC8248252

134. **Ivermectin and outcomes from Covid-19 pneumonia: A systematic review and meta-analysis of randomized clinical trial studies** Timotius Ivan Hariyanto, Devina Adella Halim, Jane Rosalind, Catherine Gunawan, Andree Kurniawan *Reviews in Medical Virology* (2021-06-06) https://doi.org/gmhmc9 DOI: 10.1002/rmv.2265 · PMCID: PMC8209939

135. **Ivermectin for prevention and treatment of COVID-19 infection: a systematic review and meta-analysis** Andrew Bryant, Theresa A Lawrie, Therese Dowswell, Edmund Fordham, Mitchell Scott, Sarah R Hill, Tony C Tham *Center for Open Science* (2021-03-11) https://doi.org/gmhmfn DOI: 10.31219/osf.io/k37ft

136. **Review of the Emerging Evidence Demonstrating the Efficacy of Ivermectin in the Prophylaxis and Treatment of COVID-19** Pierre Kory, Gianfranco Umberto Meduri, Joseph Varon, Jose Iglesias, Paul E Marik *American Journal of Therapeutics* (2021-05) https://doi.org/gjxvpr DOI: 10.1097/mjt.0000000000001377 · PMID: 34375047 · PMCID: PMC8088823

137. **The association between the use of ivermectin and mortality in patients with COVID-19: a meta-analysis** Chia Siang Kow, Hamid A Merchant, Zia Ul Mustafa, Syed Shahzad Hasan *Pharmacological Reports* (2021-03-29) https://doi.org/gmhrpr DOI: 10.1007/s43440-021-00245-z · PMID: 33779964 · PMCID: PMC8005369

138. **A Meta-analysis of Mortality, Need for ICU admission, Use of Mechanical Ventilation and Adverse Effects with Ivermectin Use in COVID-19 Patients** Smruti Karale, Vikas Bansal, Janaki Makadia, Muhammad Tayyeb, Hira Khan, Shree Spandana Ghanta, Romil Singh, Aysun Tekin, Abhishek Bhurwal, Hemant Mutneja, … Rahul Kashyap *Cold Spring Harbor Laboratory* (2021-05-04) https://doi.org/gmhmdn DOI: 10.1101/2021.04.30.21256415

139. **Does Ivermectin Work for Covid-19?** Gideon M-K; Health Nerd *Medium* (2021-06-23) https://gidmk.medium.com/does-ivermectin-work-for-covid-19-1166126c364a

140. **Meta-analysis of randomized trials of ivermectin to treat SARS-CoV-2 infection** Andrew Hill, Anna Garratt, Jacob Levi, Jonathan Falconer, Leah Ellis, Kaitlyn McCann, Victoria Pilkington, Ambar Qavi, Junzheng Wang, Hannah Wentzel *Open Forum Infectious Diseases* (2021-07-06) https://doi.org/gmh4jn DOI: 10.1093/ofid/ofab358

141. **FAQ: COVID-19 and Ivermectin Intended for Animals** Center for Veterinary Medicine *FDA* (2021-04-26) https://www.fda.gov/animal-veterinary/product-safety-information/faq-covid-19-and-ivermectin-intended-animals

142. **A Multicenter, Prospective, Adaptive, Double-blind, Randomized, Placebo-controlled Study to Evaluate the Effect of Fluvoxamine, Ivermectin, Doxasozin and Interferon Lambda 1A in Mild COVID-19 and High Risk of Complications** Cardresearch *clinicaltrials.gov* (2021-07-05) https://clinicaltrials.gov/ct2/show/NCT04727424

143. **Ivermectin to be investigated in adults aged 18+ as a possible treatment for COVID-19 in the PRINCIPLE trial — PRINCIPLE Trial** https://www.principletrial.org/news/ivermectin-to-be-investigated-as-a-possible-treatment-for-covid-19-in-oxford2019s-principle-trial

144. **Effect of Early Treatment With Hydroxychloroquine or Lopinavir and Ritonavir on Risk of Hospitalization Among Patients With COVID-19** Gilmar Reis, Eduardo Augusto dos Santos Moreira Silva, Daniela Carla Medeiros Silva, Lehana Thabane, Gurmit Singh, Jay JH Park, Jamie I Forrest, Ofir Harari, Castilho Vitor Quirino dos Santos, Ana Paula Figueiredo Guimarães de Almeida, … TOGETHER Investigators *JAMA Network Open* (2021-04-22) https://doi.org/gk4298 DOI: 10.1001/jamanetworkopen.2021.6468 · PMID: 33885775 · PMCID: PMC8063069

145. **August 6, 2021: Early Treatment of COVID-19 with Repurposed Therapies: The TOGETHER Adaptive Platform Trial (Edward Mills, PhD, FRCP)** Rethinking Clinical Trials (2021-08-11) https://rethinkingclinicaltrials.org/news/august-6-2021-early-treatment-of-covid-19-with-repurposed-therapies-the-together-adaptive-platform-trial-edward-mills-phd-frcp/

146. **Lisinopril - Drug Usage Statistics** ClinCalc DrugStats Database https://clincalc.com/DrugStats/Drugs/Lisinopril

147. **Hypertension Hot Potato — Anatomy of the Angiotensin-Receptor Blocker Recalls** JBrian Byrd, Glenn M Chertow, Vivek Bhalla *New England Journal of Medicine* (2019-04-25) https://doi.org/ggvc7g DOI: 10.1056/nejmp1901657 · PMID: 30865819 · PMCID: PMC7066505

148. **ACE Inhibitor and ARB Utilization and Expenditures in the Medicaid Fee-For-Service Program from 1991 to 2008** Boyang Bian, Christina ML Kelton, Jeff J Guo, Patricia R Wigle *Journal of Managed Care Pharmacy* (2010-11) https://doi.org/gh294c DOI: 10.18553/jmcp.2010.16.9.671 · PMID: 21067253

149. **ACE2: from vasopeptidase to SARS virus receptor** Anthony J Turner, Julian A Hiscox, Nigel M Hooper *Trends in Pharmacological Sciences* (2004-06) https://doi.org/dn77dn DOI: 10.1016/j.tips.2004.04.001 · PMID: 15165741 · PMCID: PMC7119032

150. **Structure-Based Discovery of a Novel Angiotensin-Converting Enzyme 2 Inhibitor** Matthew J Huentelman, Jasenka Zubcevic, Jose A Hernández Prada, Xiaodong Xiao, Dimiter S Dimitrov, Mohan K Raizada, David A Ostrov *Hypertension* (2004-12) https://doi.org/d5szrp DOI: 10.1161/01.hyp.0000146120.29648.36 · PMID: 15492138

151. **The Secret Life of ACE2 as a Receptor for the SARS Virus** Dimiter S Dimitrov *Cell* (2003-12) https://doi.org/d85vmw DOI: 10.1016/s0092-8674(03)00976-0 · PMID: 14675530 · PMCID: PMC7133233

152. **Hypertension, the renin–angiotensin system, and the risk of lower respiratory tract infections and lung injury: implications for COVID-19** Reinhold Kreutz, Engi Abd El-Hady Algharably, Michel Azizi, Piotr Dobrowolski, Tomasz Guzik, Andrzej Januszewicz, Alexandre Persu, Aleksander Prejbisz, Thomas Günther Riemer, Ji-Guang Wang, Michel Burnier *Cardiovascular Research* (2020-08-01) https://doi.org/ggtwpj DOI: 10.1093/cvr/cvaa097 · PMID: 32293003 · PMCID: PMC7184480

153. **Angiotensin converting enzyme 2 activity and human atrial fibrillation: increased plasma angiotensin converting enzyme 2 activity is associated with atrial fibrillation and more advanced left atrial structural remodelling** Tomos E Walters, Jonathan M Kalman, Sheila K Patel, Megan Mearns, Elena Velkoska, Louise M Burrell *Europace* (2016-10-12) https://doi.org/gbt2jw DOI: 10.1093/europace/euw246 · PMID: 27738071

154. **Cardiovascular Disease, Drug Therapy, and Mortality in Covid-19** Mandeep R Mehra, Sapan S Desai, SreyRam Kuy, Timothy D Henry, Amit N Patel *New England Journal of Medicine* (2020-06-18) https://doi.org/ggtp6v DOI: 10.1056/nejmoa2007621 · PMID: 32356626 · PMCID: PMC7206931

155. **Retraction: Cardiovascular Disease, Drug Therapy, and Mortality in Covid-19. N Engl J Med. DOI: 10.1056/NEJMoa2007621.** Mandeep R Mehra, Sapan S Desai, SreyRam Kuy, Timothy D Henry, Amit N Patel *New England Journal of Medicine* (2020-06-25) https://doi.org/ggzkpj DOI: 10.1056/nejmc2021225 · PMID: 32501665 · PMCID: PMC7274164

156. **Continuation versus discontinuation of renin–angiotensin system inhibitors in patients admitted to hospital with COVID-19: a prospective, randomised, open-label trial** Jordana B Cohen, Thomas C Hanff, Preethi William, Nancy Sweitzer, Nelson R Rosado-Santander, Carola Medina, Juan E Rodriguez-Mori, Nicolás Renna, Tara I Chang, Vicente Corrales-Medina, … Julio A Chirinos *The Lancet Respiratory Medicine* (2021-03) https://doi.org/fvgt DOI: 10.1016/s2213-2600(20)30558-0 · PMID: 33422263 · PMCID: PMC7832152

157. **Effect of Discontinuing vs Continuing Angiotensin-Converting Enzyme Inhibitors and Angiotensin II Receptor Blockers on Days Alive and Out of the Hospital in Patients Admitted With COVID-19** Renato D Lopes, Ariane VS Macedo, Pedro GM de Barros E Silva,

Renata J Moll-Bernardes, Tiago M dos Santos, Lilian Mazza, André Feldman, Guilherme D'Andréa Saba Arruda, Denílson C de Albuquerque, Angelina S Camiletti, … BRACE CORONA Investigators *JAMA* (2021-01-19) https://doi.org/gh2tw5 DOI: 10.1001/jama.2020.25864 · PMID: 33464336 · PMCID: PMC7816106

158. **Frequently Asked Questions on the Revocation of the Emergency Use Authorization for Hydroxychloroquine Sulfate and Chloroquine Phosphate** U.S. Food and Drug Administration (2020-06-19) https://www.fda.gov/media/138946/download

159. **COVID-19: chloroquine and hydroxychloroquine only to be used in clinical trials or emergency use programmes** Georgina HRABOVSZKI *European Medicines Agency* (2020-04-01) https://www.ema.europa.eu/en/news/covid-19-chloroquine-hydroxychloroquine-only-be-used-clinical-trials-emergency-use-programmes

160. **Formulation and manufacturability of biologics** Steven J Shire *Current Opinion in Biotechnology* (2009-12) https://doi.org/cjk8p6 DOI: 10.1016/j.copbio.2009.10.006 · PMID: 19880308

161. **Early Development of Therapeutic Biologics - Pharmacokinetics** A Baumann *Current Drug Metabolism* (2006-01-01) https://doi.org/bhcz79 DOI: 10.2174/138920006774832604 · PMID: 16454690

162. **Development of therapeutic antibodies for the treatment of diseases** Ruei-Min Lu, Yu-Chyi Hwang, I-Ju Liu, Chi-Chiu Lee, Han-Zen Tsai, Hsin-Jung Li, Han-Chung Wu *Journal of Biomedical Science* (2020-01-02) https://doi.org/ggqbpx DOI: 10.1186/s12929-019-0592-z · PMID: 31894001 · PMCID: PMC6939334

163. **Broadly Neutralizing Antiviral Antibodies** Davide Corti, Antonio Lanzavecchia *Annual Review of Immunology* (2013-03-21) https://doi.org/gf25g8 DOI: 10.1146/annurev-immunol-032712-095916 · PMID: 23330954

164. **Ibalizumab Targeting CD4 Receptors, An Emerging Molecule in HIV Therapy** Simona A Iacob, Diana G Iacob *Frontiers in Microbiology* (2017-11-27) https://doi.org/gcn3kh DOI: 10.3389/fmicb.2017.02323 · PMID: 29230203 · PMCID: PMC5711820

165. **Product review on the monoclonal antibody palivizumab for prevention of respiratory syncytial virus infection** Bernhard Resch *Human Vaccines & Immunotherapeutics* (2017-06-12) https://doi.org/ggqbps DOI: 10.1080/21645515.2017.1337614 · PMID: 28605249 · PMCID: PMC5612471

166. **Prophylaxis With a Middle East Respiratory Syndrome Coronavirus (MERS-CoV)–Specific Human Monoclonal Antibody Protects Rabbits From MERS-CoV Infection** Katherine V Houser, Lisa Gretebeck, Tianlei Ying, Yanping Wang, Leatrice Vogel, Elaine W Lamirande, Kevin W Bock, Ian N Moore, Dimiter S Dimitrov, Kanta Subbarao *Journal of Infectious Diseases* (2016-05-15) https://doi.org/f8pm7j DOI: 10.1093/infdis/jiw080 · PMID: 26941283 · PMCID: PMC4837915

167.     **Efficacy of antibody-based therapies against Middle East respiratory syndrome coronavirus (MERS-CoV) in common marmosets** Neeltje van Doremalen, Darryl Falzarano, Tianlei Ying, Emmie de Wit, Trenton Bushmaker, Friederike Feldmann, Atsushi Okumura, Yanping Wang, Dana P Scott, Patrick W Hanley, … Vincent J Munster *Antiviral Research* (2017-07) https://doi.org/gbh5c2 DOI: 10.1016/j.antiviral.2017.03.025 · PMID: 28389142 · PMCID: PMC6957253

168.     **IL-6 in Inflammation, Immunity, and Disease** T Tanaka, M Narazaki, T Kishimoto *Cold Spring Harbor Perspectives in Biology* (2014-09-04) https://doi.org/gftpjs DOI: 10.1101/cshperspect.a016295 · PMID: 25190079 · PMCID: PMC4176007

169.     **ACTEMRA - tocilizumab** DailyMed (2020-12-17) https://dailymed.nlm.nih.gov/dailymed/drugInfo.cfm?setid=2e5365ff-cb2a-4b16-b2c7-e35c6bf2de13

170.     **Clinical course and risk factors for mortality of adult inpatients with COVID-19 in Wuhan, China: a retrospective cohort study** Fei Zhou, Ting Yu, Ronghui Du, Guohui Fan, Ying Liu, Zhibo Liu, Jie Xiang, Yeming Wang, Bin Song, Xiaoying Gu, … Bin Cao *The Lancet* (2020-03) https://doi.org/ggnxb3 DOI: 10.1016/s0140-6736(20)30566-3

171.     **Tocilizumab in patients with severe COVID-19: a retrospective cohort study** Giovanni Guaraldi, Marianna Meschiari, Alessandro Cozzi-Lepri, Jovana Milic, Roberto Tonelli, Marianna Menozzi, Erica Franceschini, Gianluca Cuomo, Gabriella Orlando, Vanni Borghi, … Cristina Mussini *The Lancet Rheumatology* (2020-08) https://doi.org/d2pk DOI: 10.1016/s2665-9913(20)30173-9 · PMID: 32835257 · PMCID: PMC7314456

172.     **Tocilizumab Treatment for Cytokine Release Syndrome in Hospitalized Patients With Coronavirus Disease 2019** Christina C Price, Frederick L Altice, Yu Shyr, Alan Koff, Lauren Pischel, George Goshua, Marwan M Azar, Dayna Mcmanus, Sheau-Chiann Chen, Shana E Gleeson, … Maricar Malinis *Chest* (2020-10) https://doi.org/gg2789 DOI: 10.1016/j.chest.2020.06.006 · PMID: 32553536 · PMCID: PMC7831876

173.     **Impact of low dose tocilizumab on mortality rate in patients with COVID-19 related pneumonia** Ruggero Capra, Nicola De Rossi, Flavia Mattioli, Giuseppe Romanelli, Cristina Scarpazza, Maria Pia Sormani, Stefania Cossi *European Journal of Internal Medicine* (2020-06) https://doi.org/ggx4fm DOI: 10.1016/j.ejim.2020.05.009 · PMID: 32405160 · PMCID: PMC7219361

174.     **Tocilizumab therapy reduced intensive care unit admissions and/or mortality in COVID-19 patients** T Klopfenstein, S Zayet, A Lohse, J-C Balblanc, J Badie, P-Y Royer, L Toko, C Mezher, NJ Kadiane-Oussou, M Bossert, … T Conrozier *Médecine et Maladies Infectieuses* (2020-08) https://doi.org/ggvz45 DOI: 10.1016/j.medmal.2020.05.001 · PMID: 32387320 · PMCID: PMC7202806

175.     **Outcomes in patients with severe COVID-19 disease treated with tocilizumab: a case–controlled study** G Rojas-Marte, M Khalid, O Mukhtar, AT Hashmi, MA Waheed, S Ehrlich, A Aslam, S Siddiqui, C Agarwal, Y Malyshev, … J Shani *QJM: An International Journal*

*of Medicine* (2020-08) https://doi.org/gg496t DOI: 10.1093/qjmed/hcaa206 · PMID: 32569363 · PMCID: PMC7337835

176. **Effective treatment of severe COVID-19 patients with tocilizumab** Xiaoling Xu, Mingfeng Han, Tiantian Li, Wei Sun, Dongsheng Wang, Binqing Fu, Yonggang Zhou, Xiaohu Zheng, Yun Yang, Xiuyong Li, … Haiming Wei *Proceedings of the National Academy of Sciences* (2020-05-19) https://doi.org/ggv3r3 DOI: 10.1073/pnas.2005615117 · PMID: 32350134 · PMCID: PMC7245089

177. **Tocilizumab in patients admitted to hospital with COVID-19 (RECOVERY): preliminary results of a randomised, controlled, open-label, platform trial** Peter W Horby, Guilherme Pessoa-Amorim, Leon Peto, Christopher E Brightling, Rahuldeb Sarkar, Koshy Thomas, Vandana Jeebun, Abdul Ashish, Redmond Tully, David Chadwick, … RECOVERY Collaborative Group *Cold Spring Harbor Laboratory* (2021-02-11) https://doi.org/fvqj DOI: 10.1101/2021.02.11.21249258

178. **Tocilizumab in Hospitalized Patients with Severe Covid-19 Pneumonia** Ivan O Rosas, Norbert Bräu, Michael Waters, Ronaldo C Go, Bradley D Hunter, Sanjay Bhagani, Daniel Skiest, Mariam S Aziz, Nichola Cooper, Ivor S Douglas, … Atul Malhotra *New England Journal of Medicine* (2021-04-22) https://doi.org/gh5vk5 DOI: 10.1056/nejmoa2028700 · PMID: 33631066 · PMCID: PMC7953459

179. **Tocilizumab in Patients Hospitalized with Covid-19 Pneumonia** Carlos Salama, Jian Han, Linda Yau, William G Reiss, Benjamin Kramer, Jeffrey D Neidhart, Gerard J Criner, Emma Kaplan-Lewis, Rachel Baden, Lavannya Pandit, … Shalini V Mohan *New England Journal of Medicine* (2021-01-07) https://doi.org/ghp8xc DOI: 10.1056/nejmoa2030340 · PMID: 33332779 · PMCID: PMC7781101

180. **A Phase III, Randomized, Double-Blind, Multicenter Study to Evaluate the Efficacy and Safety of Remdesivir Plus Tocilizumab Compared With Remdesivir Plus Placebo in Hospitalized Patients With Severe COVID-19 Pneumonia** Hoffmann-La Roche *clinicaltrials.gov* (2021-03-09) https://clinicaltrials.gov/ct2/show/NCT04409262

181. **A Randomized, Double-Blind, Placebo-Controlled, Multicenter Study to Evaluate the Safety and Efficacy of Tocilizumab in Patients With Severe COVID-19 Pneumonia** Hoffmann-La Roche *clinicaltrials.gov* (2021-06-28) https://clinicaltrials.gov/ct2/show/NCT04320615

182. **A Randomized, Double-Blind, Placebo-Controlled, Multicenter Study to Evaluate the Efficacy and Safety of Tocilizumab in Hospitalized Patients With COVID-19 Pneumonia** Genentech, Inc. *clinicaltrials.gov* (2021-07-13) https://clinicaltrials.gov/ct2/show/NCT04372186

183. **Genentech tocilizumab Letter of Authority** U.S. Food and Drug Administration (2021-06-24) https://www.fda.gov/media/150319/download

184. **A Randomised Double-blind Placebo-controlled Trial to Determine the Safety and Efficacy of Inhaled SNG001 (IFN-β1a for Nebulisation) for the Treatment of Patients With Confirmed SARS-CoV-2 Infection** Synairgen Research Ltd. *clinicaltrials.gov* (2021-03-19) https://clinicaltrials.gov/ct2/show/NCT04385095

185. **Synairgen announces positive results from trial of SNG001 in hospitalised COVID-19 patients** Synairgen plc press release (2020-07-20) http://synairgen.web01.hosting.bdci.co.uk/umbraco/Surface/Download/GetFile?cid=1130026e-0983-4338-b648-4ac7928b9a37

186. **Safety and efficacy of inhaled nebulised interferon beta-1a (SNG001) for treatment of SARS-CoV-2 infection: a randomised, double-blind, placebo-controlled, phase 2 trial** Phillip D Monk, Richard J Marsden, Victoria J Tear, Jody Brookes, Toby N Batten, Marcin Mankowski, Felicity J Gabbay, Donna E Davies, Stephen T Holgate, Ling-Pei Ho, … Pedro MB Rodrigues *The Lancet Respiratory Medicine* (2021-02) https://doi.org/ghjzm4 DOI: 10.1016/s2213-2600(20)30511-7 · PMID: 33189161 · PMCID: PMC7836724

187. **Social Factors Influencing COVID-19 Exposure and Outcomes** COVID-19 Review Consortium *Manubot* (2021-04-30) https://greenelab.github.io/covid19-review/v/32afa309f69f0466a91acec5d0df3151fe4d61b5/#social-factors-influencing-covid-19-exposure-and-outcomes

188. **Convalescent plasma as a potential therapy for COVID-19** Long Chen, Jing Xiong, Lei Bao, Yuan Shi *The Lancet Infectious Diseases* (2020-04) https://doi.org/ggqr7s DOI: 10.1016/s1473-3099(20)30141-9 · PMID: 32113510 · PMCID: PMC7128218

189. **Convalescent Plasma to Treat COVID-19** John D Roback, Jeannette Guarner *JAMA* (2020-04-28) https://doi.org/ggqf6k DOI: 10.1001/jama.2020.4940 · PMID: 32219429

190. **Convalescent plasma transfusion for the treatment of COVID-19: Systematic review** Karthick Rajendran, Narayanasamy Krishnasamy, Jayanthi Rangarajan, Jeyalalitha Rathinam, Murugan Natarajan, Arunkumar Ramachandran *Journal of Medical Virology* (2020-05-12) https://doi.org/ggv3gx DOI: 10.1002/jmv.25961 · PMID: 32356910 · PMCID: PMC7267113

191. **Association of Convalescent Plasma Treatment With Clinical Outcomes in Patients With COVID-19** Perrine Janiaud, Cathrine Axfors, Andreas M Schmitt, Viktoria Gloy, Fahim Ebrahimi, Matthias Hepprich, Emily R Smith, Noah A Haber, Nina Khanna, David Moher, … Lars G Hemkens *JAMA* (2021-03-23) https://doi.org/gjjk4j DOI: 10.1001/jama.2021.2747 · PMID: 33635310 · PMCID: PMC7911095

192. **Convalescent Plasma Antibody Levels and the Risk of Death from Covid-19** Michael J Joyner, Rickey E Carter, Jonathon W Senefeld, Stephen A Klassen, John R Mills, Patrick W Johnson, Elitza S Theel, Chad C Wiggins, Katelyn A Bruno, Allan M Klompas, … Arturo Casadevall *New England Journal of Medicine* (2021-01-13) https://doi.org/ghs26g DOI: 10.1056/nejmoa2031893 · PMID: 33523609 · PMCID: PMC7821984

193. **Convalescent plasma in patients admitted to hospital with COVID-19 (RECOVERY): a randomised, controlled, open-label, platform trial** Peter W Horby, Lise Estcourt, Leon Peto, Jonathan R Emberson, Natalie Staplin, Enti Spata, Guilherme Pessoa-Amorim, Mark Campbell, Alistair Roddick, Nigel E Brunskill, … The RECOVERY Collaborative Group *Cold Spring Harbor Laboratory* (2021-03-10) https://doi.org/gmcq2g DOI: 10.1101/2021.03.09.21252736

194. **Chronological evolution of IgM, IgA, IgG and neutralisation antibodies after infection with SARS-associated coronavirus** P-R Hsueh, L-M Huang, P-J Chen, C-L Kao, P-C Yang *Clinical Microbiology and Infection* (2004-12) https://doi.org/cwwg87 DOI: 10.1111/j.1469-0691.2004.01009.x · PMID: 15606632

195. **Neutralizing Antibodies in Patients with Severe Acute Respiratory Syndrome-Associated Coronavirus Infection** Nie Yuchun, Wang Guangwen, Shi Xuanling, Zhang Hong, Qiu Yan, He Zhongping, Wang Wei, Lian Gewei, Yin Xiaolei, Du Liying, … Ding Mingxiao *The Journal of Infectious Diseases* (2004-09) https://doi.org/cgqj5b DOI: 10.1086/423286 · PMID: 15319862

196. **Potent human monoclonal antibodies against SARS CoV, Nipah and Hendra viruses** Ponraj Prabakaran, Zhongyu Zhu, Xiaodong Xiao, Arya Biragyn, Antony S Dimitrov, Christopher C Broder, Dimiter S Dimitrov *Expert Opinion on Biological Therapy* (2009-04-08) https://doi.org/b88kw8 DOI: 10.1517/14712590902763755 · PMID: 19216624 · PMCID: PMC2705284

197. **SARS-CoV-2 Cell Entry Depends on ACE2 and TMPRSS2 and Is Blocked by a Clinically Proven Protease Inhibitor** Markus Hoffmann, Hannah Kleine-Weber, Simon Schroeder, Nadine Krüger, Tanja Herrler, Sandra Erichsen, Tobias S Schiergens, Georg Herrler, Nai-Huei Wu, Andreas Nitsche, … Stefan Pöhlmann *Cell* (2020-04) https://doi.org/ggnq74 DOI: 10.1016/j.cell.2020.02.052 · PMID: 32142651 · PMCID: PMC7102627

198. **Structure, Function, and Antigenicity of the SARS-CoV-2 Spike Glycoprotein** Alexandra C Walls, Young-Jun Park, MAlejandra Tortorici, Abigail Wall, Andrew T McGuire, David Veesler *Cell* (2020-04) https://doi.org/dpvh DOI: 10.1016/j.cell.2020.02.058 · PMID: 32155444 · PMCID: PMC7102599

199. **SARS-CoV-2 and SARS-CoV Spike-RBD Structure and Receptor Binding Comparison and Potential Implications on Neutralizing Antibody and Vaccine Development** Chunyun Sun, Long Chen, Ji Yang, Chunxia Luo, Yanjing Zhang, Jing Li, Jiahui Yang, Jie Zhang, Liangzhi Xie *Cold Spring Harbor Laboratory* (2020-02-20) https://doi.org/ggq63j DOI: 10.1101/2020.02.16.951723

200. **Fruitful Neutralizing Antibody Pipeline Brings Hope To Defeat SARS-Cov-2** Alex Renn, Ying Fu, Xin Hu, Matthew D Hall, Anton Simeonov *Trends in Pharmacological Sciences* (2020-11) https://doi.org/gg72sv DOI: 10.1016/j.tips.2020.07.004 · PMID: 32829936 · PMCID: PMC7572790

201.    **Cross-neutralization of SARS-CoV-2 by a human monoclonal SARS-CoV antibody** Dora Pinto, Young-Jun Park, Martina Beltramello, Alexandra C Walls, MAlejandra Tortorici, Siro Bianchi, Stefano Jaconi, Katja Culap, Fabrizia Zatta, Anna De Marco, … Davide Corti *Nature* (2020-05-18) https://doi.org/dv4x DOI: 10.1038/s41586-020-2349-y · PMID: 32422645

202.    **Cryo-EM structure of the 2019-nCoV spike in the prefusion conformation** Daniel Wrapp, Nianshuang Wang, Kizzmekia S Corbett, Jory A Goldsmith, Ching-Lin Hsieh, Olubukola Abiona, Barney S Graham, Jason S McLellan *Science* (2020-03-13) https://doi.org/ggmtk2 DOI: 10.1126/science.abb2507 · PMID: 32075877

203.    **A human monoclonal antibody blocking SARS-CoV-2 infection** Chunyan Wang, Wentao Li, Dubravka Drabek, Nisreen MA Okba, Rien van Haperen, Albert DME Osterhaus, Frank JM van Kuppeveld, Bart L Haagmans, Frank Grosveld, Berend-Jan Bosch *Cold Spring Harbor Laboratory* (2020-03-12) https://doi.org/ggnw4t DOI: 10.1101/2020.03.11.987958

204.    **An update to monoclonal antibody as therapeutic option against COVID-19** Paroma Deb, MdMaruf Ahmed Molla, KM Saif-Ur-Rahman *Biosafety and Health* (2021-04) https://doi.org/gh4m7h DOI: 10.1016/j.bsheal.2021.02.001 · PMID: 33585808 · PMCID: PMC7872849

205.    **LY-CoV555, a rapidly isolated potent neutralizing antibody, provides protection in a non-human primate model of SARS-CoV-2 infection** Bryan E Jones, Patricia L Brown-Augsburger, Kizzmekia S Corbett, Kathryn Westendorf, Julian Davies, Thomas P Cujec, Christopher M Wiethoff, Jamie L Blackbourne, Beverly A Heinz, Denisa Foster, … Ester Falconer *Cold Spring Harbor Laboratory* (2020-10-09) https://doi.org/gh4sjm DOI: 10.1101/2020.09.30.318972 · PMID: 33024963 · PMCID: PMC7536866

206.    **A Randomized, Placebo-Controlled, Double-Blind, Sponsor Unblinded, Single Ascending Dose, Phase 1 First in Human Study to Evaluate the Safety, Tolerability, Pharmacokinetics and Pharmacodynamics of Intravenous LY3819253 in Participants Hospitalized for COVID-19** Eli Lilly and Company *clinicaltrials.gov* (2020-10-29) https://clinicaltrials.gov/ct2/show/NCT04411628

207.    **A Phase 1, Randomized, Placebo-Controlled Study to Evaluate the Tolerability, Safety, Pharmacokinetics, and Immunogenicity of LY3832479 Given as a Single Intravenous Dose in Healthy Participants** Eli Lilly and Company *clinicaltrials.gov* (2020-10-07) https://clinicaltrials.gov/ct2/show/NCT04441931

208.    **Effect of Bamlanivimab as Monotherapy or in Combination With Etesevimab on Viral Load in Patients With Mild to Moderate COVID-19** Robert L Gottlieb, Ajay Nirula, Peter Chen, Joseph Boscia, Barry Heller, Jason Morris, Gregory Huhn, Jose Cardona, Bharat Mocherla, Valentina Stosor, … Daniel M Skovronsky *JAMA* (2021-02-16) https://doi.org/ghvnrr DOI: 10.1001/jama.2021.0202 · PMID: 33475701 · PMCID: PMC7821080

209.    **A Randomized, Double-blind, Placebo-Controlled, Phase 2/3 Study to Evaluate the Efficacy and Safety of LY3819253 and LY3832479 in Participants With Mild to Moderate**

**COVID-19 Illness** Eli Lilly and Company *clinicaltrials.gov* (2021-09-03) https://clinicaltrials.gov/ct2/show/NCT04427501

210. **SARS-CoV-2 Neutralizing Antibody LY-CoV555 in Outpatients with Covid-19** Peter Chen, Ajay Nirula, Barry Heller, Robert L Gottlieb, Joseph Boscia, Jason Morris, Gregory Huhn, Jose Cardona, Bharat Mocherla, Valentina Stosor, … Daniel M Skovronsky *New England Journal of Medicine* (2021-01-21) https://doi.org/fgtm DOI: 10.1056/nejmoa2029849 · PMID: 33113295 · PMCID: PMC7646625

211. **Bamlanivimab and Etesevimab EUA Letter of Authorization** U.S. Food and Drug Administration (2021-02-25) https://www.fda.gov/media/145801/download

212. **Studies in humanized mice and convalescent humans yield a SARS-CoV-2 antibody cocktail** Johanna Hansen, Alina Baum, Kristen E Pascal, Vincenzo Russo, Stephanie Giordano, Elzbieta Wloga, Benjamin O Fulton, Ying Yan, Katrina Koon, Krunal Patel, … Christos A Kyratsous *Science* (2020-08-21) https://doi.org/fcqh DOI: 10.1126/science.abd0827 · PMID: 32540901 · PMCID: PMC7299284

213. **A Master Protocol Assessing the Safety, Tolerability, and Efficacy of Anti-Spike (S) SARS-CoV-2 Monoclonal Antibodies for the Treatment of Hospitalized Patients With COVID-19** Regeneron Pharmaceuticals *clinicaltrials.gov* (2021-08-20) https://clinicaltrials.gov/ct2/show/NCT04426695

214. **A Master Protocol Assessing the Safety, Tolerability, and Efficacy of Anti-Spike (S) SARS-CoV-2 Monoclonal Antibodies for the Treatment of Ambulatory Patients With COVID-19** Regeneron Pharmaceuticals *clinicaltrials.gov* (2021-08-13) https://clinicaltrials.gov/ct2/show/NCT04425629

215. **REGN-COV2, a Neutralizing Antibody Cocktail, in Outpatients with Covid-19** David M Weinreich, Sumathi Sivapalasingam, Thomas Norton, Shazia Ali, Haitao Gao, Rafia Bhore, Bret J Musser, Yuhwen Soo, Diana Rofail, Joseph Im, … George D Yancopoulos *New England Journal of Medicine* (2021-01-21) https://doi.org/gh4sjh DOI: 10.1056/nejmoa2035002 · PMID: 33332778 · PMCID: PMC7781102

216. **Coronavirus (COVID-19) Update: FDA Authorizes Monoclonal Antibodies for Treatment of COVID-19** Office of the Commissioner *FDA* (2020-11-23) https://www.fda.gov/news-events/press-announcements/coronavirus-covid-19-update-fda-authorizes-monoclonal-antibodies-treatment-covid-19

217. **An efficient method to make human monoclonal antibodies from memory B cells: potent neutralization of SARS coronavirus** Elisabetta Traggiai, Stephan Becker, Kanta Subbarao, Larissa Kolesnikova, Yasushi Uematsu, Maria Rita Gismondo, Brian R Murphy, Rino Rappuoli, Antonio Lanzavecchia *Nature Medicine* (2004-07-11) https://doi.org/b9867c DOI: 10.1038/nm1080 · PMID: 15247913 · PMCID: PMC7095806

218. **'Super-antibodies' could curb COVID-19 and help avert future pandemics** Elie Dolgin *Nature Biotechnology* (2021-06-22) https://doi.org/gmg2fx DOI: 10.1038/s41587-021-00980-x · PMID: 34158667 · PMCID: PMC8218965

219. **Passive immunotherapy of viral infections: 'super-antibodies' enter the fray** Laura M Walker, Dennis R Burton *Nature Reviews Immunology* (2018-01-30) https://doi.org/gcwgpd DOI: 10.1038/nri.2017.148 · PMID: 29379211 · PMCID: PMC5918154

220. **A Randomized, Multi-center, Double-blind, Placebo-controlled Study to Assess the Safety and Efficacy of Monoclonal Antibody VIR-7831 for the Early Treatment of Coronavirus Disease 2019 (COVID-19) in Non-hospitalized Patients** Vir Biotechnology, Inc. *clinicaltrials.gov* (2021-08-17) https://clinicaltrials.gov/ct2/show/NCT04545060

221. **Early Covid-19 Treatment With SARS-CoV-2 Neutralizing Antibody Sotrovimab** Anil Gupta, Yaneicy Gonzalez-Rojas, Erick Juarez, Manuel Crespo Casal, Jaynier Moya, Diego Rodrigues Falci, Elias Sarkis, Joel Solis, Hanzhe Zheng, Nicola Scott, … for the COMET-ICE Investigators *Cold Spring Harbor Laboratory* (2021-05-28) https://doi.org/gmg2fz DOI: 10.1101/2021.05.27.21257096

222. **Identification of SARS-CoV-2 spike mutations that attenuate monoclonal and serum antibody neutralization** Zhuoming Liu, Laura A VanBlargan, Louis-Marie Bloyet, Paul W Rothlauf, Rita E Chen, Spencer Stumpf, Haiyan Zhao, John M Errico, Elitza S Theel, Mariel J Liebeskind, … Sean PJ Whelan *Cell Host & Microbe* (2021-03) https://doi.org/gh4m7j DOI: 10.1016/j.chom.2021.01.014 · PMID: 33535027 · PMCID: PMC7839837

223. **SARS-CoV-2 variants show resistance to neutralization by many monoclonal and serum-derived polyclonal antibodies** Michael Diamond, Rita Chen, Xuping Xie, James Case, Xianwen Zhang, Laura VanBlargan, Yang Liu, Jianying Liu, John Errico, Emma Winkler, … Pavlo Gilchuk *Research Square Platform LLC* (2021-02-09) https://doi.org/gh4sjz DOI: 10.21203/rs.3.rs-228079/v1 · PMID: 33594356 · PMCID: PMC7885928

224. **Impact of the B.1.1.7 variant on neutralizing monoclonal antibodies recognizing diverse epitopes on SARS–CoV–2 Spike** Carl Graham, Jeffrey Seow, Isabella Huettner, Hataf Khan, Neophytos Kouphou, Sam Acors, Helena Winstone, Suzanne Pickering, Rui Pedro Galao, Maria Jose Lista, … Katie J Doores *Cold Spring Harbor Laboratory* (2021-02-03) https://doi.org/gh4sjq DOI: 10.1101/2021.02.03.429355 · PMID: 33564766 · PMCID: PMC7872354

225. **Antibody Resistance of SARS-CoV-2 Variants B.1.351 and B.1.1.7** Pengfei Wang, Manoj S Nair, Lihong Liu, Sho Iketani, Yang Luo, Yicheng Guo, Maple Wang, Jian Yu, Baoshan Zhang, Peter D Kwong, … David D Ho *Cold Spring Harbor Laboratory* (2021-02-12) https://doi.org/gh4sjp DOI: 10.1101/2021.01.25.428137 · PMID: 33532778 · PMCID: PMC7852271

226. **Pause in the Distribution of bamlanivimab/etesevimab** U.S. Department of Health and Human Services (2021-06-25)

https://www.phe.gov/emergency/events/COVID19/investigation-MCM/Bamlanivimab-etesevimab/Pages/bamlanivimab-etesevimab-distribution-pause.aspx

227. **HIV-1 Broadly Neutralizing Antibody Extracts Its Epitope from a Kinked gp41 Ectodomain Region on the Viral Membrane** Zhen-Yu J Sun, Kyoung Joon Oh, Mikyung Kim, Jessica Yu, Vladimir Brusic, Likai Song, Zhisong Qiao, Jia-huai Wang, Gerhard Wagner, Ellis L Reinherz *Immunity* (2008-01) https://doi.org/ftw7t3 DOI: 10.1016/j.immuni.2007.11.018 · PMID: 18191596

228. **Antibody Recognition of a Highly Conserved Influenza Virus Epitope** DC Ekiert, G Bhabha, M-A Elsliger, RHE Friesen, M Jongeneelen, M Throsby, J Goudsmit, IA Wilson *Science* (2009-04-10) https://doi.org/ffsb4r DOI: 10.1126/science.1171491 · PMID: 19251591 · PMCID: PMC2758658

229. **Identification and characterization of novel neutralizing epitopes in the receptor-binding domain of SARS-CoV spike protein: Revealing the critical antigenic determinants in inactivated SARS-CoV vaccine** Yuxian He, Jingjing Li, Lanying Du, Xuxia Yan, Guangan Hu, Yusen Zhou, Shibo Jiang *Vaccine* (2006-06) https://doi.org/b99b68 DOI: 10.1016/j.vaccine.2006.04.054 · PMID: 16725238 · PMCID: PMC7115380

230. **Escape from Human Monoclonal Antibody Neutralization Affects In Vitro and In Vivo Fitness of Severe Acute Respiratory Syndrome Coronavirus** Barry Rockx, Eric Donaldson, Matthew Frieman, Timothy Sheahan, Davide Corti, Antonio Lanzavecchia, Ralph S Baric *The Journal of Infectious Diseases* (2010-03-15) https://doi.org/cdgqjd DOI: 10.1086/651022 · PMID: 20144042 · PMCID: PMC2826557

231. **Stanford Coronavirus Antiviral & Resistance Database (CoVDB)** https://covdb.stanford.edu/page/susceptibility-data

232. **Broad and potent activity against SARS-like viruses by an engineered human monoclonal antibody** CGarrett Rappazzo, Longping V Tse, Chengzi I Kaku, Daniel Wrapp, Mrunal Sakharkar, Deli Huang, Laura M Deveau, Thomas J Yockachonis, Andrew S Herbert, Michael B Battles, … Laura M Walker *Science* (2021-02-19) https://doi.org/fsbc DOI: 10.1126/science.abf4830 · PMID: 33495307 · PMCID: PMC7963221

233. **Drug repurposing: a promising tool to accelerate the drug discovery process** Vineela Parvathaneni, Nishant S Kulkarni, Aaron Muth, Vivek Gupta *Drug Discovery Today* (2019-10) https://doi.org/gj3v46 DOI: 10.1016/j.drudis.2019.06.014 · PMID: 31238113

234. **Drug repurposing screens and synergistic drug-combinations for infectious diseases** Wei Zheng, Wei Sun, Anton Simeonov *British Journal of Pharmacology* (2018-01) https://doi.org/gj3v6j DOI: 10.1111/bph.13895 · PMID: 28685814 · PMCID: PMC5758396

235. **A critical overview of computational approaches employed for COVID-19 drug discovery** Eugene N Muratov, Rommie Amaro, Carolina H Andrade, Nathan Brown, Sean Ekins, Denis Fourches, Olexandr Isayev, Dima Kozakov, José L Medina-Franco, Kenneth M

Merz, … Alexander Tropsha *Chemical Society Reviews* (2021) https://doi.org/gmg9nm DOI: 10.1039/d0cs01065k · PMID: 34212944 · PMCID: PMC8371861

236. **Drug repurposing: a better approach for infectious disease drug discovery?** GLynn Law, Jennifer Tisoncik-Go, Marcus J Korth, Michael G Katze *Current Opinion in Immunology* (2013-10) https://doi.org/f5jvrt DOI: 10.1016/j.coi.2013.08.004 · PMID: 24011665 · PMCID: PMC4015799

237. **Origin and evolution of high throughput screening** DA Pereira, JA Williams *British Journal of Pharmacology* (2007-09) https://doi.org/brs35w DOI: 10.1038/sj.bjp.0707373 · PMID: 17603542 · PMCID: PMC1978279

238. **Developing predictive assays: The phenotypic screening "rule of 3"** Fabien Vincent, Paula Loria, Marko Pregel, Robert Stanton, Linda Kitching, Karl Nocka, Regis Doyonnas, Claire Steppan, Adam Gilbert, Thomas Schroeter, Marie-Claire Peakman *Science Translational Medicine* (2015-06-24) https://doi.org/ggp3tk DOI: 10.1126/scitranslmed.aab1201 · PMID: 26109101

239. **Phenotypic vs. Target-Based Drug Discovery for First-in-Class Medicines** DC Swinney *Clinical Pharmacology & Therapeutics* (2013-04) https://doi.org/f4q6gz DOI: 10.1038/clpt.2012.236 · PMID: 23511784

240. **Phenotypic screening in cancer drug discovery — past, present and future** John G Moffat, Joachim Rudolph, David Bailey *Nature Reviews Drug Discovery* (2014-07-18) https://doi.org/f6cnfw DOI: 10.1038/nrd4366 · PMID: 25033736

241. **ChemBridge | Screening Libraries | Diversity Libraries | DIVERSet** https://www.chembridge.com/screening_libraries/diversity_libraries/

242. **The Power of Sophisticated Phenotypic Screening and Modern Mechanism-of-Action Methods** Bridget K Wagner, Stuart L Schreiber *Cell Chemical Biology* (2016-01) https://doi.org/gfsdbh DOI: 10.1016/j.chembiol.2015.11.008 · PMID: 26933731 · PMCID: PMC4779180

243. **Drug Repurposing for Viral Infectious Diseases: How Far Are We?** Beatrice Mercorelli, Giorgio Palù, Arianna Loregian *Trends in Microbiology* (2018-10) https://doi.org/gfbp3h DOI: 10.1016/j.tim.2018.04.004 · PMID: 29759926 · PMCID: PMC7126639

244. **Systematically Prioritizing Candidates in Genome-Based Drug Repurposing** Anup P Challa, Robert R Lavieri, Judith T Lewis, Nicole M Zaleski, Jana K Shirey-Rice, Paul A Harris, David M Aronoff, Jill M Pulley *ASSAY and Drug Development Technologies* (2019-12-01) https://doi.org/gj3v6d DOI: 10.1089/adt.2019.950 · PMID: 31769998 · PMCID: PMC6921094

245. **What Are the Odds of Finding a COVID-19 Drug from a Lab Repurposing Screen?** Aled Edwards *Journal of Chemical Information and Modeling* (2020-09-11) https://doi.org/gjkv79 DOI: 10.1021/acs.jcim.0c00861 · PMID: 32914973

246. **Proteases Essential for Human Influenza Virus Entry into Cells and Their Inhibitors as Potential Therapeutic Agents** Hiroshi Kido, Yuushi Okumura, Hiroshi Yamada, Trong Quang Le, Mihiro Yano *Current Pharmaceutical Design* (2007-02-01) https://doi.org/bts3xp DOI: 10.2174/138161207780162971 · PMID: 17311557

247. **Protease inhibitors targeting coronavirus and filovirus entry** Yanchen Zhou, Punitha Vedantham, Kai Lu, Juliet Agudelo, Ricardo Carrion, Jerritt W Nunneley, Dale Barnard, Stefan Pöhlmann, James H McKerrow, Adam R Renslo, Graham Simmons *Antiviral Research* (2015-04) https://doi.org/ggr984 DOI: 10.1016/j.antiviral.2015.01.011 · PMID: 25666761 · PMCID: PMC4774534

248. **Structure of Mpro from SARS-CoV-2 and discovery of its inhibitors** Zhenming Jin, Xiaoyu Du, Yechun Xu, Yongqiang Deng, Meiqin Liu, Yao Zhao, Bing Zhang, Xiaofeng Li, Leike Zhang, Chao Peng, … Haitao Yang *Nature* (2020-04-09) https://doi.org/ggrp42 DOI: 10.1038/s41586-020-2223-y · PMID: 32272481

249. **Design of Wide-Spectrum Inhibitors Targeting Coronavirus Main Proteases** Haitao Yang, Weiqing Xie, Xiaoyu Xue, Kailin Yang, Jing Ma, Wenxue Liang, Qi Zhao, Zhe Zhou, Duanqing Pei, John Ziebuhr, … Zihe Rao *PLoS Biology* (2005-09-06) https://doi.org/bcm9k7 DOI: 10.1371/journal.pbio.0030324 · PMID: 16128623 · PMCID: PMC1197287

250. **The newly emerged SARS-Like coronavirus HCoV-EMC also has an "Achilles' heel": current effective inhibitor targeting a 3C-like protease** Zhilin Ren, Liming Yan, Ning Zhang, Yu Guo, Cheng Yang, Zhiyong Lou, Zihe Rao *Protein & Cell* (2013-04-03) https://doi.org/ggr7vh DOI: 10.1007/s13238-013-2841-3 · PMID: 23549610 · PMCID: PMC4875521

251. **Structure of Main Protease from Human Coronavirus NL63: Insights for Wide Spectrum Anti-Coronavirus Drug Design** Fenghua Wang, Cheng Chen, Wenjie Tan, Kailin Yang, Haitao Yang *Scientific Reports* (2016-03-07) https://doi.org/f8cfx9 DOI: 10.1038/srep22677 · PMID: 26948040 · PMCID: PMC4780191

252. **Structures of Two Coronavirus Main Proteases: Implications for Substrate Binding and Antiviral Drug Design** Xiaoyu Xue, Hongwei Yu, Haitao Yang, Fei Xue, Zhixin Wu, Wei Shen, Jun Li, Zhe Zhou, Yi Ding, Qi Zhao, … Zihe Rao *Journal of Virology* (2008-03) https://doi.org/b2zbhv DOI: 10.1128/jvi.02114-07 · PMID: 18094151 · PMCID: PMC2258912

253. **Ebselen, a promising antioxidant drug: mechanisms of action and targets of biological pathways** Gajendra Kumar Azad, Raghuvir S Tomar *Molecular Biology Reports* (2014-05-28) https://doi.org/f6cnq3 DOI: 10.1007/s11033-014-3417-x · PMID: 24867080

254. **Molecular characterization of ebselen binding activity to SARS-CoV-2 main protease** Cintia A Menéndez, Fabian Byléhn, Gustavo R Perez-Lemus, Walter Alvarado, Juan J de Pablo *Science Advances* (2020-09) https://doi.org/gmhshj DOI: 10.1126/sciadv.abd0345 · PMID: 32917717 · PMCID: PMC7486088

255. **Target discovery of ebselen with a biotinylated probe** Zhenzhen Chen, Zhongyao Jiang, Nan Chen, Qian Shi, Lili Tong, Fanpeng Kong, Xiufen Cheng, Hao Chen, Chu Wang, Bo Tang *Chemical Communications* (2018) https://doi.org/ggrtcm DOI: 10.1039/c8cc04258f · PMID: 30091742

256. **FDA Clears SPI's Ebselen For Phase II COVID-19 Trials** Contract Pharma
https://www.contractpharma.com/contents/view_breaking_news/2020-08-31/fda-clears-spis-ebselen-for-phase-ii-covid-19-trials/

257. **A Phase 2, Randomized, Double-Blind, Placebo-Controlled, Dose Escalation Study to Evaluate the Safety and Efficacy of SPI-1005 in Moderate COVID-19 Patients** Sound Pharmaceuticals, Incorporated *clinicaltrials.gov* (2021-06-14)
https://clinicaltrials.gov/ct2/show/NCT04484025

258. **A Phase 2, Randomized, Double-Blind, Placebo-Controlled, Dose Escalation Study to Evaluate the Safety and Efficacy of SPI-1005 in Severe COVID-19 Patients** Sound Pharmaceuticals, Incorporated *clinicaltrials.gov* (2021-06-14)
https://clinicaltrials.gov/ct2/show/NCT04483973

259. **A PHASE 1B, 2-PART, DOUBLE-BLIND, PLACEBO-CONTROLLED, SPONSOR-OPEN STUDY, TO EVALUATE THE SAFETY, TOLERABILITY AND PHARMACOKINETICS OF SINGLE ASCENDING (24-HOUR, PART 1) AND MULTIPLE ASCENDING (120-HOUR, PART 2) INTRAVENOUS INFUSIONS OF PF-07304814 IN HOSPITALIZED PARTICIPANTS WITH COVID-19** Pfizer *clinicaltrials.gov* (2021-06-23)
https://clinicaltrials.gov/ct2/show/NCT04535167

260. **AN INTERVENTIONAL EFFICACY AND SAFETY, PHASE 2/3, DOUBLE-BLIND, 2-ARM STUDY TO INVESTIGATE ORALLY ADMINISTERED PF-07321332/RITONAVIR COMPARED WITH PLACEBO IN NONHOSPITALIZED SYMPTOMATIC ADULT PARTICIPANTS WITH COVID-19 WHO ARE AT INCREASED RISK OF PROGRESSING TO SEVERE ILLNESS** Pfizer *clinicaltrials.gov* (2021-08-31)
https://clinicaltrials.gov/ct2/show/NCT04960202

261. **Use of big data in drug development for precision medicine: an update** Tongqi Qian, Shijia Zhu, Yujin Hoshida *Expert Review of Precision Medicine and Drug Development* (2019-05-20) https://doi.org/gmpgx2 DOI: 10.1080/23808993.2019.1617632 · PMID: 31286058 · PMCID: PMC6613936

262. **The COVID-19 Drug and Gene Set Library** Maxim V Kuleshov, Daniel J Stein, Daniel JB Clarke, Eryk Kropiwnicki, Kathleen M Jagodnik, Alon Bartal, John E Evangelista, Jason Hom, Minxuan Cheng, Allison Bailey, … Avi Ma'ayan *Patterns* (2020-09) https://doi.org/gg56f3 DOI: 10.1016/j.patter.2020.100090 · PMID: 32838343 · PMCID: PMC7381899

263. **Exploring the SARS-CoV-2 virus-host-drug interactome for drug repurposing** Sepideh Sadegh, Julian Matschinske, David B Blumenthal, Gihanna Galindez, Tim Kacprowski, Markus List, Reza Nasirigerdeh, Mhaned Oubounyt, Andreas Pichlmair, Tim Daniel Rose, …

Jan Baumbach *Nature Communications* (2020-07-14) https://doi.org/gg477d DOI: 10.1038/s41467-020-17189-2 · PMID: 32665542 · PMCID: PMC7360763

264. **Artificial intelligence in COVID-19 drug repurposing** Yadi Zhou, Fei Wang, Jian Tang, Ruth Nussinov, Feixiong Cheng *The Lancet Digital Health* (2020-12) https://doi.org/grs9 DOI: 10.1016/s2589-7500(20)30192-8 · PMID: 32984792 · PMCID: PMC7500917

265. **A SARS-CoV-2 protein interaction map reveals targets for drug repurposing** David E Gordon, Gwendolyn M Jang, Mehdi Bouhaddou, Jiewei Xu, Kirsten Obernier, Kris M White, Matthew J O'Meara, Veronica V Rezelj, Jeffrey Z Guo, Danielle L Swaney, … Nevan J Krogan *Nature* (2020-04-30) https://doi.org/ggvr6p DOI: 10.1038/s41586-020-2286-9 · PMID: 32353859 · PMCID: PMC7431030

266. **The pharmacology of sigma-1 receptors** Tangui Maurice, Tsung-Ping Su *Pharmacology & Therapeutics* (2009-11) https://doi.org/fhm455 DOI: 10.1016/j.pharmthera.2009.07.001 · PMID: 19619582 · PMCID: PMC2785038

267. **Comparative host-coronavirus protein interaction networks reveal pan-viral disease mechanisms.** David E Gordon, Joseph Hiatt, Mehdi Bouhaddou, Veronica V Rezelj, Svenja Ulferts, Hannes Braberg, Alexander S Jureka, Kirsten Obernier, Jeffrey Z Guo, Jyoti Batra, … Nevan J Krogan *Science (New York, N.Y.)* (2020-10-15) https://www.ncbi.nlm.nih.gov/pubmed/33060197 DOI: 10.1126/science.abe9403 · PMID: 33060197 · PMCID: PMC7808408

268. **Repurposing Sigma-1 Receptor Ligands for COVID-19 Therapy?** José Miguel Vela *Frontiers in Pharmacology* (2020-11-09) https://doi.org/gmh3mh DOI: 10.3389/fphar.2020.582310 · PMID: 33364957 · PMCID: PMC7751758

269. **The Sigma Receptor: Evolution of the Concept in Neuropsychopharmacology** T Hayashi, T-Su *Current Neuropharmacology* (2005-10-01) https://doi.org/fwpwcs DOI: 10.2174/157015905774322516 · PMID: 18369400 · PMCID: PMC2268997

270. **Drug-induced phospholipidosis confounds drug repurposing for SARS-CoV-2** Tia A Tummino, Veronica V Rezelj, Benoit Fischer, Audrey Fischer, Matthew J O'Meara, Blandine Monel, Thomas Vallet, Kris M White, Ziyang Zhang, Assaf Alon, … Brian K Shoichet *Science* (2021-07-30) https://doi.org/gmgx79 DOI: 10.1126/science.abi4708 · PMID: 34326236

271. **Emerging mechanisms of drug-induced phospholipidosis** Bernadette Breiden, Konrad Sandhoff *Biological Chemistry* (2019-12-18) https://doi.org/gjkv8x DOI: 10.1515/hsz-2019-0270 · PMID: 31408430

272. **Zebra-like bodies in COVID-19: is phospholipidosis evidence of hydroxychloroquine induced acute kidney injury?** Mohammad Obeidat, Alexandra L Isaacson, Stephanie J Chen, Marina Ivanovic, Danniele Holanda *Ultrastructural Pathology* (2020-12-04) https://doi.org/gj3v6c DOI: 10.1080/01913123.2020.1850966 · PMID: 33274661

273. **Drug repurposing screens reveal cell-type-specific entry pathways and FDA-approved drugs active against SARS-Cov-2** Mark Dittmar, Jae Seung Lee, Kanupriya Whig,

Elisha Segrist, Minghua Li, Brinda Kamalia, Lauren Castellana, Kasirajan Ayyanathan, Fabian L Cardenas-Diaz, Edward E Morrisey, … Sara Cherry *Cell Reports* (2021-04) https://doi.org/gj3v44 DOI: 10.1016/j.celrep.2021.108959 · PMID: 33811811 · PMCID: PMC7985926

274. **No shortcuts to SARS-CoV-2 antivirals** Aled Edwards, Ingo V Hartung *Science* (2021-07-29) https://doi.org/gmh3mg DOI: 10.1126/science.abj9488 · PMID: 34326222

275. **Repurposing of CNS drugs to treat COVID-19 infection: targeting the sigma-1 receptor** Kenji Hashimoto *European Archives of Psychiatry and Clinical Neuroscience* (2021-01-05) https://doi.org/ghth6q DOI: 10.1007/s00406-020-01231-x · PMID: 33403480 · PMCID: PMC7785036

276. **Fluvoxamine vs Placebo and Clinical Deterioration in Outpatients With Symptomatic COVID-19** Eric J Lenze, Caline Mattar, Charles F Zorumski, Angela Stevens, Julie Schweiger, Ginger E Nicol, JPhilip Miller, Lei Yang, Michael Yingling, Michael S Avidan, Angela M Reiersen *JAMA* (2020-12-08) https://doi.org/ghjtd5 DOI: 10.1001/jama.2020.22760 · PMID: 33180097 · PMCID: PMC7662481

277. **Prospective Cohort of Fluvoxamine for Early Treatment of Coronavirus Disease 19** David Seftel, David R Boulware *Open Forum Infectious Diseases* (2021-02-01) https://doi.org/gmj9sz DOI: 10.1093/ofid/ofab050 · PMID: 33623808 · PMCID: PMC7888564

278. **Modulation of the sigma-1 receptor–IRE1 pathway is beneficial in preclinical models of inflammation and sepsis** Dorian A Rosen, Scott M Seki, Anthony Fernández-Castañeda, Rebecca M Beiter, Jacob D Eccles, Judith A Woodfolk, Alban Gaultier *Science Translational Medicine* (2019-02-06) https://doi.org/gmj9s2 DOI: 10.1126/scitranslmed.aau5266 · PMID: 30728287 · PMCID: PMC6936250

279. **Fluvoxamine: A Review of Its Mechanism of Action and Its Role in COVID-19** Vikas P Sukhatme, Angela M Reiersen, Sharat J Vayttaden, Vidula V Sukhatme *Frontiers in Pharmacology* (2021-04-20) https://doi.org/gmj9tg DOI: 10.3389/fphar.2021.652688 · PMID: 33959018 · PMCID: PMC8094534

280. **Too Many Papers** Derek Lowe *In the Pipeline* (2021-07-19) https://blogs.sciencemag.org/pipeline/archives/2021/07/19/too-many-papers

281. **Believe it or not: how much can we rely on published data on potential drug targets?** Florian Prinz, Thomas Schlange, Khusru Asadullah *Nature Reviews Drug Discovery* (2011-08-31) https://doi.org/dfsxxb DOI: 10.1038/nrd3439-c1 · PMID: 21892149

282. **How were new medicines discovered?** David C Swinney, Jason Anthony *Nature Reviews Drug Discovery* (2011-06-24) https://doi.org/bbg5wh DOI: 10.1038/nrd3480 · PMID: 21701501

283. **Big studies dim hopes for hydroxychloroquine** Kai Kupferschmidt *Science* (2020-06-12) https://doi.org/gh7d7p DOI: 10.1126/science.368.6496.1166 · PMID: 32527806

284. **Trends in COVID-19 therapeutic clinical trials** Kevin Bugin, Janet Woodcock *Nature Reviews Drug Discovery* (2021-02-25) https://doi.org/gmj9sj DOI: 10.1038/d41573-021-00037-3 · PMID: 33633370

285. **Clinical Trial Data Sharing for COVID-19–Related Research** Louis Dron, Alison Dillman, Michael J Zoratti, Jonas Haggstrom, Edward J Mills, Jay JH Park *Journal of Medical Internet Research* (2021-03-12) https://doi.org/gk6zfj DOI: 10.2196/26718 · PMID: 33684053 · PMCID: PMC7958972

286. **The Rise and Fall of Hydroxychloroquine for the Treatment and Prevention of COVID-19** Zelyn Lee, Craig R Rayner, Jamie I Forrest, Jean B Nachega, Esha Senchaudhuri, Edward J Mills *The American Journal of Tropical Medicine and Hygiene* (2021-01-06) https://doi.org/gmj9s3 DOI: 10.4269/ajtmh.20-1320 · PMID: 33236703 · PMCID: PMC7790108

287. **Moving forward in clinical research with master protocols** Jay JH Park, Louis Dron, Edward J Mills *Contemporary Clinical Trials* (2021-07) https://doi.org/gmj9sd DOI: 10.1016/j.cct.2021.106438 · PMID: 34000408 · PMCID: PMC8120789

288. **Main protease structure and XChem fragment screen** Diamond (2020-05-05) https://www.diamond.ac.uk/covid-19/for-scientists/Main-protease-structure-and-XChem.html

289. **Non-traditional cytokines: How catecholamines and adipokines influence macrophages in immunity, metabolism and the central nervous system** Mark A Barnes, Monica J Carson, Meera G Nair *Cytokine* (2015-04) https://doi.org/f65c59 DOI: 10.1016/j.cyto.2015.01.008 · PMID: 25703786 · PMCID: PMC4590987

290. **Stress Hormones, Proinflammatory and Antiinflammatory Cytokines, and Autoimmunity** ILIA J ELENKOV, GEORGE P CHROUSOS *Annals of the New York Academy of Sciences* (2002-06) https://doi.org/fmwpx2 DOI: 10.1111/j.1749-6632.2002.tb04229.x · PMID: 12114286

291. **Recovery of the Hypothalamic-Pituitary-Adrenal Response to Stress** Arantxa García, Octavi Martí, Astrid Vallès, Silvina Dal-Zotto, Antonio Armario *Neuroendocrinology* (2000) https://doi.org/b2cq8n DOI: 10.1159/000054578 · PMID: 10971146

292. **Modulatory effects of glucocorticoids and catecholamines on human interleukin-12 and interleukin-10 production: clinical implications.** IJ Elenkov, DA Papanicolaou, RL Wilder, GP Chrousos *Proceedings of the Association of American Physicians* (1996-09) https://www.ncbi.nlm.nih.gov/pubmed/8902882 PMID: 8902882

293. **Dexamethasone for COVID-19? Not so fast.** TC Theoharides *JOURNAL OF BIOLOGICAL REGULATORS AND HOMEOSTATIC AGENTS* (2020-08-31) https://doi.org/ghfkjx DOI: 10.23812/20-editorial_1-5 · PMID: 32551464

294. **Dexamethasone in hospitalised patients with COVID-19: addressing uncertainties** Michael A Matthay, BTaylor Thompson *The Lancet Respiratory Medicine* (2020-12) https://doi.org/ftk4 DOI: 10.1016/s2213-2600(20)30503-8 · PMID: 33129421 · PMCID: PMC7598750

295. **Dexamethasone for COVID-19: data needed from randomised clinical trials in Africa** Helen Brotherton, Effua Usuf, Behzad Nadjm, Karen Forrest, Kalifa Bojang, Ahmadou Lamin Samateh, Mustapha Bittaye, Charles AP Roberts, Umberto d'Alessandro, Anna Roca *The Lancet Global Health* (2020-09) https://doi.org/gg42kx DOI: 10.1016/s2214-109x(20)30318-1 · PMID: 32679038 · PMCID: PMC7833918

296. **Experimental Treatment with Favipiravir for COVID-19: An Open-Label Control Study** Qingxian Cai, Minghui Yang, Dongjing Liu, Jun Chen, Dan Shu, Junxia Xia, Xuejiao Liao, Yuanbo Gu, Qiue Cai, Yang Yang, … Lei Liu *Engineering* (2020-10) https://doi.org/ggpprd DOI: 10.1016/j.eng.2020.03.007 · PMID: 32346491

297. **Lopinavir–ritonavir in patients admitted to hospital with COVID-19 (RECOVERY): a randomised, controlled, open-label, platform trial** Peter W Horby, Marion Mafham, Jennifer L Bell, Louise Linsell, Natalie Staplin, Jonathan Emberson, Adrian Palfreeman, Jason Raw, Einas Elmahi, Benjamin Prudon, … Martin J Landray *The Lancet* (2020-10) https://doi.org/fnx2 DOI: 10.1016/s0140-6736(20)32013-4 · PMID: 33031764 · PMCID: PMC7535623

298. **A Large, Simple Trial Leading to Complex Questions** David P Harrington, Lindsey R Baden, Joseph W Hogan *New England Journal of Medicine* (2020-12-02) https://doi.org/ghnhnx DOI: 10.1056/nejme2034294 · PMID: 33264557 · PMCID: PMC7727323

299. **Retracted coronavirus (COVID-19) papers** Retraction Watch (2020-04-29) https://retractionwatch.com/retracted-coronavirus-covid-19-papers/

300. **Clinical Outcomes and Plasma Concentrations of Baloxavir Marboxil and Favipiravir in COVID-19 Patients: An Exploratory Randomized, Controlled Trial** Yan Lou, Lin Liu, Hangping Yao, Xingjiang Hu, Junwei Su, Kaijin Xu, Rui Luo, Xi Yang, Lingjuan He, Xiaoyang Lu, … Yunqing Qiu *European Journal of Pharmaceutical Sciences* (2021-02) https://doi.org/ghx88n DOI: 10.1016/j.ejps.2020.105631 · PMID: 33115675 · PMCID: PMC7585719

301. **Efficacy of favipiravir in COVID-19 treatment: a multi-center randomized study** Hany M Dabbous, Sherief Abd-Elsalam, Manal H El-Sayed, Ahmed F Sherief, Fatma FS Ebeid, Mohamed Samir Abd El Ghafar, Shaimaa Soliman, Mohamed Elbahnasawy, Rehab Badawi, Mohamed Awad Tageldin *Archives of Virology* (2021-01-25) https://doi.org/ghx874 DOI: 10.1007/s00705-021-04956-9 · PMID: 33492523 · PMCID: PMC7829645

302. **AVIFAVIR for Treatment of Patients With Moderate Coronavirus Disease 2019 (COVID-19): Interim Results of a Phase II/III Multicenter Randomized Clinical Trial** Andrey A Ivashchenko, Kirill A Dmitriev, Natalia V Vostokova, Valeria N Azarova, Andrew A Blinow, Alina N Egorova, Ivan G Gordeev, Alexey P Ilin, Ruben N Karapetian, Dmitry V Kravchenko, … Alexandre V Ivachtchenko *Clinical Infectious Diseases* (2020-08-09) https://doi.org/ghx9c2 DOI: 10.1093/cid/ciaa1176 · PMID: 32770240 · PMCID: PMC7454388

303. **A review of the safety of favipiravir – a potential treatment in the COVID-19 pandemic?** Victoria Pilkington, Toby Pepperrell, Andrew Hill *Journal of Virus Eradication* (2020-

04) https://doi.org/ftgm DOI: 10.1016/s2055-6640(20)30016-9 · PMID: 32405421 · PMCID: PMC7331506

304. **A Randomized, Controlled Trial of Ebola Virus Disease Therapeutics** Sabue Mulangu, Lori E Dodd, Richard T Davey, Olivier Tshiani Mbaya, Michael Proschan, Daniel Mukadi, Mariano Lusakibanza Manzo, Didier Nzolo, Antoine Tshomba Oloma, Augustin Ibanda, … the PALM Writing Group *New England Journal of Medicine* (2019-12-12) https://doi.org/ggqmx4 DOI: 10.1056/nejmoa1910993 · PMID: 31774950

305. **Broad-spectrum antiviral GS-5734 inhibits both epidemic and zoonotic coronaviruses** Timothy P Sheahan, Amy C Sims, Rachel L Graham, Vineet D Menachery, Lisa E Gralinski, James B Case, Sarah R Leist, Krzysztof Pyrc, Joy Y Feng, Iva Trantcheva, … Ralph S Baric *Science Translational Medicine* (2017-06-28) https://doi.org/gc3grb DOI: 10.1126/scitranslmed.aal3653 · PMID: 28659436 · PMCID: PMC5567817

306. **Did an experimental drug help a U.S. coronavirus patient?** Jon Cohen *Science* (2020-03-13) https://doi.org/ggqm62 DOI: 10.1126/science.abb7243

307. **First 12 patients with coronavirus disease 2019 (COVID-19) in the United States** Stephanie A Kujawski, Karen K Wong, Jennifer P Collins, Lauren Epstein, Marie E Killerby, Claire M Midgley, Glen R Abedi, NSeema Ahmed, Olivia Almendares, Francisco N Alvarez, … The COVID-19 Investigation Team *Cold Spring Harbor Laboratory* (2020-03-12) https://doi.org/ggqm6z DOI: 10.1101/2020.03.09.20032896

308. **First Case of 2019 Novel Coronavirus in the United States** Michelle L Holshue, Chas DeBolt, Scott Lindquist, Kathy H Lofy, John Wiesman, Hollianne Bruce, Christopher Spitters, Keith Ericson, Sara Wilkerson, Ahmet Tural, … Satish K Pillai *New England Journal of Medicine* (2020-03-05) https://doi.org/ggjvr6 DOI: 10.1056/nejmoa2001191 · PMID: 32004427 · PMCID: PMC7092802

309. **Remdesivir for 5 or 10 Days in Patients with Severe Covid-19** Jason D Goldman, David CB Lye, David S Hui, Kristen M Marks, Raffaele Bruno, Rocio Montejano, Christoph D Spinner, Massimo Galli, Mi-Young Ahn, Ronald G Nahass, … Aruna Subramanian *New England Journal of Medicine* (2020-11-05) https://doi.org/ggz7qv DOI: 10.1056/nejmoa2015301 · PMID: 32459919 · PMCID: PMC7377062

310. **Remdesivir EUA Letter of Authorization** Denise M Hinton (2020-05-01) https://www.fda.gov/media/137564/download

311. **A Phase 3 Randomized Study to Evaluate the Safety and Antiviral Activity of Remdesivir (GS-5734™) in Participants With Moderate COVID-19 Compared to Standard of Care Treatment** Gilead Sciences *clinicaltrials.gov* (2021-01-21) https://clinicaltrials.gov/ct2/show/NCT04292730

312. **Multi-centre, adaptive, randomized trial of the safety and efficacy of treatments of COVID-19 in hospitalized adults** EU Clinical Trials Register (2020-03-09) https://www.clinicaltrialsregister.eu/ctr-search/trial/2020-000936-23/FR

313.    **A Trial of Remdesivir in Adults With Mild and Moderate COVID-19 - Full Text View - ClinicalTrials.gov** https://clinicaltrials.gov/ct2/show/NCT04252664

314.    **A Phase 3 Randomized, Double-blind, Placebo-controlled, Multicenter Study to Evaluate the Efficacy and Safety of Remdesivir in Hospitalized Adult Patients With Severe COVID-19.** Bin Cao *clinicaltrials.gov* (2020-04-13) https://clinicaltrials.gov/ct2/show/NCT04257656

315.    **FDA Approves First Treatment for COVID-19** Office of the Commissioner *FDA* (2020-10-22) https://www.fda.gov/news-events/press-announcements/fda-approves-first-treatment-covid-19

316.    **Gilead Sciences Statement on the Solidarity Trial** https://www.gilead.com/news-and-press/company-statements/gilead-sciences-statement-on-the-solidarity-trial

317.    **Conflicting results on the efficacy of remdesivir in hospitalized Covid-19 patients: comment on the Adaptive Covid-19 Treatment Trial** Leonarda Galiuto, Carlo Patrono *European Heart Journal* (2020-12-07) https://doi.org/ghp4kw DOI: 10.1093/eurheartj/ehaa934 · PMID: 33306101 · PMCID: PMC7799042

318.    **The 'very, very bad look' of remdesivir, the first FDA-approved COVID-19 drug** Jon Cohen, Kai Kupferschmidt *Science | AAAS* (2020-10-28) https://www.sciencemag.org/news/2020/10/very-very-bad-look-remdesivir-first-fda-approved-covid-19-drug

319.    **Effect of Remdesivir vs Standard Care on Clinical Status at 11 Days in Patients With Moderate COVID-19** Christoph D Spinner, Robert L Gottlieb, Gerard J Criner, José Ramón Arribas López, Anna Maria Cattelan, Alex Soriano Viladomiu, Onyema Ogbuagu, Prashant Malhotra, Kathleen M Mullane, Antonella Castagna, … for the GS-US-540-5774 Investigators *JAMA* (2020-09-15) https://doi.org/ghhz6g DOI: 10.1001/jama.2020.16349 · PMID: 32821939 · PMCID: PMC7442954

320.    **Baricitinib plus Remdesivir for Hospitalized Adults with Covid-19** Andre C Kalil, Thomas F Patterson, Aneesh K Mehta, Kay M Tomashek, Cameron R Wolfe, Varduhi Ghazaryan, Vincent C Marconi, Guillermo M Ruiz-Palacios, Lanny Hsieh, Susan Kline, … John H Beigel *New England Journal of Medicine* (2020-12-11) https://doi.org/ghpbd2 DOI: 10.1056/nejmoa2031994 · PMID: 33306283 · PMCID: PMC7745180

321.    **Letter of Authorization: EUA for baricitinib (Olumiant), in combination with remdesivir (Veklury), for the treatment of suspected or laboratory confirmed coronavirus disease 2019 (COVID-19)** Denise M Hinton *Food and Drug Administration* (2020-01-19) https://www.fda.gov/media/143822/download

322.    **Lysosomotropic agents as HCV entry inhibitors** Usman A Ashfaq, Tariq Javed, Sidra Rehman, Zafar Nawaz, Sheikh Riazuddin *Virology Journal* (2011-04-12) https://doi.org/dr5g4m DOI: 10.1186/1743-422x-8-163 · PMID: 21481279 · PMCID: PMC3090357

323. **Coagulopathy and Antiphospholipid Antibodies in Patients with Covid-19** Yan Zhang, Meng Xiao, Shulan Zhang, Peng Xia, Wei Cao, Wei Jiang, Huan Chen, Xin Ding, Hua Zhao, Hongmin Zhang, … Shuyang Zhang *New England Journal of Medicine* (2020-04-23) https://doi.org/ggrgz7 DOI: 10.1056/nejmc2007575 · PMID: 32268022 · PMCID: PMC7161262

324. **Mechanism of Action of Hydroxychloroquine in the Antiphospholipid Syndrome** Nadine Müller-Calleja, Davit Manukyan, Wolfram Ruf, Karl Lackner *Blood* (2016-12-02) https://doi.org/ggrm82 DOI: 10.1182/blood.v128.22.5023.5023

325. **14th International Congress on Antiphospholipid Antibodies Task Force Report on Antiphospholipid Syndrome Treatment Trends** Doruk Erkan, Cassyanne L Aguiar, Danieli Andrade, Hannah Cohen, Maria J Cuadrado, Adriana Danowski, Roger A Levy, Thomas L Ortel, Anisur Rahman, Jane E Salmon, … Michael D Lockshin *Autoimmunity Reviews* (2014-06) https://doi.org/ggp8r8 DOI: 10.1016/j.autrev.2014.01.053 · PMID: 24468415

326. **What is the role of hydroxychloroquine in reducing thrombotic risk in patients with antiphospholipid antibodies?** Tzu-Fei Wang, Wendy Lim *Hematology* (2016-12-02) https://doi.org/ggrn3k DOI: 10.1182/asheducation-2016.1.714 · PMID: 27913551 · PMCID: PMC6142483

327. **COVID-19: a recommendation to examine the effect of hydroxychloroquine in preventing infection and progression** Dan Zhou, Sheng-Ming Dai, Qiang Tong *Journal of Antimicrobial Chemotherapy* (2020-07) https://doi.org/ggq84c DOI: 10.1093/jac/dkaa114 · PMID: 32196083 · PMCID: PMC7184499

328. **Hydroxychloroquine treatment of patients with human immunodeficiency virus type 1** Kirk Sperber, Michael Louie, Thomas Kraus, Jacqueline Proner, Erica Sapira, Su Lin, Vera Stecher, Lloyd Mayer *Clinical Therapeutics* (1995-07) https://doi.org/cq2hx9 DOI: 10.1016/0149-2918(95)80039-5

329. **Hydroxychloroquine augments early virological response to pegylated interferon plus ribavirin in genotype-4 chronic hepatitis C patients** Gouda Kamel Helal, Magdy Abdelmawgoud Gad, Mohamed Fahmy Abd-Ellah, Mahmoud Saied Eid *Journal of Medical Virology* (2016-12) https://doi.org/f889nt DOI: 10.1002/jmv.24575 · PMID: 27183377 · PMCID: PMC7167065

330. **Making the Best Match: Selecting Outcome Measures for Clinical Trials and Outcome Studies** WJ Coster *American Journal of Occupational Therapy* (2013-02-22) https://doi.org/f4rf5s DOI: 10.5014/ajot.2013.006015 · PMID: 23433270 · PMCID: PMC3628620

331. **No evidence of rapid antiviral clearance or clinical benefit with the combination of hydroxychloroquine and azithromycin in patients with severe COVID-19 infection** JM Molina, C Delaugerre, J Le Goff, B Mela-Lima, D Ponscarme, L Goldwirt, N de Castro *Médecine et Maladies Infectieuses* (2020-06) https://doi.org/ggqzrb DOI: 10.1016/j.medmal.2020.03.006 · PMID: 32240719

332. **Efficacy of hydroxychloroquine in patients with COVID-19: results of a randomized clinical trial** Zhaowei Chen, Jijia Hu, Zongwei Zhang, Shan Jiang, Shoumeng Han, Dandan Yan, Ruhong Zhuang, Ben Hu, Zhan Zhang *Cold Spring Harbor Laboratory* (2020-04-10) https://doi.org/ggqm4v DOI: 10.1101/2020.03.22.20040758

333. **Therapeutic effect of hydroxychloroquine on novel coronavirus pneumonia (COVID-19)** Chinese Clinical Trial Registry (2020-02-12) http://www.chictr.org.cn/showprojen.aspx?proj=48880

334. **A pilot study of hydroxychloroquine in treatment of patients with common coronavirus disease-19 (COVID-19)** CHEN Jun, LIU Danping, LIU Li, LIU Ping, XU Qingnian, XIA Lu, LING Yun, HUANG Dan, SONG Shuli, ZHANG Dandan, … LU Hongzhou *Journal of Zhejiang University (Medical Sciences)* (2020-03) https://doi.org/10.3785/j.issn.1008-9292.2020.03.03 DOI: 10.3785/j.issn.1008-9292.2020.03.03

335. **Breakthrough: Chloroquine phosphate has shown apparent efficacy in treatment of COVID-19 associated pneumonia in clinical studies** Jianjun Gao, Zhenxue Tian, Xu Yang *BioScience Trends* (2020-02-29) https://doi.org/ggm3mv DOI: 10.5582/bst.2020.01047 · PMID: 32074550

336. **Targeting the Endocytic Pathway and Autophagy Process as a Novel Therapeutic Strategy in COVID-19** Naidi Yang, Han-Ming Shen *International Journal of Biological Sciences* (2020) https://doi.org/ggqspm DOI: 10.7150/ijbs.45498 · PMID: 32226290 · PMCID: PMC7098027

337. **SARS-CoV-2: an Emerging Coronavirus that Causes a Global Threat** Jun Zheng *International Journal of Biological Sciences* (2020) https://doi.org/ggqspr DOI: 10.7150/ijbs.45053 · PMID: 32226285 · PMCID: PMC7098030

338. **RETRACTED: Hydroxychloroquine or chloroquine with or without a macrolide for treatment of COVID-19: a multinational registry analysis** Mandeep R Mehra, Sapan S Desai, Frank Ruschitzka, Amit N Patel *The Lancet* (2020-05) https://doi.org/ggwzsb DOI: 10.1016/s0140-6736(20)31180-6 · PMID: 32450107 · PMCID: PMC7255293

339. **Retraction—Hydroxychloroquine or chloroquine with or without a macrolide for treatment of COVID-19: a multinational registry analysis** Mandeep R Mehra, Frank Ruschitzka, Amit N Patel *The Lancet* (2020-06) https://doi.org/ggzqng DOI: 10.1016/s0140-6736(20)31324-6 · PMID: 32511943 · PMCID: PMC7274621

340. **Life Threatening Severe QTc Prolongation in Patient with Systemic Lupus Erythematosus due to Hydroxychloroquine** John P O'Laughlin, Parag H Mehta, Brian C Wong *Case Reports in Cardiology* (2016) https://doi.org/ggqzrc DOI: 10.1155/2016/4626279 · PMID: 27478650 · PMCID: PMC4960328

341. **Keep the QT interval: It is a reliable predictor of ventricular arrhythmias** Dan M Roden *Heart Rhythm* (2008-08) https://doi.org/d5rchx DOI: 10.1016/j.hrthm.2008.05.008 · PMID: 18675237 · PMCID: PMC3212752

342. **Safety of hydroxychloroquine, alone and in combination with azithromycin, in light of rapid wide-spread use for COVID-19: a multinational, network cohort and self-controlled case series study** Jennifer C.E.Lane, James Weaver, Kristin Kostka, Talita Duarte-Salles, Maria Tereza F Abrahao, Heba Alghoul, Osaid Alser, Thamir M Alshammari, Patricia Biedermann, Edward Burn, … Daniel Prieto-Alhambra *Cold Spring Harbor Laboratory* (2020-04-10) https://doi.org/ggrn7s DOI: 10.1101/2020.04.08.20054551

343. **Chloroquine diphosphate in two different dosages as adjunctive therapy of hospitalized patients with severe respiratory syndrome in the context of coronavirus (SARS-CoV-2) infection: Preliminary safety results of a randomized, double-blinded, phase IIb clinical trial (CloroCovid-19 Study)** Mayla Gabriela Silva Borba, Fernando Fonseca Almeida Val, Vanderson Souza Sampaio, Marcia Almeida Araújo Alexandre, Gisely Cardoso Melo, Marcelo Brito, Maria Paula Gomes Mourão, José Diego Brito-Sousa, Djane Baía-da-Silva, Marcus Vinitius Farias Guerra, … CloroCovid-19 Team *Cold Spring Harbor Laboratory* (2020-04-16) https://doi.org/ggr3nj DOI: 10.1101/2020.04.07.20056424

344. **Heart risk concerns mount around use of chloroquine and hydroxychloroquine for Covid-19 treatment** Jacqueline Howard, Elizabeth Cohen, Nadia Kounang, Per Nyberg *CNN* (2020-04-14) https://www.cnn.com/2020/04/13/health/chloroquine-risks-coronavirus-treatment-trials-study/index.html

345. **WHO Director-General's opening remarks at the media briefing on COVID-19** World Health Organization (2020-05-25) https://www.who.int/dg/speeches/detail/who-director-general-s-opening-remarks-at-the-media-briefing-on-covid-19---25-may-2020

346. **Hydroxychloroquine in patients mainly with mild to moderate COVID–19: an open–label, randomized, controlled trial** Wei Tang, Zhujun Cao, Mingfeng Han, Zhengyan Wang, Junwen Chen, Wenjin Sun, Yaojie Wu, Wei Xiao, Shengyong Liu, Erzhen Chen, … Qing Xie *Cold Spring Harbor Laboratory* (2020-05-07) https://doi.org/ggr68m DOI: 10.1101/2020.04.10.20060558

347. **Detection of SARS-CoV-2 in Different Types of Clinical Specimens** Wenling Wang, Yanli Xu, Ruqin Gao, Roujian Lu, Kai Han, Guizhen Wu, Wenjie Tan *JAMA* (2020-03-11) https://doi.org/ggpp6h DOI: 10.1001/jama.2020.3786 · PMID: 32159775 · PMCID: PMC7066521

348. **Outcomes of hydroxychloroquine usage in United States veterans hospitalized with Covid-19** Joseph Magagnoli, Siddharth Narendran, Felipe Pereira, Tammy Cummings, James W Hardin, SScott Sutton, Jayakrishna Ambati *Cold Spring Harbor Laboratory* (2020-04-21) https://doi.org/ggspt6 DOI: 10.1101/2020.04.16.20065920 · PMID: 32511622 · PMCID: PMC7276049

349. **Hydroxychloroquine for Early Treatment of Adults With Mild Coronavirus Disease 2019: A Randomized, Controlled Trial** Oriol Mitjà, Marc Corbacho-Monné, Maria Ubals, Cristian Tebé, Judith Peñafiel, Aurelio Tobias, Ester Ballana, Andrea Alemany, Núria Riera-Martí, Carla A Pérez, … Martí Vall-Mayans *Clinical Infectious Diseases* (2020-07-16) https://doi.org/gg5f9x DOI: 10.1093/cid/ciaa1009 · PMID: 32674126 · PMCID: PMC7454406

350. **A Randomized Trial of Hydroxychloroquine as Postexposure Prophylaxis for Covid-19** David R Boulware, Matthew F Pullen, Ananta S Bangdiwala, Katelyn A Pastick, Sarah M Lofgren, Elizabeth C Okafor, Caleb P Skipper, Alanna A Nascene, Melanie R Nicol, Mahsa Abassi, … Kathy H Hullsiek *New England Journal of Medicine* (2020-08-06) https://doi.org/dxkv DOI: 10.1056/nejmoa2016638 · PMID: 32492293 · PMCID: PMC7289276

351. **Efficacy and Safety of Hydroxychloroquine vs Placebo for Pre-exposure SARS-CoV-2 Prophylaxis Among Health Care Workers** Benjamin S Abella, Eliana L Jolkovsky, Barbara T Biney, Julie E Uspal, Matthew C Hyman, Ian Frank, Scott E Hensley, Saar Gill, Dan T Vogl, Ivan Maillard, … Prevention and Treatment of COVID-19 With Hydroxychloroquine (PATCH) Investigators *JAMA Internal Medicine* (2021-02-01) https://doi.org/ghd6nj DOI: 10.1001/jamainternmed.2020.6319 · PMID: 33001138 · PMCID: PMC7527945

352. **Mechanisms of action of hydroxychloroquine and chloroquine: implications for rheumatology** Eva Schrezenmeier, Thomas Dörner *Nature Reviews Rheumatology* (2020-02-07) https://doi.org/ggzjnh DOI: 10.1038/s41584-020-0372-x · PMID: 32034323

353. **Elimination or Prolongation of ACE Inhibitors and ARB in Coronavirus Disease 2019 - Full Text View - ClinicalTrials.gov** https://clinicaltrials.gov/ct2/show/NCT04338009

354. **Stopping ACE-inhibitors in COVID-19 - Full Text View - ClinicalTrials.gov** https://clinicaltrials.gov/ct2/show/NCT04353596

355. **Losartan for Patients With COVID-19 Not Requiring Hospitalization - Full Text View - ClinicalTrials.gov** https://clinicaltrials.gov/ct2/show/NCT04311177

356. **Losartan for Patients With COVID-19 Requiring Hospitalization - Full Text View - ClinicalTrials.gov** https://clinicaltrials.gov/ct2/show/NCT04312009

357. **The CORONAvirus Disease 2019 Angiotensin Converting Enzyme Inhibitor/Angiotensin Receptor Blocker InvestigatiON (CORONACION) Randomized Clinical Trial** Prof John William McEvoy *clinicaltrials.gov* (2020-06-26) https://clinicaltrials.gov/ct2/show/NCT04330300

358. **Ramipril for the Treatment of COVID-19 - Full Text View - ClinicalTrials.gov** https://clinicaltrials.gov/ct2/show/NCT04366050

359. **Suspension of Angiotensin Receptor Blockers and Angiotensin-converting Enzyme Inhibitors and Adverse Outcomes in Hospitalized Patients With Coronavirus Infection (COVID-19). A Randomized Trial** D'Or Institute for Research and Education *clinicaltrials.gov* (2020-07-01) https://clinicaltrials.gov/ct2/show/NCT04364893

360. **Response by Cohen et al to Letter Regarding Article, "Association of Inpatient Use of Angiotensin-Converting Enzyme Inhibitors and Angiotensin II Receptor Blockers With Mortality Among Patients With Hypertension Hospitalized With COVID-19"** Jordana B Cohen, Thomas C Hanff, Andrew M South, Matthew A Sparks, Swapnil Hiremath, Adam P Bress, JBrian Byrd, Julio A Chirinos *Circulation Research* (2020-06-05) https://doi.org/gg3xsg DOI: 10.1161/circresaha.120.317205 · PMID: 32496917 · PMCID: PMC7265880

361. **Sound Science before Quick Judgement Regarding RAS Blockade in COVID-19** Matthew A Sparks, Andrew South, Paul Welling, JMatt Luther, Jordana Cohen, James Brian Byrd, Louise M Burrell, Daniel Batlle, Laurie Tomlinson, Vivek Bhalla, … Swapnil Hiremath *Clinical Journal of the American Society of Nephrology* (2020-05-07) https://doi.org/ggq8gn DOI: 10.2215/cjn.03530320 · PMID: 32220930 · PMCID: PMC7269218

362. **The Coronavirus Conundrum: ACE2 and Hypertension Edition** Matthew Sparks, Swapnil Hiremath *NephJC* http://www.nephjc.com/news/covidace2

363. **Hall of Fame among Pro-inflammatory Cytokines: Interleukin-6 Gene and Its Transcriptional Regulation Mechanisms** Yang Luo, Song Guo Zheng *Frontiers in Immunology* (2016-12-19) https://doi.org/ggqmgv DOI: 10.3389/fimmu.2016.00604 · PMID: 28066415 · PMCID: PMC5165036

364. **IL-6 Trans-Signaling via the Soluble IL-6 Receptor: Importance for the Pro-Inflammatory Activities of IL-6** Stefan Rose-John *International Journal of Biological Sciences* (2012) https://doi.org/f4c4hf DOI: 10.7150/ijbs.4989 · PMID: 23136552 · PMCID: PMC3491447

365. **Interleukin-6 and its receptor: from bench to bedside** Jürgen Scheller, Stefan Rose-John *Medical Microbiology and Immunology* (2006-05-31) https://doi.org/ck8xch DOI: 10.1007/s00430-006-0019-9 · PMID: 16741736

366. **Plasticity and cross-talk of Interleukin 6-type cytokines** Christoph Garbers, Heike M Hermanns, Fred Schaper, Gerhard Müller-Newen, Joachim Grötzinger, Stefan Rose-John, Jürgen Scheller *Cytokine & Growth Factor Reviews* (2012-06) https://doi.org/f3z743 DOI: 10.1016/j.cytogfr.2012.04.001 · PMID: 22595692

367. **Soluble receptors for cytokines and growth factors: generation and biological function** S Rose-John, PC Heinrich *Biochemical Journal* (1994-06-01) https://doi.org/ggqmgd DOI: 10.1042/bj3000281 · PMID: 8002928 · PMCID: PMC1138158

368. **Interleukin-6; pathogenesis and treatment of autoimmune inflammatory diseases** Toshio Tanaka, Masashi Narazaki, Kazuya Masuda, Tadamitsu Kishimoto *Inflammation and Regeneration* (2013) https://doi.org/ggqmgt DOI: 10.2492/inflammregen.33.054

369. **Into the Eye of the Cytokine Storm** JR Tisoncik, MJ Korth, CP Simmons, J Farrar, TR Martin, MG Katze *Microbiology and Molecular Biology Reviews* (2012-03-05) https://doi.org/f4n9h2 DOI: 10.1128/mmbr.05015-11 · PMID: 22390970 · PMCID: PMC3294426

370. **Systematic Review and Meta-Analysis of Case-Control Studies from 7,000 COVID-19 Pneumonia Patients Suggests a Beneficial Impact of Tocilizumab with Benefit Most Evident in Non-Corticosteroid Exposed Subjects.** Abdulla Watad, Nicola Luigi Bragazzi, Charlie Bridgewood, Muhammad Mansour, Naim Mahroum, Matteo Riccò, Ahmed Nasr, Amr Hussein, Omer Gendelman, Yehuda Shoenfeld, … Dennis McGonagle *SSRN Electronic Journal* (2020) https://doi.org/gg62hz DOI: 10.2139/ssrn.3642653

371. **The efficacy of IL-6 inhibitor Tocilizumab in reducing severe COVID-19 mortality: a systematic review** Avi Gurion Kaye, Robert Siegel *PeerJ* (2020-11-02) https://doi.org/ghx8r4 DOI: 10.7717/peerj.10322 · PMID: 33194450 · PMCID: PMC7643559

372. **Rationale and evidence on the use of tocilizumab in COVID-19: a systematic review** A Cortegiani, M Ippolito, M Greco, V Granone, A Protti, C Gregoretti, A Giarratano, S Einav, M Cecconi *Pulmonology* (2021-01) https://doi.org/gg5xv3 DOI: 10.1016/j.pulmoe.2020.07.003 · PMID: 32713784 · PMCID: PMC7369580

373. **New insights and long-term safety of tocilizumab in rheumatoid arthritis** Graeme Jones, Elena Panova *Therapeutic Advances in Musculoskeletal Disease* (2018-10-07) https://doi.org/gffsdt DOI: 10.1177/1759720x18798462 · PMID: 30327685 · PMCID: PMC6178374

374. **Tocilizumab during pregnancy and lactation: drug levels in maternal serum, cord blood, breast milk and infant serum** Jumpei Saito, Naho Yakuwa, Kayoko Kaneko, Chinatsu Takai, Mikako Goto, Ken Nakajima, Akimasa Yamatani, Atsuko Murashima *Rheumatology* (2019-08) https://doi.org/ggzhks DOI: 10.1093/rheumatology/kez100 · PMID: 30945743

375. **Short-course tocilizumab increases risk of hepatitis B virus reactivation in patients with rheumatoid arthritis: a prospective clinical observation** Le-Feng Chen, Ying-Qian Mo, Jun Jing, Jian-Da Ma, Dong-Hui Zheng, Lie Dai *International Journal of Rheumatic Diseases* (2017-07) https://doi.org/f9pbc5 DOI: 10.1111/1756-185x.13010 · PMID: 28160426

376. **Why tocilizumab could be an effective treatment for severe COVID-19?** Binqing Fu, Xiaoling Xu, Haiming Wei *Journal of Translational Medicine* (2020-04-14) https://doi.org/ggv5c8 DOI: 10.1186/s12967-020-02339-3 · PMID: 32290839 · PMCID: PMC7154566

377. **Risk of adverse events including serious infections in rheumatoid arthritis patients treated with tocilizumab: a systematic literature review and meta-analysis of randomized controlled trials** L Campbell, C Chen, SS Bhagat, RA Parker, AJK Ostor *Rheumatology* (2010-11-14) https://doi.org/crqn7c DOI: 10.1093/rheumatology/keq343 · PMID: 21078627

378. **Risk of serious infections in tocilizumab versus other biologic drugs in patients with rheumatoid arthritis: a multidatabase cohort study** Ajinkya Pawar, Rishi J Desai, Daniel H Solomon, Adrian J Santiago Ortiz, Sara Gale, Min Bao, Khaled Sarsour, Sebastian Schneeweiss, Seoyoung C Kim *Annals of the Rheumatic Diseases* (2019-04) https://doi.org/gg62hx DOI: 10.1136/annrheumdis-2018-214367 · PMID: 30679153

379. **Risk of infections in rheumatoid arthritis patients treated with tocilizumab** Veronika R Lang, Matthias Englbrecht, Jürgen Rech, Hubert Nüsslein, Karin Manger, Florian Schuch, Hans-Peter Tony, Martin Fleck, Bernhard Manger, Georg Schett, Jochen Zwerina *Rheumatology* (2012-05) https://doi.org/d3b3rh DOI: 10.1093/rheumatology/ker223 · PMID: 21865281

380. **Use of Tocilizumab for COVID-19-Induced Cytokine Release Syndrome** Jared Radbel, Navaneeth Narayanan, Pinki J Bhatt *Chest* (2020-07) https://doi.org/ggtxvs DOI: 10.1016/j.chest.2020.04.024 · PMID: 32343968 · PMCID: PMC7195070

381. **Incidence of ARDS and outcomes in hospitalized patients with COVID-19: a global literature survey** Susan J Tzotzos, Bernhard Fischer, Hendrik Fischer, Markus Zeitlinger *Critical Care* (2020-08-21) https://doi.org/gh294r DOI: 10.1186/s13054-020-03240-7 · PMID: 32825837 · PMCID: PMC7441837

382. **The Efficacy of IL-6 Inhibitor Tocilizumab in Reducing Severe COVID-19 Mortality: A Systematic Review** Avi Kaye, Robert Siegel *Cold Spring Harbor Laboratory* (2020-07-14) https://doi.org/gg62hv DOI: 10.1101/2020.07.10.20150938

383. **Utilizing tocilizumab for the treatment of cytokine release syndrome in COVID-19** Ali Hassoun, Elizabeth Dilip Thottacherry, Justin Muklewicz, Qurrat-ul-ain Aziz, Jonathan Edwards *Journal of Clinical Virology* (2020-07) https://doi.org/ggx359 DOI: 10.1016/j.jcv.2020.104443 · PMID: 32425661 · PMCID: PMC7229471

384. **The antiviral effect of interferon-beta against SARS-Coronavirus is not mediated by MxA protein** Martin Spiegel, Andreas Pichlmair, Elke Mühlberger, Otto Haller, Friedemann Weber *Journal of Clinical Virology* (2004-07) https://doi.org/cmc3ds DOI: 10.1016/j.jcv.2003.11.013 · PMID: 15135736

385. **Coronavirus virulence genes with main focus on SARS-CoV envelope gene** Marta L DeDiego, Jose L Nieto-Torres, Jose M Jimenez-Guardeño, Jose A Regla-Nava, Carlos Castaño-Rodriguez, Raul Fernandez-Delgado, Fernando Usera, Luis Enjuanes *Virus Research* (2014-12) https://doi.org/f6wm24 DOI: 10.1016/j.virusres.2014.07.024 · PMID: 25093995 · PMCID: PMC4261026

386. **Synairgen to start trial of SNG001 in COVID-19 imminently** Synairgen plc press release (2020-03-18) http://synairgen.web01.hosting.bdci.co.uk/umbraco/Surface/Download/GetFile?cid=23c9b12c-508b-48c3-9081-36605c5a9ccd

387. **Nebulised interferon beta-1a for patients with COVID-19** Nathan Peiffer-Smadja, Yazdan Yazdanpanah *The Lancet Respiratory Medicine* (2021-02) https://doi.org/ftmj DOI: 10.1016/s2213-2600(20)30523-3 · PMID: 33189160 · PMCID: PMC7833737

388. **Effect of Intravenous Interferon β-1a on Death and Days Free From Mechanical Ventilation Among Patients With Moderate to Severe Acute Respiratory Distress Syndrome** VMarco Ranieri, Ville Pettilä, Matti K Karvonen, Juho Jalkanen, Peter Nightingale, David Brealey, Jordi Mancebo, Ricard Ferrer, Alain Mercat, Nicolò Patroniti, … for the INTEREST Study Group *JAMA* (2020-02-25) https://doi.org/ghzkww DOI: 10.1001/jama.2019.22525 · PMID: 32065831

389. **A Randomized Clinical Trial of the Efficacy and Safety of Interferon β-1a in Treatment of Severe COVID-19** Effat Davoudi-Monfared, Hamid Rahmani, Hossein Khalili,

Mahboubeh Hajiabdolbaghi, Mohamadreza Salehi, Ladan Abbasian, Hossein Kazemzadeh, Mir Saeed Yekaninejad *Antimicrobial Agents and Chemotherapy* (2020-08-20) https://doi.org/gg5xvm DOI: 10.1128/aac.01061-20 · PMID: 32661006 · PMCID: PMC7449227

390. **A Multicenter, Adaptive, Randomized Blinded Controlled Trial of the Safety and Efficacy of Investigational Therapeutics for the Treatment of COVID-19 in Hospitalized Adults (ACTT-3)** National Institute of Allergy and Infectious Diseases (NIAID) *clinicaltrials.gov* (2021-02-04) https://clinicaltrials.gov/ct2/show/NCT04492475

391. **Tocilizumab (Actemra): Adult Patients with Moderately to Severely Active Rheumatoid Arthritis** Canadian Agency for Drugs and Technologies in Health *CADTH Common Drug Reviews* (2015-08) https://www.ncbi.nlm.nih.gov/books/NBK349513/table/T43/

392. **A Cost Comparison of Treatments of Moderate to Severe Psoriasis** Cheryl Hankin, Steven Feldman, Andy Szczotka, Randolph Stinger, Leslie Fish, David Hankin *Drug Benefit Trends* (2005-05) https://escholarship.umassmed.edu/meyers_pp/385

393. **TNF-α inhibition for potential therapeutic modulation of SARS coronavirus infection** Edward Tobinick *Current Medical Research and Opinion* (2008-09-22) https://doi.org/bq4cx2 DOI: 10.1185/030079903125002757 · PMID: 14741070

394. **Sanofi and Regeneron begin global Kevzara® (sarilumab) clinical trial program in patients with severe COVID-19** Sanofi (2020-03-16) http://www.news.sanofi.us/2020-03-16-Sanofi-and-Regeneron-begin-global-Kevzara-R-sarilumab-clinical-trial-program-in-patients-with-severe-COVID-19

395. **Sarilumab COVID-19 - Full Text View - ClinicalTrials.gov** https://clinicaltrials.gov/ct2/show/NCT04327388

396. **COVID-19: combining antiviral and anti-inflammatory treatments** Justin Stebbing, Anne Phelan, Ivan Griffin, Catherine Tucker, Olly Oechsle, Dan Smith, Peter Richardson *The Lancet Infectious Diseases* (2020-04) https://doi.org/dph5 DOI: 10.1016/s1473-3099(20)30132-8 · PMID: 32113509 · PMCID: PMC7158903

397. **Baricitinib as potential treatment for 2019-nCoV acute respiratory disease** Peter Richardson, Ivan Griffin, Catherine Tucker, Dan Smith, Olly Oechsle, Anne Phelan, Michael Rawling, Edward Savory, Justin Stebbing *The Lancet* (2020-02) https://doi.org/ggnrsx DOI: 10.1016/s0140-6736(20)30304-4 · PMID: 32032529 · PMCID: PMC7137985

398. **Lilly Begins Clinical Testing of Therapies for COVID-19 | Eli Lilly and Company** https://investor.lilly.com/news-releases/news-release-details/lilly-begins-clinical-testing-therapies-covid-19

399. **Baricitinib Combined With Antiviral Therapy in Symptomatic Patients Infected by COVID-19: an Open-label, Pilot Study** Fabrizio Cantini *clinicaltrials.gov* (2020-04-19) https://clinicaltrials.gov/ct2/show/NCT04320277

400. **Design and Synthesis of Hydroxyferroquine Derivatives with Antimalarial and Antiviral Activities** Christophe Biot, Wassim Daher, Natascha Chavain, Thierry Fandeur, Jamal Khalife, Daniel Dive, Erik De Clercq *Journal of Medicinal Chemistry* (2006-05) https://doi.org/db4n83 DOI: 10.1021/jm0601856 · PMID: 16640347

401. **An orally bioavailable broad-spectrum antiviral inhibits SARS-CoV-2 in human airway epithelial cell cultures and multiple coronaviruses in mice** Timothy P Sheahan, Amy C Sims, Shuntai Zhou, Rachel L Graham, Andrea J Pruijssers, Maria L Agostini, Sarah R Leist, Alexandra Schäfer, Kenneth H Dinnon, Laura J Stevens, … Ralph S Baric *Science Translational Medicine* (2020-04-29) https://doi.org/ggrqd2 DOI: 10.1126/scitranslmed.abb5883 · PMID: 32253226 · PMCID: PMC7164393

402. **Antiviral Monoclonal Antibodies: Can They Be More Than Simple Neutralizing Agents?** Mireia Pelegrin, Mar Naranjo-Gomez, Marc Piechaczyk *Trends in Microbiology* (2015-10) https://doi.org/f7vzrf DOI: 10.1016/j.tim.2015.07.005 · PMID: 26433697 · PMCID: PMC7127033

403. **Molecular biology of the cell** Bruce Alberts (editor) *Garland Science* (2002) ISBN: 9780815332183

404. **Intranasal Treatment with Poly(I{middle dot}C) Protects Aged Mice from Lethal Respiratory Virus Infections** J Zhao, C Wohlford-Lenane, J Zhao, E Fleming, TE Lane, PB McCray, S Perlman *Journal of Virology* (2012-08-22) https://doi.org/f4bzfp DOI: 10.1128/jvi.01410-12 · PMID: 22915814 · PMCID: PMC3486278